\documentclass[aps,prd,reprint,superscriptaddress,nobibnotes,nofootinbib,floatfix]{revtex4-2}
\usepackage{graphicx}
\usepackage{color}
\usepackage{orcidlink}
\usepackage{amsmath}
\usepackage{amssymb}
\usepackage{mathtools}
\usepackage{bbm}
\usepackage{braket}
\usepackage{xspace}
\usepackage[caption=false]{subfig}
\makeatletter
\AtBeginDocument{\let\sf@orig@arrayparboxrestore\@arrayparboxrestore
 	\def\@arrayparboxrestore{\let\sf@saved@label\label \sf@orig@arrayparboxrestore
		\ifx\sf@saved@label\subfloat@label \let\label\subfloat@label \fi}}
\makeatother
\usepackage{slashed}
\usepackage{pgfplots}
\usepackage{cleveref}
\crefname{equation}{Eq.}{Eqs.}
\Crefname{equation}{Equation}{Equations}
\crefname{figure}{Fig.}{Figs.}
\Crefname{figure}{Figure}{Figures}
\crefname{table}{Table}{Tables}
\Crefname{table}{Table}{Tables}
\crefname{section}{Sec.}{Secs.}
\Crefname{section}{Section}{Sections}
\crefname{subsection}{Sec.}{Secs.}
\Crefname{subsection}{Section}{Sections}
\crefname{appendix}{Appendix}{Appendixes}
\Crefname{appendix}{Appendix}{Appendixes}

\AddToHook{cmd/appendix/before}{%
	\crefalias{section}{appendix}%
	\crefalias{subsection}{appendix}%
}
\newcommand{\matel}{\ensuremath{\tilde{\mathcal{M}}}\xspace}
\newcommand{\matelpdf}{\ensuremath{\matel_\text{PDF}}\xspace}
\newcommand{\matellcda}{\ensuremath{\matel_\text{LCDA}}\xspace}
\newcommand{\mel}{\ensuremath{\mathcal{M}}\xspace}
\newcommand{\mellat}{\ensuremath{\mathcal{M}^\text{lat}}\xspace}
\newcommand{\op}{\ensuremath{\widehat{O}}\xspace}
\newcommand{\oplat}{\ensuremath{O^\text{lat}}\xspace}
\newcommand{\psilat}{\ensuremath{\varphi}\xspace}
\newcommand{\te}[1][]{\ensuremath{e^{-i H {#1}}}\xspace}
\newcommand{\finalsite}{\ensuremath{w}\xspace}
\newcommand{\startsite}{\ensuremath{v}\xspace}
\newcommand{\site}{\ensuremath{j}\xspace}
\newcommand{\sitesum}{\ensuremath{l}\xspace}
\newcommand{\siteeven}{\ensuremath{n}\xspace}
\newcommand{\z}{\siteeven}
\newcommand{\knum}{\ensuremath{q}\xspace}
\newcommand{\normfact}{\ensuremath{{\alpha}}\xspace}
\newcommand{\normfactpdf}{\ensuremath{\normfact^2}\xspace}
\newcommand{\normfactlcda}{\ensuremath{\normfact}\xspace}
\newcommand{\spinind}{\ensuremath{s}\xspace}
\newcommand{\dtrotter}{\ensuremath{\tau}\xspace}
\newcommand{\Nt}{\ensuremath{N_{\dtrotter}}\xspace}
\newcommand{\dt}{\ensuremath{\Delta t}\xspace}
\newcommand{\Lcut}{\ensuremath{L_{\text{cut}}}\xspace}
\newcommand{\energy}{\ensuremath{M}\xspace}
\newcommand{\encont}{\ensuremath{P^0}\xspace}
\newcommand{\bjorx}{{\ensuremath{\xi}}\xspace}
\newcommand{\pdf}[1][]{\ensuremath{f^{#1}_{\psi}\left(\bjorx\right)}\xspace}
\newcommand{\pdfp}[1][]{\ensuremath{f^{#1}_{p}\left(\bjorx\right)}\xspace}
\newcommand{\pdfbare}{\ensuremath{f_{\psi}}\xspace}
\newcommand{\antipdf}{\ensuremath{f_{\overline{\psi}}\left(\bjorx\right)}\xspace}
\newcommand{\pdfantipdf}{\ensuremath{f_{\psi/\overline{\psi}}\left(\bjorx\right)}\xspace}
\newcommand{\uscale}{\ensuremath{u}\xspace}
\newcommand{\zlight}{\ensuremath{z^{-}}\xspace}
\newcommand{\zmax}{\ensuremath{z_m}\xspace}
\newcommand{\Plight}{\ensuremath{P^+}\xspace}
\newcommand{\lcda}{\ensuremath{\phi}\xspace}
\newcommand{\lcdau}{\ensuremath{\phi\left(\uscale\right)}\xspace}
\newcommand{\meson}{\ensuremath{h}\xspace}
\newcommand{\deconst}{\ensuremath{F_\meson}\xspace}
\newcommand{\normpsi}{\ensuremath{N_{\psi}}\xspace}
\newcommand{\normpsibar}{\ensuremath{N_{\overline{\psi}}}\xspace}
\newcommand{\normpsipsibar}{\ensuremath{N_{\psi / \overline{\psi}}}\xspace}
\newcommand{\mass}{\ensuremath{\tilde{m}}\xspace}
\newcommand{\vol}{\ensuremath{\tilde{V}}\xspace}
\newcommand{\N}{\ensuremath{N}\xspace}
\renewcommand{\Re}{\operatorname{Re}}
\renewcommand{\Im}{\operatorname{Im}}
\newcommand{\wilson}[2]{\ensuremath{W_{{#1} \leftarrow {#2}}}}
\newcommand{\wilsondagger}[2]{\ensuremath{W^{*}_{{#1} \leftarrow {#2}}}}
\newcommand{\gammas}{\ensuremath{\gamma_s}\xspace}
\newcommand{\Charge}{\ensuremath{\hat{C}}\xspace}
\newcommand{\Parity}{\ensuremath{\hat{P}}\xspace}
\newcommand{\Time}{\ensuremath{\hat{T}}\xspace}
\newcommand{\PT}{\ensuremath{\widehat{PT}}\xspace}
\newcommand{\eye}{\mathrm{1}\mkern-4mu\text{l}}
\newcommand{\order}[1]{\ensuremath{\mathcal{O}\left(#1\right)}\xspace}
\newcommand{\ordersym}{\ensuremath{\mathcal{O}}\xspace}
\newcommand{\trace}[1]{\ensuremath{\mathrm{Tr}\left(#1\right)}\xspace}
\newcommand{\marktext}[1]{\textbf{\underline{#1}}}
\usetikzlibrary{external}
\usepgfplotslibrary{external}
\usepgfplotslibrary{fillbetween}

\pgfplotsset{compat=newest}
\usetikzlibrary{arrows.meta}

\tikzset{every node/.style={sloped,allow upside down},baseline={([yshift=-0.5ex]current bounding box.center)},inner sep=.3mm,x=1cm,y=1cm}

\newcommand{\tikzlinewidth}{1.5pt}
\newcommand{\tikzlinewidtholdpdf}{\tikzlinewidth}
\newcommand{\tikzmarksize}{3.0pt}

\definecolor{mycolor2}{HTML}{E40303}%red
\definecolor{mycolor3}{HTML}{FF8C00}%orange
\definecolor{mycolor1}{HTML}{008026}%green
\definecolor{mycolor4}{HTML}{004DFF}%blue
\definecolor{mycolor5}{HTML}{750787}%purple
\definecolor{mycolor6}{HTML}{613915}%brown
\definecolor{mycolor7}{rgb}{0.400,0.612,0.086}%green
\definecolor{mycolor8}{HTML}{00C0C0}%dark bright blue
\definecolor{mycolor9}{HTML}{000000}%black
\definecolor{mycolor10}{HTML}{CC8CA0}%dark pink
\definecolor{mycolor11}{HTML}{FFAFC8}%pink
\definecolor{mycolorblack}{rgb}{0,0,0}%black

\newcommand{\lightercolor}[3]{% Reference Color, Percentage, New Color Name
	\colorlet{#3}{#1!#2!white}
}
\lightercolor{mycolor1}{50}{mycolor1Light}
\lightercolor{mycolor2}{50}{mycolor2Light}
\lightercolor{mycolor3}{50}{mycolor3Light}
\lightercolor{mycolor4}{50}{mycolor4Light}
\lightercolor{mycolor5}{50}{mycolor5Light}
\lightercolor{mycolor7}{50}{mycolor7Light}
\lightercolor{mycolor11}{50}{mycolor11Light}
\tikzset{myline1/.style={color=mycolor1, line width=\tikzlinewidth, mark size=\tikzmarksize, mark=x, mark options={solid, mycolor1}}}
\tikzset{myline2/.style={color=mycolor2, line width=\tikzlinewidth, mark size=\tikzmarksize, mark=o, mark options={solid, mycolor2}}}
\tikzset{myline3/.style={color=mycolor3, line width=\tikzlinewidth, mark size=\tikzmarksize, mark=square, mark options={solid, mycolor3}}}
\tikzset{myline4/.style={color=mycolor4, line width=\tikzlinewidth, mark size=\tikzmarksize, mark=diamond, mark options={solid, mycolor4}}}
\tikzset{myline5/.style={color=mycolor5, line width=\tikzlinewidth, mark size=\tikzmarksize, mark=triangle, mark options={solid, mycolor5}}}
\tikzset{myline6/.style={color=mycolor6, line width=\tikzlinewidth, mark size=\tikzmarksize, mark=+, mark options={solid, mycolor6}}}
\tikzset{myline7/.style={color=mycolor7, dotted, line width=\tikzlinewidth, mark size=\tikzmarksize, mark=pentagon, mark options={solid, mycolor7}}}
\tikzset{myline8/.style={color=mycolor8, dashed, line width=\tikzlinewidth, mark size=\tikzmarksize, mark=+, mark options={solid, mycolor8}}}
\tikzset{myline9/.style={mycolor9, solid, line width=\tikzlinewidth, mark size=\tikzmarksize, mark=triangle, mark options={solid, mycolor9, rotate=180}}}
\tikzset{myline10/.style={color=mycolor10, dotted, line width=\tikzlinewidth, mark size=\tikzmarksize, mark=pentagon, mark options={solid, mycolor10}}}
\tikzset{myline11/.style={color=mycolor11, solid, line width=\tikzlinewidth, mark size=\tikzmarksize, mark=pentagon, mark options={solid, mycolor11, rotate=180}}}
\tikzset{mylineblack/.style={color=mycolorblack, line width=\tikzlinewidth, mark size=\tikzmarksize, mark=x, mark options={solid, mycolorblack}}}

\tikzset{myphantomline1/.style={color=mycolor1Light, line width=0pt, mark size=0.0pt, mark=none}}
\tikzset{myphantomline2/.style={color=mycolor2Light, line width=0pt, mark size=0.0pt, mark=none}}
\tikzset{myphantomline3/.style={color=mycolor3Light, line width=0pt, mark size=0.0pt, mark=none}}
\tikzset{myphantomline4/.style={color=mycolor4Light, line width=0pt, mark size=0.0pt, mark=none}}
\tikzset{myphantomline5/.style={color=mycolor5Light, line width=0pt, mark size=0.0pt, mark=none}}
\tikzset{myphantomline11/.style={color=mycolor11Light, line width=0pt, mark size=0.0pt, mark=none}}
\tikzset{myarea1/.style={color=mycolor1Light}}
\tikzset{myarea2/.style={color=mycolor2Light}}
\tikzset{myarea3/.style={color=mycolor3Light}}
\tikzset{myarea4/.style={color=mycolor4Light}}
\tikzset{myarea5/.style={color=mycolor5Light}}
\tikzset{myarea11/.style={color=mycolor11Light}}
\pgfplotsset{
	/pgfplots/error bar legend/.style={
		legend image code/.code={
			\draw[sharp plot,mark=-,mark repeat=2,mark phase=1,##1]
			plot coordinates { (0.3cm, -0.15cm) (0.3cm,0cm) (0.3cm, 0.15cm) };%
			\draw[mark repeat=2,mark phase=2,##1]
			plot coordinates {(0cm,0cm) (0.3cm,0cm) (0.6cm,0cm)};%
}}}

\pgfplotsset{
	/pgfplots/error band legend/.style={
		legend image code/.code={
			\draw[fill=#1, draw=none] (-0cm,-0.15cm) rectangle ++(0.6cm,0.3cm);
			\draw[sharp plot,mark=-##1] (0cm,0cm) -- (0.6cm,0cm);%
}}}

\tikzset{timearrow/.style={->,line width=0.5mm,color=mycolor2}}
\tikzset{spacearrow/.style={->,line width=0.5mm,color=mycolor4}}
\tikzset{spacedoublearrow/.style={<->,line width=0.5mm,color=mycolor4}}
\tikzset{fermion/.style={circle,draw=black,thick,fill=mycolor1,minimum size=2mm}}
\tikzset{antifermion/.style={circle,draw=black,thick,fill=mycolor3,minimum size=2mm}}
\begin{document}

%%%%%%%%%%%%%%%%%%%%%%%%%%%%%%%%%%%%%%%%%%%%%%%%%%%%%%%%%
%                           TITLE PAGE
%%%%%%%%%%%%%%%%%%%%%%%%%%%%%%%%%%%%%%%%%%%%%%%%%%%%%%%%%
\title{
Parton Physics in the Schwinger Model from Matrix Product States
}
\author{\firstname{Mari~Carmen} \surname{Ba\~{n}uls}\orcidlink{0000-0001-6419-6610}}
\affiliation{Max Planck Institute of Quantum Optics, Garching, Germany}

\author{\firstname{Krzysztof} \surname{Cichy}\orcidlink{0000-0002-5705-3256}}
\affiliation{Faculty of Physics and Astronomy, Adam Mickiewicz University, Pozna\'{n}, Poland}

\author{\firstname{C.-J.~David} \surname{Lin}\orcidlink{0000-0003-3743-0840}}
\affiliation{Institute of Physics, National Yang Ming Chiao Tung University, Hsinchu, Taiwan}
\affiliation{Center for Theoretical and Computational Physics, National Yang Ming Chiao Tung University, Hsinchu, Taiwan}
\affiliation{Center for High Energy Physics, Chung-Yuan Christian University, Taoyuan, Taiwan}
\affiliation{Physics Division, National Center for Theoretical Sciences, Taipei, Taiwan}

\author{\firstname{Manuel} \surname{Schneider}\orcidlink{0000-0001-9348-8700}}
\email{manuel.schneider@nycu.edu.tw}
\affiliation{Institute of Physics, National Yang Ming Chiao Tung University, Hsinchu, Taiwan}
\affiliation{Center for Theoretical and Computational Physics, National Yang Ming Chiao Tung University, Hsinchu, Taiwan}

\date{November 23, 2026}

\begin{abstract}
	Parton distribution functions (PDFs) and light-cone distribution amplitudes (LCDAs) encode the non-perturbative structure of hadrons in terms of their partonic constituents. They are defined through matrix elements containing parton operators that are connected by a Wilson line along a light-front direction, which makes them difficult to access in Euclidean lattice calculations. We demonstrate how parton functions can be directly obtained in Minkowski space using the Hamiltonian framework. We compute fermion and antifermion PDFs, LCDAs, and decay constants of the vector meson in the Schwinger model for several fermion masses. States in Hilbert space are efficiently represented by tensor networks, with an expressivity that is systematically controlled by a bond dimension. The light-front Wilson line is implemented as a stepwise evolution on the lattice that alternates spatial shifts of an external charge with Trotterized time-evolution steps acting on a matrix product state. All systematic uncertainties are quantified by varying and extrapolating in the lattice spacing, volume, bond dimension, gauge-field truncation, and Trotter step width. The slow growth of the entanglement entropy makes moderate bond dimensions sufficient, which supports the feasibility of using tensor networks for computing light-front matrix elements in more general settings. We confirm that the computed distributions fulfill the expected normalizations, momentum sum rules, and symmetry properties.
\end{abstract}


\maketitle

%%%%%%%%%%%%%%%%%%%%%%%%%%%%%%%%%%%%%%%%%%%%%%%%%%%%%%%%%
%                           BODY
%%%%%%%%%%%%%%%%%%%%%%%%%%%%%%%%%%%%%%%%%%%%%%%%%%%%%%%%%

\section{Introduction}
\label{sec:introduction}
%!TEX root = ./Schwinger_PDF_LCDA.tex

Hadrons are not elementary particles but are bound states of quarks and gluons. These constituents, the partons, interact by the strong force. Hadron structure can, therefore, not be addressed by an expansion in a small coupling constant with perturbation theory. This non-perturbative inner structure of hadrons can be described by parton functions. These provide information about the parton content itself as well as momentum and spin contributions of each constituent to the hadron. Parton functions arise in the factorization of scattering cross sections and decays, where the process-dependent perturbative part is convolved with the remaining non-perturbative part, which is described by universal (process-independent) parton functions~\cite{collins_2011}.

The collinear parton distribution function (PDF) describes the momentum distribution of a parton in a hadron. It depends on the Bjorken parameter \bjorx and corresponds, in the parton limit, to the probability density of a parton inside the hadron with momentum fraction \bjorx along the interaction direction in a scattering event. The precise knowledge of PDFs is paramount for hadron colliders, where they enter the data analysis at a fundamental level and where systematics can be dominated by PDF uncertainties~\cite{CMS2025,Bozzi2025}. At the same time, PDFs can be extracted from collider data. The upcoming Electron-Ion Collider (EIC)~\cite{AbdulKhalek:2021gbh} will measure them at unprecedented precision. Theoretical efforts are therefore needed to match experimental advances.

Another relevant parton-physics quantity is the meson light-cone distribution amplitude (LCDA), which arises in the theoretical description of decays and in the factorization of exclusive scattering processes~\cite{Lepage:1980fj}. It is generically denoted as $\lcdau$ and can be interpreted (in the light-cone gauge and the parton limit) as the overlap of the hadron with a state consisting of a quark and an antiquark carrying $\uscale$ and $1-\uscale$ fractions of the meson momentum, respectively.

Due to their non-perturbative nature, parton functions are a natural aim for calculations within lattice field theory. This is, however, a highly nontrivial task because contemporary lattice quantum chromodynamics (QCD) numerical computations are implemented using Monte Carlo methods for the evaluation of the path integral, leading to the need to formulate the theory in Euclidean space. Since parton functions such as PDFs and LCDAs involve non-perturbative QCD dynamics on the light front, it is challenging to extract them from calculations carried out in Euclidean space.

An alternative route to parton functions is the direct evaluation of light-front matrix elements in Minkowski space. This is possible in the Hamiltonian approach, where time dependence is accessible, and early works have computed parton distributions in the Schwinger model with exact diagonalization on small lattices and in strongly truncated approximations~\cite{BERGKNOFF1977215,MoPerry,KROGER199858,Eller1987,Eller1989,Jirari1999}. However, these methods cannot be systematically improved to large system sizes and fully controlled uncertainties. In more recent years, new, so-called quantum-inspired methods have emerged, in particular tensor networks and quantum computing/quantum simulations, that can address the Hamiltonian framework and potentially overcome these challenges.

This work tests the applicability of tensor networks for calculating light-front matrix elements and parton functions. As a first demonstration in a gauge theory, we calculate PDFs and LCDAs in the massive Schwinger model, or quantum electrodynamics in one spatial dimension ($\text{QED}_2$)~\cite{SchwingerOriginal,SchwingerModelMassive}. The model describes Dirac fermions coupled to gauge fields with a $U(1)$ symmetry. While significantly simpler than QCD, both theories share important features such as dynamical mass generation, confinement, and asymptotic freedom (the latter in a trivial sense, since the coupling constant is dimensionful)~\cite{SchwingerOriginal,SchwingerModelMassive,Coleman1976,Casher1974,Kogut1974}. The Schwinger model has therefore become a testbed for new algorithms and methods.

Tensor network states (TNS) are families of many-body states with an efficient parametrization in terms of local tensors that are connected by mutual indices~\cite{Verstraete2008,SCHOLLWOCK201196,Orus2014annphys,Bridgeman_2017,silvi2019tensor,Cirac2021,Okunishi2022,Banuls_TNA_RouteMap}. A bond dimension parameter $D$ controls the number of elements in the tensors and governs the expressivity of the ansatz. This way, the Hilbert space is truncated in a systematically improvable way, interpolating between a mean-field approximation ($D=1$) and an exact representation (where $D$ grows exponentially with the system size). In particular, ground and low-energy eigenstates of local Hamiltonians, as well as their thermal equilibrium states, host a bounded amount of entanglement, often satisfying an area law~\cite{Eisert2010,Hastings2007,wolf2008area,Arad2016lowE}, or small corrections thereof~\cite{Calabrese2004}, which makes them, in principle, amenable to efficient TNS descriptions~\cite{Cirac2021}. In contrast, under time evolution the amount of entanglement in the states may grow fast~\cite{Calabrese_2007,Eisert2006time,Schuch2008}, limiting the representability by TNS, especially at long times. Yet, efficient TNS algorithms exist to simulate time evolution that can be used to study regimes of moderate entanglement~\cite{garcia-ripoll2006,Paeckel2019,bauerfeind2020ttn,barthel2026cost,czarnik2019ipeps,dziamarga2022tangent,wu2026realtime}. Remarkably, symmetries of the problem can be naturally incorporated in the tensors~\cite{PerezGarcia2008string,Schuch2010peps,Singh2010global,Tagliacozzo:2014bta,Zohar2016,Haegeman2015gauge,kelman2024gpeps}. 

In the last decade, multiple works have applied TNS for the study of low-dimensional lattice gauge theories, mostly for ground states, spectra and thermal equilibrium properties, but addressing also real-time dynamics in specific setups [see, e.g., references in the reviews~\cite{banuls2020ropp,banuls2020epjd,meurice2022trg}, or more recent works in Refs.~\cite{Kadoh:2022Ia,felser2020u1,robaina2021z3,magnifico2021qed3d,emonts2023z2,kelman2024gpeps,Akiyama_2024,Magnifico:2024eiy,Wu:2025aly}]. Early works have demonstrated the feasibility of TNS to extract equilibrium properties of low-dimensional continuum theories. More recently, TNS have been used to explore various dynamical aspects of the lattice Schwinger model~\cite{Buyens:2016hhu,Rigobello:2021fxw,Belyansky:2023rgh,Papaefstathiou:2024zsu,Florio:2024aix,Janik:2025bbz,Florio:2025hoc,Barata:2025hgx,Shao:2025obi,Shao:2025ygy,Ikeda:2025bjb,2025arXiv251016101A,Grieninger2026}. 

Here, we work with matrix product states (MPS), which encode the entanglement structure of the one-dimensional area law. We use them to explore parton physics in the Schwinger model and extract the continuum limit of a dynamical TNS calculation, thus verifying that time-dependent quantities of field theories can also be obtained with TNS. Our approach is based on the calculation of light-front matrix elements from time-evolved MPS to directly extract parton functions through a Fourier transform. We note that TNS have been recently used to address parton physics, notably in the framework of quasi-distribution functions~\cite{Grieninger2024quasifragmentation,Grieninger2026}, and also for an effective model~\cite{Kang2025}. We build on our own earlier work~\cite{Schneider:2025gL,SchwingerPDFletter2025}, where we presented a method to obtain the vector meson PDF in the massive Schwinger model using MPS. In this paper, we provide more details on our method and present new results on this quantity, which help us further control systematic errors and validate our computation. The PDF is obtained with fully controlled uncertainties for several fermion masses. In addition, here we report our work on the meson LCDA.

This paper is structured as follows: \Cref{sec:partons} discusses the status of parton function determinations, introduces the PDF and the LCDA in general and in one spatial dimension in particular, and explains their properties and symmetries. \Cref{sec:model} develops how they are calculated in the Schwinger model from light-front matrix elements. The implementation of the light-like Wilson line by a stepwise evolution on the lattice is described in \cref{sec:Wilson_line}. Further simulation details are given in \cref{sec:simulation}, where we discuss how we obtain the spectrum, calculate matrix elements from time-evolved states, and obtain the parton functions by a Fourier transform with controlled uncertainties. The numerical results are shown in \cref{sec:results}, where we present matrix elements, PDFs, LCDAs, and decay constants for several fermion masses, and examine the entanglement structure during the time evolution. Finally, we summarize our work and give an outlook in \cref{sec:conclusion}.

\section{Parton functions}
\label{sec:partons}
%!TEX root = ./Schwinger_PDF_LCDA.tex
\subsection{Current status}

Parton physics has been one of the central subjects in particle and nuclear physics since the discovery of Bjorken scaling~\cite{Bjorken:1968dy} in the experiments of deep inelastic scattering (DIS). Research on this topic has seen an impressive amount of progress in the last six decades. In recent years, with the construction of the EIC~\cite{AbdulKhalek:2021gbh}, it has been attracting a particularly significant amount of attention. In addition to the longitudinal PDFs, the EIC experiments will allow for more detailed information on hadron structure. This includes the generalized parton distributions (GPDs)~\cite{Diehl:2003ny}, which describe the profile of the longitudinal momentum in the transverse plane, and the transverse-momentum-dependent PDFs (TMDPDFs)~\cite{Boussarie:2023izj}. The determination of these universal parton functions will allow for tomography of hadrons in the next few decades.

Although the EIC experiments will substantially enhance our knowledge of hadron structure, it should be noted that both perturbative and non-perturbative QCD theoretical calculations will remain essential for this research direction~\cite{Abir:2023fpo}. On the one hand, perturbation theory is a crucial tool for establishing factorization of processes, allowing for clean extractions of PDFs~\cite{collins_2011}. On the other hand, non-perturbative computations are often needed to complement experiments. For instance, the experimental determination of the pion PDF at the EIC will be performed using the Sullivan process~\cite{Sullivan:1971kd} in semi-inclusive DIS (SIDIS). In this process, the pion is in general off-shell. Therefore, a sound theoretical calculation of this PDF will serve as a good independent validation of results obtained from experimental methods. For such non-perturbative calculations, lattice field theory is the most reliable, first-principles approach.

The LCDA plays a central role in the investigation of exclusive processes, such as the study of pion and kaon form factors in meson-lepton elastic scattering~\cite{Lepage:1980fj} using QCD factorization. Measuring these form factors is one of the important aspects of hadron-structure research programs in future experiments at JLab~\cite{Arrington:2021alx} and the EIC~\cite{Aguilar:2019teb}. The LCDA is also an important quantity in the examination of $B$-meson non-leptonic decays that are relevant to the extraction of parameters in the Cabibbo-Kobayashi-Maskawa matrix~\cite{Beneke:1999br}.

Conventionally, lattice QCD practitioners proceed with the determination of Mellin moments of the PDFs and LCDAs by computing matrix elements of local operators resulting from a light-cone operator product expansion (OPE). However, since the rotational symmetry is explicitly broken by the lattice regularization, spin is no longer a good quantum number on the lattice. This leads to the mixing among twist-2 operators with different mass dimensions under renormalization. In particular, when an operator mixes with its lower-dimensional counterparts, power divergences can arise, making the subtraction difficult. Moreover, higher moments suffer from exponentially decaying signal-to-noise ratios. In practice, these obstacles limit the application of this conventional strategy to the determination of only the first four and two Mellin moments of the PDFs and the LCDAs, respectively.

In the past two decades, novel approaches for determining PDFs and LCDAs from Euclidean lattice QCD were proposed. Most notably, the quasi-distribution~\cite{Ji:2013dva} and pseudo-distribution~\cite{Radyushkin:2017cyf} methods in principle allow for the construction of the Bjorken-variable dependence in parton functions. These approaches overcome the spacetime metric limitation by exploiting the relation between momentum-boosted matrix elements that probe spatial correlations and the physically required ones on the light front. Namely, they have the same infrared structure, i.e., they encode the same infrared divergences, and thus their difference lies only in the ultraviolet regime. As such, it can be calculated in perturbation theory, leading to a matching procedure between the Euclidean lattice objects and the physical ones. In recent years, there has been tremendous theoretical and numerical progress in these approaches, leading to substantial information for parton physics. We refer to Refs.~\cite{Holligan:2024umc,Holligan:2024wpv,Ding:2024saz,LatticeParton:2024zko,LatticeParton:2024vck,Bhattacharya:2025yba,Good:2025daz,NieMiera:2025inn,Chu:2025kew,Chu:2025jsi,Miller:2025wgr,Gao:2026wlz} for an incomplete list of recent works and to the reviews in Refs.~\cite{Cichy:2018mum,Ji:2020ect,Constantinou:2020pek,Cichy:2021ewm,Lin:2025hka,Boer:2025ixc,Constantinou:2026wii,Cichy:2026xrh}. Furthermore, strategies for extracting higher moments were also suggested~\cite{Detmold:2005gg,Braun:2007wv,Davoudi:2012ya,Chambers:2017dov,Karpie:2018zaz,Detmold:2021uru,Shindler:2023xpd}, overcoming the limitations of standard local OPE-based extractions, and numerical results have started to appear~\cite{Chambers:2017dov,Detmold:2021qln,Bhattacharya:2023ays,Detmold:2025lyb,Bhattacharya:2024wtg,Francis:2025rya,Francis:2025pgf,Gao:2025inf,Dutrieux:2026grg,Miller:2026hza}.

Although the abundant research in this direction has seen significant progress, all these approaches are multi-step procedures that are still limited by various sources of systematic uncertainties, of both theoretical and lattice origins. As a consequence, the quasi- and pseudo-distribution methods, as well as other approaches, presently produce results of limited quantitative reliability. In practice, this means a limited reliable range of moderate Bjorken parameter values in the quasi-distribution method~\cite{Ji:2022ezo} and at most only two higher Mellin moments than what could be achieved by the conventional approach. Overcoming these limitations remains very difficult in all approaches. For example, the methods based on boosted hadron states, such as quasi- and pseudo-distributions~\cite{Ji:2013dva,Radyushkin:2017cyf}, need the hadron momentum to be large to suppress power corrections in the factorization between lattice and physical objects. However, the noise increases exponentially with each additional unit of momentum. In turn, the gradient flow method~\cite{Shindler:2023xpd} does not require a momentum boost, but needs a window in gradient flow time, which should be large enough to suppress discretization effects, but small enough for perturbative matching to hold. A similar window restriction needs to be satisfied also in the heavy OPE method~\cite{Detmold:2005gg}, where the fictitious heavy quark needs to be significantly heavier than the non-perturbative QCD scale, but remain well below the lattice cutoff. While improvements of these Euclidean-space calculations are expected in the foreseeable future, it is still essential to consider alternative strategies, in particular those enabling lattice computations performed directly in Minkowski space, i.e., requiring access to real time.

\subsection{Definitions}

The unpolarized collinear PDF of the fermions $\psi$ in a hadron is defined as~\cite{collins_2011}
\begin{align}
	\pdf = \int_{-\infty}^{\infty} &\frac{d\lambda}{2 \pi} e^{-i \lambda \bjorx n \cdot P} \nonumber\\
	&\braket{P| \overline{\psi}(\lambda n) \wilson{\lambda n}{0} \frac{n \cdot \gamma}{2} \psi(0)|P}_c .
\end{align}
The $\psi$ are anti-commuting Dirac fermion operators with $\overline{\psi} = \psi^\dagger \gamma^0$, $\Ket{P}$ is the hadronic state with momentum $P$, $\wilson{\lambda n}{0}$ is a straight Wilson line from $0$ to $\lambda n$, $\gamma$ is the vector of Dirac matrices, and $n$ is a vector in a light-front direction. The relevant range of $\bjorx$ is between 0 and 1, where $\bjorx = 0$ corresponds to fermions that carry no collinear momentum, while fermions with $\bjorx=1$ would carry all the momentum of the hadron. The subscript $c$ of expectation values denotes the subtraction of vacuum contributions~\cite{collins_2011}:
\begin{equation}
	\Braket{P' | \hat{O} | P}_c = \Braket{P' | \hat{O} | P} - \Braket{0 | \hat{O} | 0} \Braket{P' | P} . \label{eq:vacuum_subtraction}
\end{equation}
The vacuum state $\Ket{0}$ is normalized, $\Braket{0|0} = 1$. Momentum eigenstates are normalized to $\braket{P'|P} = (2 \pi)^{\mathrm{dim}} 2 \encont \delta^{\mathrm{dim}}(p-p')$, where dim is the number of spatial dimensions, $p$ is the spatial momentum, and $\encont$ is the energy of the state $\ket{P}$. The subtracted part of the matrix elements corresponds to disconnected diagrams from the Wick contractions of fermionic operators. These have to be removed so that the PDF only probes differences in the parton content of the state relative to the vacuum.

The antifermion PDF is defined in a similar way, with a charge-conjugated operator in the expectation value~\cite{collins_2011}:
\begin{align}
	\antipdf = \int_{-\infty}^{\infty} &\frac{d\lambda}{2 \pi} e^{-i \lambda \bjorx n \cdot P} \nonumber\\
	&\trace{\frac{n \cdot \gamma}{2} \ \braket{P| \psi(\lambda n) \wilsondagger{\lambda n}{0} \overline{\psi}(0)|P}_c} . \label{eq:antifermion_PDF}
\end{align}
The trace goes over spinor indices.

The internal structure of mesons is further described by LCDAs. They encode how the hadron's collinear momentum is split among its constituents. For a meson $\meson$ represented by the momentum eigenstate $\ket{P}$ and with valence fermions $\psi$ and antifermions $\overline{\psi}$, the LCDA \lcdau is defined by~\cite{Ball:1996tb}
\begin{align}
	(n \cdot P) \deconst &\int_{0}^{1} d \uscale e^{-i (1-\uscale) \lambda n \cdot P} \lcdau \nonumber\\
 	&= \Braket{0| \overline{\psi}(\lambda n) \wilson{\lambda n}{0} (n \cdot \gamma) \gammas \psi(0) |P} .
\end{align}
The LCDA corresponds to the overlap of the meson state with the vacuum on which a fermion-antifermion pair has been inserted. The parameter \uscale, which can vary between 0 and 1, describes the momentum fractions of the fermion and the antifermion, where $\uscale=0$ corresponds to the antifermion carrying all the momentum, $\uscale = \frac{1}{2}$ to an equal splitting, and $\uscale = 1$ to the fermion carrying all the momentum. The matrix $\gammas$ depends on the current that the meson couples to. For scalar and vector mesons $\gammas = \eye$, while for pseudoscalar and pseudovector mesons $\gammas = \gamma^5$ projects to the correct parity. The LCDA is chosen to be normalized to $\int_{0}^{1} d\uscale \lcdau = 1$. The decay constant \deconst is then given by
\begin{equation}
	(n \cdot P) \deconst = \Braket{0| \overline{\psi}(0) (n \cdot \gamma) \gammas \psi(0) |P} .
\end{equation}
The phases of the states are arbitrary, and we choose them so that $\deconst$ is real and non-negative.\footnote{A transformation $\Ket{0} \rightarrow e^{i \theta} \Ket{0}$ or the equivalent transformation $\Ket{P} \rightarrow e^{-i \theta} \Ket{P}$ can change the phase of the matrix elements in \deconst arbitrarily without changing the physics.}

\subsection{Parton functions in 1+1 dimensions}

In this work, we address PDFs and LCDAs in a 1+1-dimensional theory.  We choose the light-like vector $n$ to have the components $n^+ = 0$, $n^- = 1$ such that $n^0 = n^1 = \frac{1}{\sqrt{2}}$. With $\zlight = \lambda n^-$ we obtain\footnote{We use the metric $g_{\mu\nu} = \begin{pmatrix}1 & 0 \\ 0 & -1\end{pmatrix}$ in 1+1 dimensions and define the light-cone coordinates as $z^\pm = \frac{1}{\sqrt{2}} \left( z_0 \pm z_1 \right) = \frac{1}{\sqrt{2}} \left( t \mp z \right) = z_\mp$. The light-front vector $n$ points towards positive $t$ and $z$ (time and space coordinates) in this convention. Note that the orthogonal light-front direction is also used in the literature, with the same standard notation \zlight for the relevant component but with a different convention for light-cone coordinates.}
\begin{align}
	\pdf  = \int_{-\infty}^{\infty} \frac{d\zlight}{4 \pi} & e^{-i \bjorx \zlight \Plight} \nonumber\\
	&\Braket{P| \overline{\psi}(\zlight) \wilson{\zlight}{0} \gamma^+ \psi(0)|P}_c ,
\end{align}
while the LCDA is described by
\begin{align}
	\lcdau = \frac{1}{\deconst} \int_{-\infty}^{\infty} & \frac{d\zlight}{2\pi} e^{i (1-\uscale) \zlight \Plight} \nonumber\\
	&\Braket{0| \overline{\psi}(\zlight) \wilson{\zlight}{0} \gamma^+ \psi(0) |P} .
\end{align}
The decay constant in this light-front frame is 
\begin{equation}
	\deconst = \frac{1}{\Plight} \Braket{0| \overline{\psi}(0) \gamma^+ \psi(0) |P} .
\end{equation}
The cases $\gammas = \gamma^5 = \sigma_x$ and $\gammas = \eye$ are equivalent, because spin and, therefore, helicity do not play a role in one spatial dimension\footnote{We choose $\gamma^0 = \sigma_z = \begin{pmatrix} 1 & 0 \\ 0 & -1 \end{pmatrix}$ and $\gamma^1 = i\sigma_y = \begin{pmatrix} 0 & 1 \\ -1 & 0 \end{pmatrix}$.}. The only effect of $\gamma^5$ is an irrelevant overall sign factor. The absence of transverse directions and angular momentum also means that scalar and pseudovector mesons are equivalent and characterized by the parity quantum number $\eta_P = +1$, while pseudoscalar and vector mesons equally describe $\eta_P = -1$ hadrons~\cite{Coleman1976,Banks1976}. In this work, we consider chargeless states, where the quantum numbers for charge conjugation and parity are equal: $\eta_C = \eta_P$.

For both the PDF and the LCDA, we need to evaluate matrix elements of the form\footnote{We use either general vectors $z_i$, $z_f$ as arguments of operators and matrix elements, or the light-front coordinates $\zlight_f$, $\zlight_i$ to indicate that both points should be on the same light front.}
\begin{equation}
 \matel(\zlight_f, \zlight_i) = \Braket{F| \op\left(\zlight_f, \zlight_i\right) |I} , \label{eq:matel}
\end{equation}
where
\begin{equation}
	\op\left(\zlight_f, \zlight_i\right) = \psi^\dagger \left(\zlight_f\right) \wilson{\zlight_f}{\zlight_i} \Gamma \psi\left(\zlight_i\right) . \label{eq:operator}
\end{equation}
The gamma structure is $\Gamma = \gamma^0 \gamma^+$ in both cases. The initial state $\Ket{I} \equiv \Ket{P}$ is the hadronic state at a given momentum $P$. The final state $\Ket{F}$ corresponds to the same hadron $\Ket{P}$ for the PDF, and to the vacuum $\Ket{0}$ for the LCDA. We note that vacuum contributions of the form $\Braket{0| \op\left(\zlight_f, \zlight_i\right) |0} $ must be subtracted only in the PDF case [see \cref{eq:vacuum_subtraction}] from the matrix elements in \cref{eq:matel}.

\subsection{Properties of parton functions and matrix elements}
The PDF and the LCDA are real and have to vanish outside the physical regions of their parameters:
\begin{align}
	\pdf \equiv& 0 \qquad \text{for}~|\bjorx| \ge 1 , \\
	\lcdau \equiv& 0 \qquad \text{for}~\uscale \le 0~\text{or}~\uscale \ge 1 . \label{eq:LCDA_domain}
\end{align}
Their normalization is given by
\begin{align}
	\int_0^1 d\bjorx \pdf \equiv& \normpsi , \\
 	\int_0^1 d\uscale \lcdau \equiv& 1 ,
\end{align}
where \normpsi is the average number of fermions in the hadron.

PDFs are probability density functions in the parton limit and, as such, are non-negative. In QCD, however, this property can be violated depending on the renormalization scheme~\cite{ALTARELLI1998277,Collins:2021vke,Candido2024}.

The total momentum of all partons must be equal to the hadron momentum. In a theory with partons $\{p\}$ (including fermions, antifermions, bosons), this means
\begin{equation}
	\sum_p \int_0^1 d\bjorx \pdfp \bjorx = 1 . \label{eq:momentum_conservation_PDF}
\end{equation}

In the following, we discuss the symmetry properties of matrix elements, PDFs, and LCDAs. Their derivation is detailed in \cref{app:symmetries}. The initial and final coordinates of matrix elements and operators are denoted by vectors $z_i$ and $z_f$ which are assumed to be on the same light front, but the relations in this section hold for general vectors $z_i$ and $z_f$.

The PDF matrix elements are translationally invariant, which can be seen by applying the translation operator $e^{-i r \cdot \hat{P}}$ that shifts the coordinates of operators by a distance $r$:
\begin{align}
	\matelpdf(z_f, z_i) =& \Braket{P| e^{i r \cdot \hat{P}} \op(z_f, z_i) e^{-i r \cdot \hat{P}} |P}_c \nonumber\\
	=& \Braket{P| \op(z_f + r, z_i + r) |P}_c \nonumber\\
	=& \matelpdf(z_f + r, z_i + r) \nonumber\\
	\equiv& \matelpdf(z_f - z_i) . \label{eq:matel_PDF_twoargument}
\end{align}
PDF matrix elements thus only depend on the separation $\Delta z = z_f - z_i$. They further fulfill~\cite{SchwingerPDFletter2025}
\begin{align}
	\matelpdf(\Delta z) = \left(\matelpdf(-\Delta z) \right)^* . \label{eq:matel_symmetry}
\end{align}
Therefore, the real part has to be symmetric with respect to the origin at $\Delta z = 0$, while the imaginary part has to be antisymmetric.

While $\bjorx \in [0,1]$ corresponds to the fermion distribution, \pdf is well defined for negative values of $\bjorx$ as well, where it represents the antifermion distribution. The antifermion PDF in \cref{eq:antifermion_PDF} is then related to \pdf by~\cite{collins_2011}
\begin{equation}
	\antipdf = - \pdfbare (-\bjorx) . \label{eq:antifermion_from_fermion_PDF}
\end{equation}
For a meson that is its own antiparticle, such that $\antipdf \equiv \pdf$, this means that \pdf is antisymmetric with respect to the origin $\bjorx = 0$. In this case, one can derive
\begin{align}
	\matelpdf(-\Delta z) = - \matelpdf(\Delta z) .
\end{align}
Together with \cref{eq:matel_symmetry}, we conclude that the PDF matrix elements of such mesons are purely imaginary.

The translation of LCDA matrix elements by a vector $r$ results in a phase shift:
\begin{align}
	\matellcda(z_f, z_i) =& e^{i r \cdot P} \Braket{0| e^{i r \cdot \hat{P}} \op(z_f, z_i) e^{-i r \cdot \hat{P}} |P} \nonumber\\
	=& e^{i r \cdot P} \matellcda(z_f + r, z_i + r) . \label{eq:matel_LCDA_twoargument}
\end{align}

We choose the arbitrary phase of the LCDA matrix elements such that $\matellcda(0, 0)$ is real and positive. Then, the real part of the light-front matrix elements $\matellcda\left(\frac{\zlight}{2}, -\frac{\zlight}{2}\right)$ is symmetric, while the imaginary part is antisymmetric with respect to $\zlight=0$.

A chargeless meson is an eigenstate of the charge conjugation operator with eigenvalues $\eta_C = \pm 1$. In this case, the LCDA and the corresponding matrix elements fulfill the symmetry properties:
\begin{align}
	\matellcda \left(\frac{\zlight}{2}, -\frac{\zlight}{2}\right) =& -\eta_C \matellcda \left(-\frac{\zlight}{2}, \frac{\zlight}{2}\right), \nonumber\\
	\lcda\left(\frac{1}{2} + \uscale\right) =& -\eta_C \lcda\left(\frac{1}{2} - \uscale\right) .
\end{align}
Matrix elements are therefore either real and symmetric (for $\eta_C = -1$) or purely imaginary and antisymmetric (for $\eta_C = 1$). Since the Fourier transform is parity-even, it follows that the LCDA is symmetric (for $\eta_C = -1$) or antisymmetric (for $\eta_C = 1$) with respect to $\uscale = \frac{1}{2}$.

\section{Parton physics in the Schwinger model}
\label{sec:model}
%!TEX root = ./Schwinger_PDF_LCDA.tex

\subsection{Light-front operators on the lattice}

\Cref{app:SchwingerModel} contains more details on the transformations from the Lagrangian density to a spin-system Hamiltonian that reproduces the Schwinger model in the continuum limit. Here, we focus on the formulation of parton functions and the relevant matrix elements on the lattice.

The Hamiltonian density of the Schwinger model in the temporal gauge is
\begin{equation}
	{\mathcal{H}} = -i \overline{\psi} \gamma^1(\partial_1 +igA_1) \psi + m \overline{\psi} \psi +\frac{1}{2} E^2 , \label{eq:HamiltonDensity}
\end{equation}
with the electric field $E = \partial_0 A_1$.

Physical states need to fulfill the constraint of Gauss's law, where we include a coupling of the gauge field to an external charge density $\rho$:
\begin{equation}
	(g \psi^\dagger \psi + \rho - \nabla E)\Ket{\Phi}_\text{phys} = 0 . \label{eq:Gauss_law}
\end{equation}

For the discretized lattice field theory, we employ Kogut-Susskind staggered fermions~\cite{staggeredFermions}, where only the upper (lower) component of the two-component Dirac fermions is kept on even (odd) sites of the lattice with lattice spacing $a$. The two-component fermions $\psi(z)$ are recovered in the continuum limit from the one-component fields $\psilat_\site$ on the lattice as
\begin{equation}
	\frac{1}{\sqrt{2a}} \psilat_{\site} \rightarrow
	\begin{cases*}
		\psi_\text{upper}(\site a) & \text{if}~\site~\text{even} \\
		\psi_\text{lower}((\site-1)a) & \text{if}~\site~\text{odd}.
	\end{cases*} \label{eq:Schwinger_phi_def}
\end{equation}

The light-front operators in \cref{eq:operator} consist of four contributions in the staggered lattice description, corresponding to the four possibilities to combine the initial even $(\startsite=0)$ or odd $(\startsite=1)$ sites to the final even $(\finalsite=0)$ or odd $(\finalsite=1)$ sites:
\begin{align}
	\oplat(\siteeven) =& \Gamma_{0,0} \oplat_{0,0} + \Gamma_{0,1} \oplat_{0,1} + \Gamma_{1,0} \oplat_{1,0} + \Gamma_{1,1} \oplat_{1,1}, \nonumber\\
	\oplat_{\finalsite,\startsite}(\siteeven) =& \psilat^\dagger_{(\siteeven, \siteeven + \finalsite)} \wilson{(\siteeven, \siteeven + \finalsite)}{(0, \startsite)} \psilat_{(0,\startsite)}. \label{eq:operator_lattice_phi}
\end{align}
On the right-hand side of the second equation, we use the notation $(t,z)$ for the time and spatial coordinates.\footnote{We choose to apply $\psilat$ and $\psilat^\dagger$ at the same pair of times $0$ and $\siteeven$, independently of \startsite and \finalsite; see \cref{sec:Wilson_line} below for details.} The origin $0$ is taken to be the even site in the center of the one-dimensional lattice. The site index $\siteeven$ is assumed to be even. In the continuum limit, the spatial separation $\siteeven a$ corresponds to a light-front distance $\zlight = \sqrt{2} \siteeven a $. This leads to $\frac{1}{2a} \oplat(\siteeven) \xrightarrow[\zlight = \sqrt{2} \siteeven a]{a \rightarrow 0} \op(\zlight,0)$. In one spatial dimension, the gamma matrices for PDFs and LCDAs are identical, $\Gamma = \frac{1}{\sqrt{2}} \begin{pmatrix} \phantom{+}1 & -1 \\ -1 & \phantom{+}1 \end{pmatrix}$.

We replace fermionic degrees of freedom with spin variables through a Jordan-Wigner transformation $\psilat_\site = \prod_{\sitesum<\site}\left(-i\sigma_{\sitesum}^{z}\right)\sigma_\site^{-}$. Furthermore, gauge degrees of freedom are eliminated by using Gauss's law~\cite{Hamer1982}. This leads to the following spin Hamiltonian acting on a Hilbert space that consists only of physical states, i.e., fulfilling Gauss's law:
\begin{align}
	H =& x \sum_{\site=0}^{N-2} \left [ \sigma_\site^+\sigma_{\site+1}^- + \sigma_\site^-\sigma_{\site+1}^+ \right ] \nonumber\\
	&+ \frac{\mu}{2} \sum_{\site=0}^{N-1} \left [ 1 + (-1)^\site \sigma_\site^z \right ] \nonumber\\
	&+ \sum_{\site=0}^{N-2} \left [\frac{1}{2}\sum_{\sitesum=0}^\site \left((-1)^\sitesum+\sigma_\sitesum^{z} + 2 Q_\sitesum\right)\right ]^2 . \label{eq:Hdimless}
\end{align}
We use the dimensionless variables $x = \frac{1}{a^2 g^2}$ and $\mu = 2 \sqrt{x} \frac{m}{g} - \frac{1}{4}$, where an additive mass renormalization is included~\cite{Schwinger_latticeMass,MassRenormalizationNumerical,MassRenormalizationNumerical_PoS}. The external charges $Q_\site$ correspond to the continuum charge density, where $\frac{g}{a} Q_\site \xrightarrow[z = \site a]{a \rightarrow 0} \rho(z)$. The dimensionful energy of a state $\Ket{\Phi}$ is $\encont = \frac{ag^2}{2} \Braket{\Phi | H | \Phi}_c \equiv \frac{g}{2 \sqrt{x}} \energy$, where $\energy = \Braket{\Phi | H | \Phi}_c$ is the expectation value of the lattice Hamiltonian with subtracted vacuum energy.

In the massless limit, the dynamically generated mass of the vector meson is $\encont_V = \frac{g}{\sqrt{\pi}}$. We normalize fermion masses by this value. Volumes are measured in terms of physical volumes $N a = \frac{N}{\sqrt{x} g}$ multiplied by the fermion mass, because the required system size typically scales with the inverse of the energy scale. Throughout this paper, we thus describe the fermion mass and the volume by
\begin{align}
	\mass &\equiv \frac{m}{\encont_V} = \frac{m}{g}\sqrt{\pi}, \nonumber\\
	\vol &\equiv \mass \frac{N}{\sqrt{x}} =
	\sqrt{\pi} m N a .
\end{align}

In the spin model, the light-front operators become
\begin{align}
	\oplat_{\finalsite,\startsite}(\siteeven) =&
	e^{i H t_n} \left( \prod_{\sitesum<{\siteeven + \finalsite}} i\sigma_\sitesum^{z} \right) \sigma_{\siteeven + \finalsite}^{+} \nonumber\\
	&e^{-i H t_n} \wilson{(\siteeven, \siteeven + \finalsite)}{(0,\startsite)} \left(\prod_{\sitesum'<\startsite} -i\sigma_{\sitesum'}^{z}\right)\sigma_\startsite^{-} , \label{eq:operator_lattice}
\end{align}
with $t_n = \frac{1}{2x} n$. The Wilson line goes in a light-front direction.

We evaluate the matrix elements for the first excited state $\Ket{h}$, which in the continuum and thermodynamic limits corresponds to the first hadron in the spectrum (vector meson) at rest. The momentum of the state can thus be neglected, and only the energy contributes to $P^+ = \frac{\encont}{\sqrt{2}} = \frac{1}{\sqrt{2}} \frac{1}{2 x a} \energy$. On the lattice, we define the matrix elements\footnote{Compared to $\mathcal{M}(\Delta z = n)$ in Ref.~\cite{SchwingerPDFletter2025}, this differs by a factor of $\frac{1}{\sqrt{2}}$.}
\begin{align}
	\mellat_{\finalsite,\startsite}(\siteeven) =& \Braket{f | \oplat_{\finalsite,\startsite}(\siteeven) | h} , \nonumber\\
	\mellat(\siteeven) =& \sum_{\finalsite,\startsite} \Gamma_{\finalsite,\startsite} \mellat_{\finalsite,\startsite}(\siteeven) . \label{eq:matel_lattice}
\end{align}
The state $\Ket{f}$ corresponds to the hadronic state $\Ket{h}$ for the PDF, and to the vacuum $\Ket{0}$ for the LCDA. All states on the lattice are normalized to $\Braket{f|f} = 1$. Vacuum contributions are understood to be subtracted from PDF matrix elements.

The combined matrix elements
\begin{align}
	\mel_{\text{PDF}}(\siteeven) =& \frac{\sqrt{x}}{2} \normfactpdf \mellat_\text{PDF}(\siteeven), \nonumber\\
	\mel_{\text{LCDA}}(\siteeven) =& \frac{\sqrt{x}}{2} \normfactlcda \mellat_\text{LCDA}(\siteeven)
\end{align}
converge in the continuum and thermodynamic limits to $\matel(\zlight, 0) / g$ with $\zlight = \sqrt{2} n a$ [see \cref{eq:matel}]. The normalization factors $\normfact$ depend on the boundary conditions and account for the different state normalizations on the lattice and in the continuum. We discuss them in more detail in \cref{sec:boundary_conditions} below.

The PDF, the LCDA, and the decay constant can finally be calculated from the lattice matrix elements:
\begin{align}
	\pdf =& \frac{1}{\pi \sqrt{2x}} \sum_{\siteeven=\text{even}} e^{-i \frac{\energy}{2x} \bjorx \siteeven} \mel_{\text{PDF}}(\siteeven) , \nonumber\\
	\lcdau =& \frac{\sqrt{2}}{\pi \sqrt{x} \deconst} \sum_{\siteeven=\text{even}} e^{i \frac{\energy}{2 x} (1-\uscale) \siteeven} \mel_{\text{LCDA}}(\siteeven) , \nonumber\\
	\deconst =& \frac{2 \sqrt{2x}}{\energy} \mel_{\text{LCDA}}(0) . \label{eq:parton_functions}
\end{align}

\subsection{State normalization and boundary conditions}
\label{sec:boundary_conditions}

The factors $\normfact$ that enter the normalization of the PDF and the decay constant depend on the boundary conditions.

\subsubsection{Periodic boundary conditions}

For periodic boundary conditions (pbc), the factors $\normfact$ correspond to the different state normalizations on the lattice and in the continuum. We demand that the rescaled state norm $\normfactpdf \Braket{k'|k}$ converges to the relativistic continuum normalization $\Braket{P'|P}$. Since the vacuum state is normalized in the same way on the lattice as in the continuum, the LCDA obtains only a normalization factor from the initial state. The discrete lattice momenta are $k_{\knum} = \frac{2 \pi}{V} {\knum}$, with volume $V=Na$ and integer values $\knum$. The spacing between consecutive momenta is therefore $\Delta k = \frac{2 \pi}{N a}$. Combining these relations, we obtain:
\begin{align}
	\normfactpdf \braket{k'|k} &\xrightarrow[k \rightarrow p]{a \rightarrow 0} \braket{P'|P} \xrightarrow{V \rightarrow \infty} 4 \pi \encont \delta(p' - p) , \nonumber\\
	\frac{1}{\Delta k} \braket{k'|k} &= \frac{1}{\Delta k} \delta_{k',k} \xrightarrow[V \rightarrow \infty]{a \rightarrow 0} \delta(p' - p) , \nonumber\\
	\rightarrow \normfactlcda_\text{pbc} &=
	\sqrt{\frac{4 \pi \encont}{\Delta k}} = \sqrt{2 \encont V} = \sqrt{\frac{\energy N}{x}} .
\end{align}

\subsubsection{Open boundary conditions}

While open boundary conditions (obc) do not allow for a well-defined momentum, the first excited state in the spectrum still captures the structure of the hadronic state at rest in the continuum and infinite volume limits. However, obc require that the envelope of the wave function vanishes at the boundaries. The center-of-mass wave function of the meson in its lowest mode has the envelope $\epsilon(z) = \sqrt{2} \cos{\frac{\pi z}{V}}$ (with normalization $\frac{1}{V} \int_{-V/2}^{V/2} dz \epsilon^2(z) = 1$). Therefore, the amplitudes of states at the center of the lattice at $z=0$ differ compared to pbc by $\epsilon(0) = \sqrt{2}$. Matrix elements close to the center therefore need to be rescaled by the adapted normalization factor
\begin{equation}
	\normfactlcda_\text{obc} = \frac{1}{\sqrt{2}} \normfactlcda_\text{pbc} = \sqrt{\frac{\energy N}{2x}} .
\end{equation}
We use this scaling factor for all matrix elements, which causes $\order{\left(z/V\right)^2}$ artifacts away from the center. The PDF depends on $\normfact^2$ and is therefore expected to have stronger finite-volume effects with obc than the LCDA. Since the matrix elements vanish for large separations, the boundary effects are removed in the thermodynamic limit.

The matrix product state formalism used in this work is particularly efficient with obc~\cite{Verstraete2008,SCHOLLWOCK201196,Orus2014annphys}. We therefore perform all calculations with these boundary conditions and use the scaling factors $\normfact_\text{obc}$ for the PDF and the decay constant. The LCDA does not directly depend on $\normfact$, because it contains a ratio between two matrix elements that cancels the scaling factor. We finally obtain the obc matrix elements
\begin{align}
	\mel_{\text{PDF}}(\siteeven) =& \frac{\energy N}{4\sqrt{x}} \mellat_\text{PDF}(\siteeven) , \nonumber\\
	\mel_{\text{LCDA}}(\siteeven) =& \frac{\sqrt{\energy N}}{2\sqrt{2}} \mellat_\text{LCDA}(\siteeven) . \label{eq:matel_obc}
\end{align}
These converge to their continuum counterparts $\matel / g$ in \cref{eq:matel}. Therefore, the scaling of the bare matrix elements in terms of lattice parameters at fixed mass $\encont = \frac{g}{2 \sqrt{x}} \energy$ is $\mellat_\text{PDF}(n) \propto  N^{-1}$, and $\mellat_\text{LCDA}(n) \propto  N^{-1/2} x^{-1/4}$.

\section{Wilson line on the light front}
\label{sec:Wilson_line}
%!TEX root = ./Schwinger_PDF_LCDA.tex

\begin{figure}[tbp]
	\centering
	\def\figWidth{.7\columnwidth}
	%!TEX root =  ../Schwinger_PDF_LCDA.tex
\begin{tikzpicture}
	\begin{axis}[
		width=\figWidth,
		height=\figWidth,
		scale only axis,
		scale only axis,
		xmin=-6,
		xmax=6,
		xlabel style={font=\color{white!15!black},yshift=-1ex},
		xlabel={position [lattice units]},
		ymin=-6,
		ymax=6,
		ylabel style={font=\color{white!15!black},xshift=1ex},
		ylabel={time [lattice units]},
		axis background/.style={fill=white},
		xmajorgrids,
		ymajorgrids,
		legend style={at={(1.03,1)}, anchor=south west, legend cell align=left, align=left, draw=white!15!black},
		ylabel style={{rotate=-90}},
		scaled ticks=false,
		tick label style={/pgf/number format/fixed},
		xtick={-6,-4,...,6},
		ytick={-6,-4,...,6},
		minor tick num=1,
		xminorgrids,
		yminorgrids,
		xticklabel style={yshift=-1ex},
		yticklabel style={xshift=-1ex}
		]
		\draw[mylineblack,dashed] (-6,-6) -- (6,6); %
		%positive light-front direction
		\node[fermion] at (0,0) {};
        \node[fermion] at (1,0) {};
        \node[antifermion] at (2,2) {};
        \node[antifermion] at (3,2) {};
        \node[antifermion] at (4,4) {};
        \node[antifermion] at (5,4) {};
        \node[antifermion] at (6,6) {};
        \node[antifermion] at (-1,-2) {};
        \node[antifermion] at (-2,-2) {};
        \node[antifermion] at (-3,-4) {};
        \node[antifermion] at (-4,-4) {};
        \node[antifermion] at (-5,-6) {};
        \node[antifermion] at (-6,-6) {};
		\draw[spacedoublearrow] (0,0) -- (1,0);
		\draw[timearrow] (1,0) -- (1,2);
		\draw[spacearrow] (1,2) -- (2,2);
        \draw[spacearrow] (2,2) -- (3,2);
		\draw[timearrow] (3,2) -- (3,4);
		\draw[spacearrow] (3,4) -- (4,4);
		\draw[spacearrow] (4,4) -- (5,4);
        \draw[timearrow] (5,4) -- (5,6);
        \draw[spacearrow] (5,6) -- (6,6);
		%negative light-front direction
		\draw[spacearrow] (0,0) -- (-1,0);
		\draw[timearrow] (-1,0) -- (-1,-2);
		\draw[spacearrow] (-1,-2) -- (-2,-2);
		\draw[spacearrow] (-2,-2) -- (-3,-2);
		\draw[timearrow] (-3,-2) -- (-3,-4);
		\draw[spacearrow] (-3,-4) -- (-4,-4);
		\draw[spacearrow] (-4,-4) -- (-5,-4);
        \draw[timearrow] (-5,-4) -- (-5,-6);
        \draw[spacearrow] (-5,-6) -- (-6,-6);
	\end{axis}
\end{tikzpicture}%
	\caption{Stepwise evolution implementing the Wilson line in a light-front direction on the lattice. Blue horizontal arrows: spatial evolution of the Wilson line, implemented as moving an external charge; red vertical arrows: temporal evolution, corresponding to a time evolution of the state; green circles at positions 0 and 1: insertion of initial fermionic operator; all circles: locations where expectation values with antifermionic operators are calculated; dashed line: light-front direction.}
	\label{fig:zigzag}
\end{figure}
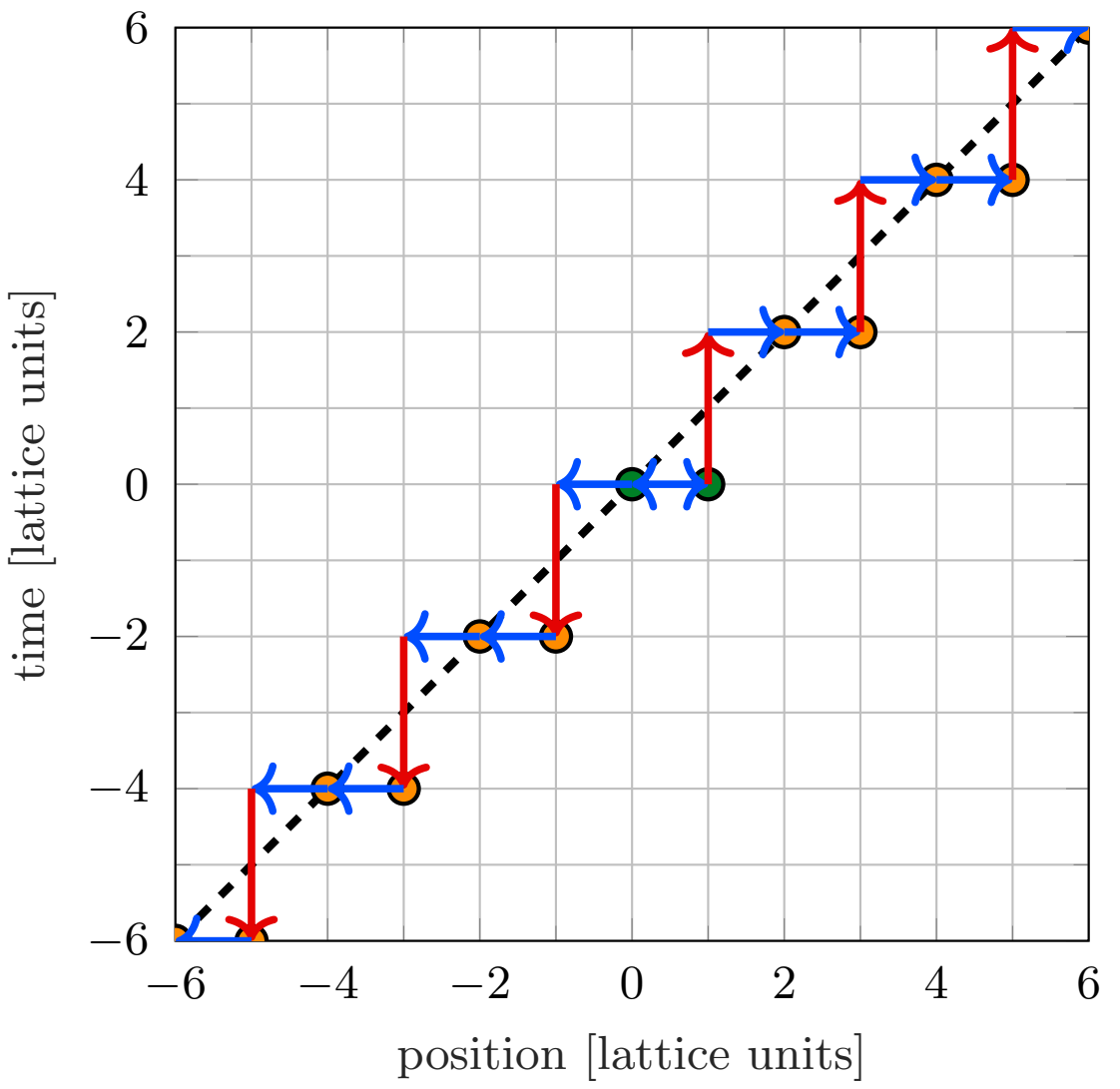

The Wilson line in the light-front direction can be approximated on the lattice by a stepwise evolution~\cite{SchwingerPDFletter2025,Schneider:2025gL} as sketched in \cref{fig:zigzag}. Matrix elements of the form given in \cref{eq:matel_lattice} are implemented by first applying the fermionic operator $\psilat_\startsite(0) = \left(\prod_{\sitesum<\startsite} -i\sigma_{\sitesum}^{z}\right)\sigma^{-}_\startsite$ to the initial state. Two independent calculations with $\startsite=0$ and $\startsite=1$ are necessary to reconstruct the combined matrix elements $\mellat(\siteeven)$. The site $\siteeven=0$ is an even site at the center of the obc lattice in this convention. The Wilson line is then evolved stepwise in the positive (towards the upper right in \cref{fig:zigzag}) or negative direction. At each position indicated by circles in \cref{fig:zigzag}, the conjugate fermion operator $\psilat^\dagger_{\finalsite}(\siteeven) = \left( \prod_{\sitesum<{\siteeven + \finalsite}} i\sigma_{\sitesum}^{z} \right) \sigma^{+}_{\siteeven+\finalsite}$ is applied, and the overlap with the final state is calculated.

The spatial Wilson line affects a state by changing the electric flux on the corresponding links by one unit. In the spin formulation that we employ here, this is not directly possible because the gauge degrees of freedom are integrated out and all states fulfill Gauss's law. The flux at each link can be reconstructed by summing all the charges to the left (or right) of the corresponding link. A stepwise evolution of the Wilson line requires intermediate states that are not in the physical Hilbert space. Changing the electric flux on a link can be traded for placing a pair of an external charge and its conjugate on both ends of the link. This changes Gauss's law accordingly and has the same effect on the Hamiltonian as changing the flux~\cite{SchwingerPDFletter2025,Schneider:2025gL,Echevarria2021,Pisarski2022}. The external charges of the Wilson line finally cancel with the charges of the initial and final operators, making matrix elements gauge invariant.
 
The time-like Wilson line is trivially the identity operator in the temporal gauge. However, the spatial Wilson lines and operators are applied on different time slices, implying a time evolution of the state. As detailed in Ref.~\cite{SchwingerPDFletter2025}, the Wilson line can be expressed as a stepwise time evolution with external charges that are moved in each evolution step as
\begin{align}
	e^{-i H t_\siteeven} \wilson{\siteeven + \finalsite}{\startsite} =& \te[_{\siteeven-1} \dt] \dots \te[_{3} \dt] \te[_{1} \dt] \label{eq:Wilson_zigzag}
\end{align}
for positive $\siteeven$, and similarly for the negative direction. The index of the Hamiltonian $H_{\site}$ indicates the inclusion of a single positive external charge at position $\site$. Red vertical arrows in \cref{fig:zigzag} thus correspond to time-evolution steps with a Hamiltonian containing an external charge. Blue horizontal arrows move the external charge to a new position for the next time evolution.

The time step $\dt$ needs to be chosen such that the stepwise evolution follows the light-front direction. Determining the speed of light is a nontrivial task when the Hamiltonian limit $a_t \rightarrow 0$ of an anisotropic Euclidean lattice gauge theory is considered, where different couplings and lattice spacings in spatial and temporal directions are present~\cite{Hasenfratz1981,Burgers:1987mb,Hamer_scales,SU2_universality,Lamm2020,Renormalization_quantum}. Here, we only consider a light-front direction in the Schwinger model when taking the continuum limit. The relation is given by the transformations of time-like and space-like quantities. The desired speed of light $c=1$ is achieved by setting $\dt = \text{sgn}(\siteeven) 2a \cdot \frac{ag^2}{2} = \text{sgn}(\siteeven) \frac{1}{x}$ in \cref{eq:Wilson_zigzag}. This results from the rescaling factor $\frac{2}{ag^2}$ of the Hamiltonian [see \cref{appeq:Hdimless} in \cref{app:SchwingerModel}], and the fact that each zigzag step corresponds to a distance of two lattice units $2a$. We numerically confirm that correlations spread within the causal light cone with this choice~\cite{SchwingerPDFletter2025}.

We note that the path of the Wilson line on the lattice is not unique. A stepwise evolution with a step width of $a$ instead of $2a$ is similarly possible. We verified in lattice perturbation theory that the error of a stepwise Wilson line in this $U(1)$ gauge theory scales with the step width. Therefore, the PDF and LCDA contain $\order{a}$ lattice artifacts, even if the energy is improved by an additive mass renormalization~\cite{Schwinger_latticeMass,MassRenormalizationNumerical,MassRenormalizationNumerical_PoS} to $\order{a^2}$ errors.

Stepwise Wilson lines can generally lead to cusp singularities~\cite{CuspDivergence}. However, no such difficulties and, more generally, no ultraviolet (UV) divergences occur in light-front matrix elements of the 1+1 dimensional gauge theory considered here.

\section{Simulation details}
\label{sec:simulation}
%!TEX root = ./Schwinger_PDF_LCDA.tex

For one-dimensional, local Hamiltonians, the ground state can be efficiently approximated by MPS~\cite{Hastings2007,Verstraete2006,Eisert2010,Cirac2021}, a property that extends to low-energy eigenstates~\cite{Arad2016lowE}. Even though the spin Hamiltonian in~\cref{eq:Hdimless} is not explicitly local, it has been shown that the vacuum and bound states can be accurately represented by MPS with moderate bond dimensions~\cite{Schwinger_massSpectrum}.

\subsection{Spectrum}

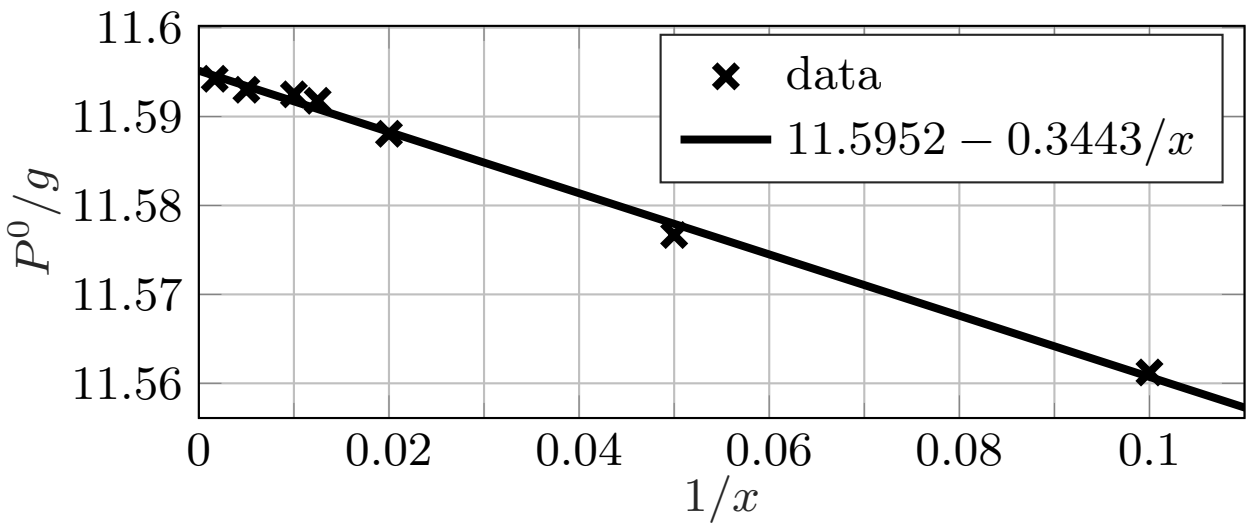
\begin{figure}[tbp]
	\centering
	\def\figWidth{.8*\columnwidth}
	\def\figHeight{.3*\columnwidth}
	%!TEX root =  ../Schwinger_PDF_LCDA.tex
% written by Make_masses_tikz.m -- M/g vs 1/x with the linear a^2 extrapolation
% parameters: mtilde = 10, N/sqrt(x) ~ 20, D = 80
\begin{tikzpicture}

\begin{axis}[%
width=\figWidth,
height=\figHeight,
at={(0\figWidth,0\figHeight)},
scale only axis,
xmin=0,
xmax=0.11,
xlabel style={font=\color{white!15!black}},
xlabel={$1/x$},
ymin=11.5561,
ymax=11.6002,
ylabel style={font=\color{white!15!black}},
ylabel={$\encont/g$},
axis background/.style={fill=white},
title style={font=\bfseries},
xmajorgrids,
ymajorgrids,
xminorgrids=true,
minor x tick num=1,
legend style={at={(0.98,0.98)}, anchor=north east, legend cell align=left, align=left, draw=white!15!black},
ylabel style={{rotate=-90}}, scaled ticks=false, tick label style={/pgf/number format/fixed},
xticklabel style={yshift=-.5ex,/pgf/number format/fixed,/pgf/number format/precision=3},
yticklabel style={xshift=-.5ex,/pgf/number format/fixed,/pgf/number format/precision=3}
]

\addplot [mylineblack, only marks]
table[row sep=crcr]{%
0.1	11.56122388\\
0.05	11.5766829365\\
0.02	11.5880659301\\
0.0125	11.5917763954\\
0.01	11.5925\\
0.005	11.5930156776\\
0.001666666667	11.5942514492\\
};
\addlegendentry{data}

\addplot [mylineblack, mark=none]
table[row sep=crcr]{%
0	11.5951569265\\
1	11.2508110692\\
};
\addlegendentry{$11.5952 -0.3443 / x$}

\end{axis}
\end{tikzpicture}%
	\caption{Mass of the vector meson when approaching the continuum limit $(ag)^2 = \frac{1}{x} \rightarrow 0$; parameters: $\mass = 10$, $D=80$, at fixed volume $\vol = 200$; solid line: extrapolation in $a^2$; an additive mass renormalization is used to remove $\order{a}$ lattice artifacts.}
	\label{fig:masses}
\end{figure}

The Hamiltonian in~\cref{eq:Hdimless}, as a sum of few-body terms, can be efficiently treated by variational and time-evolution MPS algorithms. Following the strategy in Ref.~\cite{Schwinger_massSpectrum}, we use a variational optimization of the MPS tensors to find the ground state $\ket{0}$, employing the exact matrix product operator (MPO)~\cite{Pirvu_2010} expression of $H$ (shown in~\cref{app:mpo}). The first excited state is similarly obtained by including a penalty term in the Hamiltonian with the projector $\ket{0} \bra{0}$, which ensures that the MPS of the excited state has no significant overlap with the ground state. An additional penalty term with the charge operator $\frac{1}{2} \sum_{\site=0}^{N-1} \sigma_{\site}^z$ forces the excited state to be in the chargeless sector. These variational searches are performed without external charges, $Q_{\site} \equiv 0$.

When the lattice mass $\mu$ includes the additive mass shift~\cite{Schwinger_latticeMass,MassRenormalizationNumerical,MassRenormalizationNumerical_PoS}, lattice artifacts of order $a$ are removed, and the error of the meson mass scales with $a^2 \propto \frac{1}{x}$ when taking the continuum limit. An example of this scaling can be seen in \cref{fig:masses}. Nevertheless, as discussed above in \cref{sec:Wilson_line}, $\order{a}$ effects are still present for the PDF and LCDA matrix elements even with renormalized masses, because the stepwise Wilson line introduces further lattice artifacts.

\subsection{Initial operators}

In order to calculate the matrix elements [see \cref{eq:matel_lattice}], we start from the first excited state as the initial state. To obtain the PDF, we also run a similar simulation for the ground state, which is required for the vacuum subtraction [see \cref{eq:vacuum_subtraction}]. In each simulation, the initial fermionic operator $\psilat_\startsite(0) = \left(\prod_{\sitesum<\startsite} -i\sigma_{\sitesum}^{z}\right)\sigma_\startsite^{-}$ from \cref{eq:operator_lattice_phi,eq:operator_lattice} is applied to the state. The Pauli matrices act locally and can therefore be applied exactly to the MPS without truncation. Each simulation is run twice with $\startsite=0$ and $\startsite=1$, such that the initial fermionic operator acts on an even or an odd site [see \cref{eq:operator_lattice}]. For each initial position, the Wilson line needs to be evolved in the positive ($n \ge 0$) and negative ($n<0$) light-front direction. Thus, for each fixed setup (e.g., $\mass$, $D$, $\Lcut$, $\Nt$, $x$, $\vol$), the PDF calculation requires eight simulations. The LCDA would require four runs, but we evaluate it simultaneously in the PDF calculations with negligible overhead.

\subsection{Time evolution}
\label{subsec:time_evolution}

In each simulation, the stepwise Wilson line is applied by a time evolution with a Hamiltonian containing an external charge [see \cref{eq:Wilson_zigzag}]. Initially, the external charge is inserted at site $1$ for the positive light-front direction, or at site $-1$ for the negative direction, to prepare for the time evolution of the state [see \cref{fig:zigzag}; site $0$ is the even site in the middle of the spin chain]. In each subsequent zigzag step, the external charge is moved by two sites.

In order to apply the evolution steps $U_{\siteeven}(\dt) = \te[_{\siteeven} \dt]$ onto the MPS, we use a second-order Suzuki-Trotter decomposition~\cite{Trotter,Suzuki1990FractalDO}, a standard strategy in TNS evolution algorithms. The time step $\dt$ is split into $\Nt = \frac{\dt}{\dtrotter}$ Trotter steps:
\begin{align}
	U_{\siteeven}(\dt) =& \left[ U^{e} \left(\frac{\dtrotter}{2}\right) U^{o} \left(\frac{\dtrotter}{2}\right) U^{L}_{\siteeven} (\dtrotter) U^{o} \left(\frac{\dtrotter}{2}\right) U^{e} \left(\frac{\dtrotter}{2} \right) \right]^{\Nt} \nonumber\\
	&+ \order{\dtrotter^2} .
\end{align}
The time-evolution operators $U^{e/o}(t) = e^{-i H^{e/o} t}$ correspond to the hopping terms $H^{e/o} = x \sum_{\site=\text{even}/\text{odd}}\allowbreak \left [ \sigma_\site^+\sigma_{\site+1}^- + \sigma_\site^-\sigma_{\site+1}^+ \right ]$, acting respectively on the even and odd bonds, and can each be expressed by an MPO with a maximal bond dimension of four~\cite{Schwinger_thermalEvolution}.

For the time-evolution operator $U^{L}_{\siteeven}$, which corresponds to the mass term and the electric term, we employ the strategy described in the supplemental material of Ref.~\cite{Schwinger_thermalEvolution}. The operator can be exactly expressed as an MPO with a bond dimension of size $\order{N}$. At site $\site$, the left internal index of the MPO tensor encodes the electric field $L_{\site-1}$, given by the sum of all charges to the left of site $\site$. The right index encodes the electric field $L_\site$, which includes the dynamical and static charges at site $\site$. Because working with the exact MPO becomes too expensive for large system sizes, we instead truncate these MPO tensors and effectively restrict the electric flux per link to $|L_\site| \le \Lcut$. The MPO then has a maximum bond dimension of $2 \Lcut +1$. The validity of this truncation is motivated by the observation that a configuration with several charges of the same sign next to each other, without charges of the opposite sign in between them, increases the electric field and thus the energy. These configurations therefore do not contribute significantly to the low-energy regime, which we confirm numerically by varying \Lcut.

The different MPOs are applied successively to the MPS. In each step, an iterative optimization is performed to find a new optimal MPS of bond dimension $\chi$. This MPS minimizes the distance (in the 2-norm) to the exact action of the MPO on the previous MPS. We set $\chi = D$ in all our calculations, such that the bond dimension is kept bounded during the time evolution.

The costs of the algorithm are dominated by the time evolution with the electric term. Its computing time scales as $\order{N_\text{steps} \N \Nt \chi^3 \Lcut d}$, where $d=2$ is the local Hilbert space dimension.\footnote{A more detailed scaling is $\ordersym (N_\text{steps} \cdot \Nt \cdot (N_\text{it} + 1) \cdot \N \cdot (2 \chi^3 (2\Lcut +1) d + \chi^2 (2\Lcut +1)^2 d^2))$. In each Trotter step, the overlap between the previous MPS, the MPO describing the time evolution with the electric term, and the new MPS is optimized iteratively. $N_\text{it}$ is the required iteration number.} We set the number of zigzag steps $N_\text{steps}$ proportional to $\N$ in our simulations and calculate the matrix elements for separations up to the boundaries, leading to an $\order{\N^2}$ scaling. The costs could be reduced in cases where the matrix elements decay to zero faster than the total lattice size that is used to accurately describe the initial state.

\subsection{Calculating matrix elements}

To finish the calculation of the matrix elements in \cref{eq:matel_lattice}, after the step to $\siteeven-1$, we contract the evolved states with the final state on which the remaining fermionic operator has been applied, $\bra{f} \psilat^\dagger_{\finalsite}(\siteeven)=\bra{f}\left( \prod_{\sitesum<{\siteeven + \finalsite}} i\sigma_{\sitesum}^{z} \right) \sigma_{\siteeven + \finalsite}^{+} $, both for the meson state and the vacuum. This is done separately for $\finalsite=0,\,1$, to obtain $\mellat_{0,\startsite}(\siteeven)$ and $\mellat_{1,\startsite}(\siteeven)$. The matrix element is finally multiplied by the phase factor $e^{i \braket{f|H|f} t_\siteeven}$. For the next iteration, the evolution of the states continues in the same manner, with $\siteeven$ shifted by two sites.

\subsection{Obtaining parton functions from matrix elements}

The PDF and the LCDA are obtained by Fourier transforms of the matrix elements according to \cref{eq:parton_functions}. We omit data points which lie within $2\%$ of the lattice size from a boundary, where artifacts from the obc can occur. From the lattice, discrete data points $\mellat (\siteeven)$ are calculated at even distances $\siteeven$. A discrete Fourier transform would lead to parton functions evaluated at specific values of \bjorx and $1-\uscale$ at the points $4 \pi \frac{x}{\energy N} \knum$, with integers $\knum \in \left[ -\frac{N}{4}, \frac{N-2}{4} \right]$. If the lattice parameters are chosen such that $\pi x > \energy$, the whole range of $\bjorx \in \left[ -1, 1 \right]$ can be covered.  A finer spacing in the discretized values of \bjorx and \uscale would require matrix elements at larger separations. However, the matrix elements decrease with increasing distance on the light front and vanish for large separations, which we ensure in our calculations by sufficiently large volumes in all simulations. We therefore replace the discrete Fourier transform with a continuous one. We can then evaluate the parton functions at any value of \bjorx and \uscale. We note that this is only justified because the matrix elements become compatible with zero at large separations; otherwise artifacts at short distances in \bjorx and \uscale would arise. Both conditions, $\pi x > \energy$ and $\mellat (\siteeven) \xrightarrow{|\siteeven| \rightarrow N/2} 0$, have to be carefully checked in order to obtain information on the whole range of kinematic variables. We only include data in our analysis for which these conditions are met.

\subsection{Extrapolations, error analysis and final results}
\label{subsec:data_analysis}

Computed quantities depend on the simulation parameters $D$, $\Lcut$, $\Nt$, $x$, and $\vol$. To obtain information about the relevant limit where all of these approach $\infty$, we use the following procedure, which is further detailed in \cref{app:extrapolations}.

We choose a set of base parameters which are all large enough such that the results are close to convergence to the relevant limit. From this point in parameter space, we vary one parameter at a time while keeping all others fixed. We extrapolate to the relevant limit in the corresponding parameter with several functional forms. This way, we estimate the uncertainties at the base point for each parameter from the spread between the results at the base value and other data points (including the extrapolated values).

We then choose one of the parameters, for which we further reduce the uncertainties. At the base parameters in our simulations, the volume effects typically dominate the uncertainties for PDFs, while lattice artifacts are more relevant for LCDAs and the decay constants. Therefore, we apply the following procedure in $\vol$ and $x$, respectively: we consider the simulation results at the base parameter and those closer to the relevant limit, including extrapolations. Among these, we take the midpoint between the two results that differ the most from each other, and assign uncertainties that cover the whole range of results. The midpoint is reported as a final result. Since this point is always compatible with the base point (within uncertainty), and the effects of the other parameters on the latter are known, we can add the uncertainties. The final error bars combine (as the square root of the sum of squares) the uncertainty of the midpoint and the uncertainties related to all the other parameters at the base point.

As described in \cref{subsec:time_evolution}, the electric flux is truncated during the time evolution. Earlier works showed that a two-particle ansatz describes the meson for heavy fermion masses well~\cite{BERGKNOFF1977215,Eller1987,MoPerry}, which means that large values of the electric flux do not occur. Contributions with higher fermion numbers and therefore slightly increased ranges of the electric flux are expected for masses $\mass < 0.5$~\cite{MoPerry}. We find that a restriction of the electric field to values between $-\Lcut$ and $\Lcut$ already leads to accurate results with $\Lcut=2$ for all fermion masses studied in this work, and the effect of the truncation is negligible for the choice of $\Lcut=10$ in our simulations. We further find that the dominant error from Trotterization in our simulations is a phase error of matrix elements, with a constant phase factor in each evolution step as discussed in \cref{app:extrapolations}.

\section{Results}
\label{sec:results}
%!TEX root = ./Schwinger_PDF_LCDA.tex

\subsection{Matrix elements}

\begin{figure}[tbp]
	\centering
	\def\figWidth{.85*\columnwidth}
	\def\figHeight{.6*\figWidth}
	\subfloat[Real part; $\Nt$-dependence\label{sfig:pdf_matrix_elements_real}]{%
		\begin{minipage}{\columnwidth}
			\input{tikz/PDF_M_real_tau_selection_m2_5}
		\end{minipage}%
	}\hfill
	\subfloat[Imaginary part; $x$-dependence\label{sfig:pdf_matrix_elements_imag}]{%
		\begin{minipage}{\columnwidth}
			\input{tikz/PDF_M_imag_x_selection_m2_5}
		\end{minipage}%
	}
	\caption{PDF matrix elements as defined in \cref{eq:matel_obc}; lattice parameters: $\mass = 2.5$; $D = 120$; $\Lcut=10$; $\vol = 100$. \protect\subref{sfig:pdf_matrix_elements_real} The real part at $x=25$ decreases as the Trotterization becomes finer. \protect\subref{sfig:pdf_matrix_elements_imag} Dependence of the imaginary part on the lattice spacing with $\Nt=100$. Data for the largest parameters are shown as continuous lines for clarity.}
	\label{fig:pdf_matrix_elements}
\end{figure}

\begin{figure}[tbp]
	\centering
	\def\figWidth{.85*\columnwidth}
	\def\figHeight{.6*\figWidth}
	\subfloat[Real part; $x$-dependence\label{sfig:lcda_matrix_elements_real}]{%
		\begin{minipage}{\columnwidth}
			\input{tikz/LCDA_M_real_x_selection_m10}
		\end{minipage}%
	}\hfill
	\def\figWidth{.8*\columnwidth}
	\subfloat[Imaginary part; $\Nt$-dependence\label{sfig:lcda_matrix_elements_imag}]{%
		\begin{minipage}{\columnwidth}
			%!TEX root =  ../Schwinger_PDF_LCDA.tex
% Generated by writeTikzPlot.m -- do not edit data by hand.
\begin{tikzpicture}
\begin{axis}[%
width=\figWidth,
height=\figHeight,
at={(0\figWidth,0\figHeight)},
scale only axis,
xmin=-6.78822509939,
xmax=6.78822509939,
xlabel style={font=\color{white!15!black}},
xlabel={$\zlight g$},
ylabel={$\Im\left(\mathcal{M}_\mathrm{LCDA}\left(\frac{\zlight}{2}, -\frac{\zlight}{2}\right)\right)$},
ymin=-0.244715018268,
ymax=0.233531987046,
ylabel style={font=\color{white!15!black}},
ylabel style={{rotate=-90}}, scaled ticks=false, tick label style={/pgf/number format/fixed},
axis background/.style={fill=white},
xmajorgrids,
ymajorgrids,
legend pos=north east,
legend style={legend cell align=left, cells={align=left}, align=left, draw=white!15!black},
xticklabel style={yshift=-.5ex, /pgf/number format/fixed, /pgf/number format/precision=2},
yticklabel style={xshift=-.5ex, /pgf/number format/fixed, /pgf/number format/precision=3}
]
\addplot [myline2, only marks]
  table[row sep=crcr]{%
-6.78822509939086	0.000118008393828598\\
-6.50538238691624	-0.000147638574457817\\
-6.22253967444162	0.000229987636025422\\
-5.939696961967	8.41752656928904e-05\\
-5.65685424949238	0.000307257382949919\\
-5.37401153701776	0.000370417904988792\\
-5.09116882454314	0.000636398529720376\\
-4.80832611206852	0.00165206746723032\\
-4.5254833995939	0.00296614435827562\\
-4.24264068711928	0.00586165533023493\\
-3.95979797464467	0.0101326620418712\\
-3.67695526217005	0.0173336682893228\\
-3.39411254969543	0.0276622825868117\\
-3.11126983722081	0.0425743883323647\\
-2.82842712474619	0.0623081278234158\\
-2.54558441227157	0.0873478620938607\\
-2.26274169979695	0.116619079857599\\
-1.97989898732233	0.148765608710124\\
-1.69705627484771	0.179335453774536\\
-1.41421356237309	0.2027505240202\\
-1.13137084989848	0.211793486804252\\
-0.848528137423857	0.196222274914323\\
-0.565685424949238	0.151899696470539\\
-0.282842712474619	0.081788898737041\\
0	2.64173952338595e-16\\
0.282842712474619	-0.0837163548407083\\
0.565685424949238	-0.158459526882635\\
0.848528137423857	-0.206364023720032\\
1.13137084989848	-0.222976518026671\\
1.41421356237309	-0.213630903625771\\
1.69705627484771	-0.188827457446029\\
1.97989898732233	-0.155681298772211\\
2.26274169979695	-0.120654597805683\\
2.54558441227157	-0.0891543858554548\\
2.82842712474619	-0.0626696573089131\\
3.11126983722081	-0.0421363665611836\\
3.39411254969543	-0.027336333633584\\
3.67695526217005	-0.0167609436117449\\
3.95979797464467	-0.0100736467672773\\
4.24264068711928	-0.0058183514460962\\
4.5254833995939	-0.00322854084360524\\
4.80832611206852	-0.00174560922826639\\
5.09116882454314	-0.000682720964173478\\
5.37401153701776	-0.000539384028253401\\
5.65685424949238	-0.000244359938924244\\
5.939696961967	-9.62931844980532e-05\\
6.22253967444162	0.000217359653196946\\
6.50538238691624	-0.000188942039681086\\
};
\addlegendentry{$\Nt = 20$}
\addplot [myline1, only marks]
  table[row sep=crcr]{%
-6.78822509939086	1.76196116617874e-05\\
-6.50538238691624	-1.52544786192664e-05\\
-6.22253967444162	-5.72537493501648e-05\\
-5.939696961967	-8.86628464195251e-05\\
-5.65685424949238	-0.000205544248146104\\
-5.37401153701776	-0.000422822514871468\\
-5.09116882454314	-0.000785323059547122\\
-4.80832611206852	-0.00135051295846605\\
-4.5254833995939	-0.00222677506430505\\
-4.24264068711928	-0.003472017402473\\
-3.95979797464467	-0.0052566249128146\\
-3.67695526217005	-0.00767804805326168\\
-3.39411254969543	-0.0109886820666649\\
-3.11126983722081	-0.0153976848128217\\
-2.82842712474619	-0.020881235950022\\
-2.54558441227157	-0.0272177638987003\\
-2.26274169979695	-0.0335365584885304\\
-1.97989898732233	-0.0383869710451812\\
-1.69705627484771	-0.0398871676819624\\
-1.41421356237309	-0.0363479346093893\\
-1.13137084989848	-0.0278627277948626\\
-0.848528137423857	-0.0172621295168691\\
-0.565685424949238	-0.00826979697344088\\
-0.282842712474619	-0.0029651616188815\\
0	-1.32086976169298e-16\\
0.282842712474619	-0.00538033990010167\\
0.565685424949238	-0.0191436785725201\\
0.848528137423857	-0.0300154555919231\\
1.13137084989848	-0.0330564788223788\\
1.41421356237309	-0.030429454695508\\
1.69705627484771	-0.0252674484899756\\
1.97989898732233	-0.0191097787745973\\
2.26274169979695	-0.0130940803600394\\
2.54558441227157	-0.00820143264402565\\
2.82842712474619	-0.00469366564183636\\
3.11126983722081	-0.00232558960379048\\
3.39411254969543	-0.00086964994621754\\
3.67695526217005	-8.732309043568e-05\\
3.95979797464467	0.000246219892560105\\
4.24264068711928	0.000343459461436689\\
4.5254833995939	0.000319086460865634\\
4.80832611206852	0.000267169990144114\\
5.09116882454314	0.000167413374218828\\
5.37401153701776	0.000107742311840117\\
5.65685424949238	8.27940286534416e-05\\
5.939696961967	5.39485652101164e-05\\
6.22253967444162	2.63140316185897e-05\\
6.50538238691624	1.64358635149396e-05\\
};
\addlegendentry{$\Nt = 100$}
\addplot [mylineblack, mark=none]
  table[row sep=crcr]{%
-6.78822509939086	-9.49281275004606e-06\\
-6.50538238691624	-2.67692108403265e-05\\
-6.22253967444162	-4.06290538366824e-05\\
-5.939696961967	-0.000113541005028399\\
-5.65685424949238	-0.000219557272047147\\
-5.37401153701776	-0.000433134997562296\\
-5.09116882454314	-0.000830822950955917\\
-4.80832611206852	-0.00145165209491667\\
-4.5254833995939	-0.00238406147847339\\
-4.24264068711928	-0.00377431063893297\\
-3.95979797464467	-0.0057269651119962\\
-3.67695526217005	-0.00845576451045306\\
-3.39411254969543	-0.0122132707618862\\
-3.11126983722081	-0.0171996645390053\\
-2.82842712474619	-0.0234951829982754\\
-2.54558441227157	-0.0308080920160728\\
-2.26274169979695	-0.0382415581454509\\
-1.97989898732233	-0.0442500112980414\\
-1.69705627484771	-0.0467507729282335\\
-1.41421356237309	-0.0438323274898633\\
-1.13137084989848	-0.0353605243414246\\
-0.848528137423857	-0.0239362682630864\\
-0.565685424949238	-0.0132746907211081\\
-0.282842712474619	-0.00561293172368926\\
0	2.64173952338595e-16\\
0.282842712474619	-0.00293294728800164\\
0.565685424949238	-0.0147888158345149\\
0.848528137423857	-0.0244963005290312\\
1.13137084989848	-0.0271012053417687\\
1.41421356237309	-0.0246723109218686\\
1.69705627484771	-0.0201161814568591\\
1.97989898732233	-0.0147954368165379\\
2.26274169979695	-0.00968422553306487\\
2.54558441227157	-0.00563391064491259\\
2.82842712474619	-0.00284054876343406\\
3.11126983722081	-0.00105683410721149\\
3.39411254969543	-2.09359013678689e-05\\
3.67695526217005	0.000451010335370636\\
3.95979797464467	0.000576461516019051\\
4.24264068711928	0.000543141572855123\\
4.5254833995939	0.000436834960331895\\
4.80832611206852	0.000315803255567473\\
5.09116882454314	0.000212492731882933\\
5.37401153701776	0.00012619058381291\\
5.65685424949238	7.87817677553035e-05\\
5.939696961967	4.65221507930476e-05\\
6.22253967444162	1.42283759783859e-05\\
6.50538238691624	2.26250959546605e-05\\
};
\addlegendentry{$\Nt = 200$}
\end{axis}
\end{tikzpicture}%
		\end{minipage}%
	}
	\caption{LCDA matrix elements from \cref{eq:LCDA_matel_plots}; lattice parameters: $\mass = 10$; $D = 80$; $\Lcut=10$; $\vol = 100$. \protect\subref{sfig:lcda_matrix_elements_real} Dependence of the real part on the lattice spacing with $\Nt=100$. \protect\subref{sfig:lcda_matrix_elements_imag} The imaginary part at $x=100$ decreases as the Trotterization becomes finer. Data for the largest parameters are shown as continuous lines for clarity.}
	\label{fig:lcda_matrix_elements}
\end{figure}

Examples of PDF matrix elements as defined in \cref{eq:matel_obc} are shown in \cref{fig:pdf_matrix_elements} [see also Ref.~\cite{SchwingerPDFletter2025} for more examples]. The real part of these matrix elements is small in magnitude compared to the imaginary part. It decreases towards zero with finer Trotterization and when approaching the continuum limit, in agreement with the expectation that the real part vanishes because of charge conjugation symmetry. We numerically verify that the imaginary part of this matrix element is antisymmetric with respect to the origin when the simulation parameters are chosen appropriately (close to the continuum and thermodynamic limits). This is in agreement with the expected symmetry properties [see \cref{eq:matel_symmetry}].

\Cref{fig:lcda_matrix_elements} shows examples of $\mel_{\text{LCDA}}$. In order to see the expected symmetry properties, we make use of \cref{eq:matel_LCDA_twoargument} and plot
\begin{equation}
	\mel_{\text{LCDA}}\left(\frac{\zlight}{2}, -\frac{\zlight}{2}\right)
	= e^{i \frac{\energy}{4 x} \siteeven} \mel_{\text{LCDA}}(\siteeven) , \label{eq:LCDA_matel_plots}
\end{equation}
where $\zlight = \frac{1}{g} \sqrt{\frac{2}{x}} \siteeven$ and $\mel_{\text{LCDA}}(\siteeven)$ as defined in \cref{eq:matel_obc}. We observe that these matrix elements become real and symmetric as expected when the parameters are chosen close enough to the continuum limit. The Trotterization has the strongest effect on the violation of the symmetry properties in our finite lattice calculations [see \cref{sfig:lcda_matrix_elements_imag}], because a finite Trotter step width introduces a constant phase error in each evolution step that violates the symmetries and introduces a bias between fermions and antifermions [see \cref{app:extrapolations}].

\subsection{PDF}

\begin{figure*}[tbp]
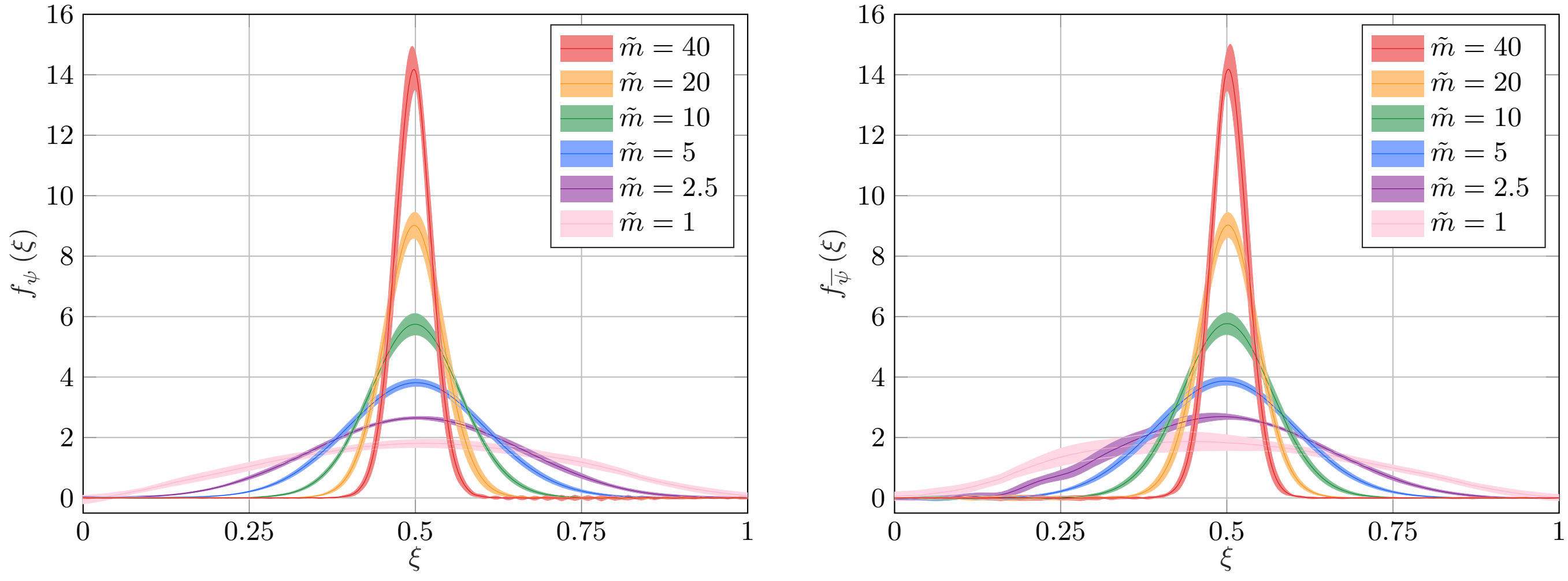

	\centering
	\def\figWidth{.88*\columnwidth}
	\def\figHeight{.75*\figWidth}
	\renewcommand{\tikzlinewidth}{0.2pt}
	\subfloat[fermion PDF\label{sfig:PDF}]{%
		\begin{minipage}{\columnwidth}
			%!TEX root =  ../Schwinger_PDF_LCDA.tex
% fermion PDF bands; the DATA lives in PDF_fermionmass_allsys_data.tex (updated by Evaluate_PDF_extrapolations.m)
\begin{tikzpicture}

\begin{axis}[%
width=\figWidth,
height=\figHeight,
at={(0\figWidth,0\figHeight)},
scale only axis,
xmin=0,
xmax=1,
xtick={0,0.25,...,1},
xlabel style={font=\color{white!15!black}},
xlabel={\bjorx},
ymin=-0.5,
ymax=16,
ytick={0,2,...,16},
ylabel style={font=\color{white!15!black}},
ylabel={\pdf},
axis background/.style={fill=white},
title style={font=\bfseries},
xmajorgrids,
ymajorgrids,
legend style={at={(0.98,0.98)}, anchor=north east, legend cell align=left, align=left, draw=white!15!black},
ylabel style={{rotate=-90}}, scaled ticks=false, tick label style={/pgf/number format/fixed},
xticklabel style={yshift=-.5ex},
yticklabel style={xshift=-.5ex}
]
\input{tikz/PDF_fermionmass_allsys_data}
\end{axis}
\end{tikzpicture}%
		\end{minipage}%
	}\hfill
	\subfloat[antifermion PDF\label{sfig:antiPDF}]{%
		\begin{minipage}{\columnwidth}
			%!TEX root =  ../Schwinger_PDF_LCDA.tex
% antifermion PDF = -f(-xi): renders the SAME data file, mirrored via pgfplots x/y filters
\begin{tikzpicture}

\begin{axis}[%
x filter/.expression={-x},
y filter/.expression={-y},
width=\figWidth,
height=\figHeight,
at={(0\figWidth,0\figHeight)},
scale only axis,
xmin=0,
xmax=1,
xtick={0,0.25,...,1},
xlabel style={font=\color{white!15!black}},
xlabel={\bjorx},
ymin=-0.5,
ymax=16,
ytick={0,2,...,16},
ylabel style={font=\color{white!15!black}},
ylabel={\antipdf},
axis background/.style={fill=white},
title style={font=\bfseries},
xmajorgrids,
ymajorgrids,
legend style={at={(0.98,0.98)}, anchor=north east, legend cell align=left, align=left, draw=white!15!black},
ylabel style={{rotate=-90}}, scaled ticks=false, tick label style={/pgf/number format/fixed},
xticklabel style={yshift=-.5ex},
yticklabel style={xshift=-.5ex}
]
\input{tikz/PDF_fermionmass_allsys_data}
%\legend{};	%omit legend
\end{axis}
\end{tikzpicture}%
		\end{minipage}%
	}
	\renewcommand{\tikzlinewidth}{\tikzlinewidtholdpdf}
	\caption{Fermion and antifermion PDFs of the vector meson and their uncertainties in the one-flavor Schwinger model for different fermion masses $\mass$. The antifermion PDF is obtained from the negative $\bjorx$-range according to \cref{eq:antifermion_from_fermion_PDF}.}
	\label{fig:PDF_fermionmass}
\end{figure*}

Fermion and antifermion PDFs for different masses and their uncertainty bands, as obtained using the analy\-sis procedure described in \cref{subsec:data_analysis}, are shown in \cref{fig:PDF_fermionmass}. The PDFs peak at $\bjorx = \frac{1}{2}$. The distributions become wider for smaller fermion masses and sharper for heavier fermions. This is consistent with previous works~\cite{Eller1987,Eller1989,Jirari1999} and with the massless limit, where the PDF is constant and therefore flat~\cite{SchwingerPartonPerturbative}, as well as with the infinite mass limit, where the PDF becomes a delta function at $\bjorx=\frac{1}{2}$.

Within error bands, the PDFs fulfill the expected symmetries: the imaginary part vanishes, and $\pdf$ becomes zero outside the range $|\bjorx| \le 1$. Because the real part of $\mel_{\text{PDF}}$ vanishes and the imaginary part is antisymmetric, $\pdf$ is also an antisymmetric function. The antifermion PDF in \cref{sfig:antiPDF} is calculated according to \cref{eq:antifermion_from_fermion_PDF} from the negative $\bjorx$-range. Because of the observed antisymmetry of $\pdf$, we find that the fermion PDFs are equal to the antifermion PDFs within uncertainties, as expected for a meson that is its own antiparticle. The method's applicability is therefore confirmed by these symmetry properties. The uncertainties could be further reduced by imposing the antisymmetry properties of the matrix elements and PDFs: $\mel_{\text{PDF}}(\siteeven) \rightarrow \frac{1}{2}(\mel_{\text{PDF}}(\siteeven) - \mel_{\text{PDF}}(-\siteeven))$ and $\pdf \rightarrow \frac{1}{2}(\pdf - \pdfbare (-\bjorx))$. Moreover, data from a direct calculation of the antifermion PDF with charge-conjugated operators according to \cref{eq:antifermion_PDF} could be included.

\begin{table*}[t]
	\centering
	%!TEX root =  ../Schwinger_PDF_LCDA.tex
% one row per mass; cells are updated in place by updateScalarSummary.m
% (PDF run: N_psi, N_psibar, P_psi, P_psibar; LCDA run: F_M)
\begin{tabular}{|c|c|c|c|c|c|}
	\hline
	$\mass$ & $\deconst$ & $\normpsi$ & $\normpsibar$ & $P_{\psi}$ & $P_{\overline{\psi}}$ \\
	\hline
	1 & $0.5245 \pm 0.0085$ & $1.008 \pm 0.044$ & $1.028 \pm 0.058$ & $0.518 \pm 0.026$ & $0.500 \pm 0.022$ \\ %KEY m1
	2.5 & $0.4682 \pm 0.0053$ & $1.011 \pm 0.019$ & $1.019 \pm 0.027$ & $0.509 \pm 0.016$ & $0.5042 \pm 0.0065$ \\ %KEY m2_5
	5 & $0.4004 \pm 0.0065$ & $1.001 \pm 0.017$ & $1.003 \pm 0.018$ & $0.502 \pm 0.011$ & $0.5001 \pm 0.0084$ \\ %KEY m5
	10 & $0.3327 \pm 0.0051$ & $0.988 \pm 0.022$ & $0.988 \pm 0.023$ & $0.494 \pm 0.010$ & $0.495 \pm 0.012$ \\ %KEY m10
	20 & $0.2680 \pm 0.0035$ & $0.990 \pm 0.039$ & $0.990 \pm 0.037$ & $0.493 \pm 0.022$ & $0.497 \pm 0.017$ \\ %KEY m20
	40 & $0.2191 \pm 0.0039$ & $0.977 \pm 0.044$ & $0.978 \pm 0.045$ & $0.484 \pm 0.020$ & $0.493 \pm 0.025$ \\ %KEY m40
	\hline
\end{tabular}
% asymmetry columns commented out for now (LCDA production values, 2026-09):
%   A_phi(m=1)   = 0.1286 +- 0.09367
%   A_phi(m=2.5) = 0.1194 +- 0.06434
%   A_phi(m=5)   = 0.1436 +- 0.1204
%   A_phi(m=10)  = 0.1149 +- 0.08549
%   A_phi(m=20)  = 0.1069 +- 0.06928
%   A_phi(m=40)  = 0.2432 +- 0.1289

	\caption{Scalar observables for different fermion masses $\mass$; see \cref{app:extrapolations} for the data analysis and uncertainty estimation; $\deconst$: decay constant; $\normpsi$: norm of fermion PDF; $\normpsibar$: norm of antifermion PDF; $P_{\psi} = \int_0^1 d\bjorx \pdf \bjorx$: average momentum fraction of fermions; $P_{\overline{\psi}} = \int_{-1}^0 d\bjorx \pdf \bjorx$: average momentum fraction of antifermions.}
	\label{tab:scalars}
\end{table*}

We find numerically that the PDF normalization for both the fermion and the antifermion is compatible with 1 for all masses considered, as shown in \cref{tab:scalars}. This is consistent with the previous observation that the wave function is dominated by two-fermion contributions (fermion and antifermion) for masses $\mass > 0.5$~\cite{MoPerry}.

The fermion PDF is symmetric with respect to $\bjorx = \frac{1}{2}$ in our data: $\pdf = \pdfbare(1-\bjorx)$ for $0 \le \bjorx \le 1$, as also observed in previous works~\cite{Eller1987,Eller1989,KROGER199858,Jirari1999}. Using this symmetry, the momentum carried by the fermions is
\begin{align}
	\int_0^1 d\bjorx \pdf \bjorx
		=& \int_0^1 d\bjorx \pdfbare(1-\bjorx) \bjorx \label{eq:momentum_PDF_symmetry1} \\
		=& -\int_1^0 d\bjorx \pdfbare(\bjorx) (1-\bjorx) \label{eq:momentum_PDF_symmetry2} \\
		=& \int_0^1 d\bjorx \pdfbare(\bjorx) (1-\bjorx) \label{eq:momentum_PDF_symmetry3} , \\
		\Rightarrow \int_0^1 d\bjorx \pdf \bjorx =& \frac{1}{2} \int_0^1 d\bjorx \pdfbare(\bjorx) = \frac{1}{2} ,
\end{align}
where we made the substitution $\bjorx \rightarrow 1 - \bjorx$ in \cref{eq:momentum_PDF_symmetry2} and used the fact that the left-hand side in \cref{eq:momentum_PDF_symmetry1} is the negative of the second term in \cref{eq:momentum_PDF_symmetry3}. The last equality is the previously discussed normalization of the PDF to 1. We conclude that the fermions carry half of the meson's momentum. From symmetry, it follows that the antifermions carry the other half. This is numerically confirmed by calculating the average fermion and antifermion momentum, which are both consistent with $0.5$ as shown in the last two columns of \cref{tab:scalars}.

The momentum conservation in \cref{eq:momentum_conservation_PDF} is therefore fulfilled if the gauge bosons do not carry momentum. The momentum of a photon is in the direction of its Poynting vector $T^{0i} = F^{0 \nu} {F_{\nu}}^{i}$ and its magnitude is proportional to both the electric and magnetic fields~\cite{Bjorken:1965zz,Kaku1993}. Since $F^{00}=F^{ii}=0$ and no transverse directions or magnetic fields exist in one spatial dimension, the gauge field carries no momentum. Therefore, the gauge-boson PDF vanishes~\cite{PhysRevD.98.054011}.

The relative uncertainties of the fermion and antifermion PDFs at $\bjorx = \frac{1}{2}$ range between $1.8\%$ and $6.3\%$. Only for $\mass=1$ does the uncertainty reach $7.6\%$ for the fermion PDF and $14.5\%$ for the antifermion PDF. The reason is that small masses require larger volumes and finer lattices to attain results sufficiently close to the thermodynamic and continuum limits. The contributions of the individual lattice parameters to the uncertainties are shown in \cref{app:extrapolations}.

\subsection{LCDA}

\begin{figure}[tbp]
	\centering
	\def\figWidth{.85*\columnwidth}
	\def\figHeight{.75*\figWidth}
	\renewcommand{\tikzlinewidth}{0.2pt}
	\input{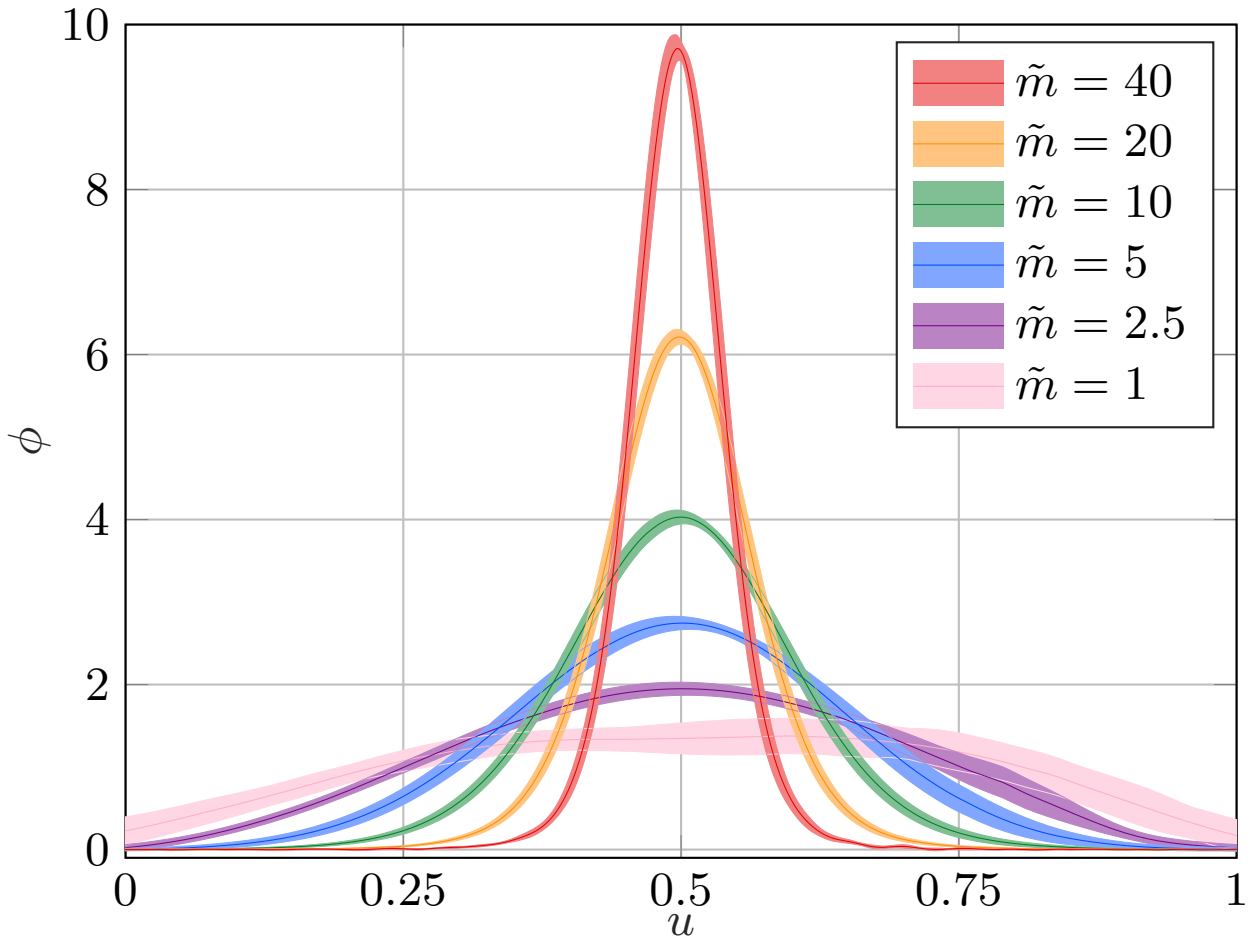}
	\renewcommand{\tikzlinewidth}{\tikzlinewidtholdpdf}
	\caption{LCDAs of the vector meson in the Schwinger model for different fermion masses $\mass$ and their uncertainties.}
	\label{fig:LCDA_fermionmass}
\end{figure}

The LCDAs for different fermion masses in the Schwinger model are shown in \cref{fig:LCDA_fermionmass}. The broadening of the LCDAs with decreasing mass is observed similarly to the PDF and to previous works~\cite{BERGKNOFF1977215,MoPerry}. It is also consistent with the extremal cases: in the massless limit, the LCDA becomes constant; in the heavy-mass limit, the meson is formed by a static pair of a heavy fermion and its antifermion with mass $\encont = 2m$, and the LCDA is a delta function at $\uscale = \frac{1}{2}$~\cite{BERGKNOFF1977215,MoPerry}.

The LCDA is real, up to small deviations close to the boundaries, and vanishes outside the domain $0 \le \uscale \le 1$ within uncertainties, as required [see \cref{eq:LCDA_domain}]. The LCDA is further symmetric with respect to $\uscale = \frac{1}{2}$, in agreement with the charge parity $\eta_C = -1$ of the vector meson.

The uncertainties from all simulation parameters are included in the error bands; see \cref{fig:LCDA_errors} in \cref{app:extrapolations} for the individual contributions. The uncertainties at $\uscale = \frac{1}{2}$ range between $1.2\%$ and $4.1\%$ and increase slightly with decreasing fermion mass, indicating that larger lattice sizes and bond dimensions, and therefore more numerical resources, are required for smaller masses. Only at $\mass=1$ does the uncertainty reach $14\%$ because of the small volume and coarse lattice that affect the PDF uncertainties similarly.

The decay constant is reported in \cref{tab:scalars}. Its extraction only requires the matrix elements at the origin and no time evolution.

\subsection{Entanglement entropy and time evolution}

\begin{figure}[tbp]
	\centering
	\def\figWidth{.85*\columnwidth}
	\def\figHeight{.5*\figWidth}
	\subfloat[ground state\label{sfig:entanglement_gs}]{%
		\begin{minipage}{\columnwidth}
			%!TEX root =  ../Schwinger_PDF_LCDA.tex
% written by Make_entropy_panels_tikz.m -- S_max vs t*g (positive
% branch, linear axes), ground state, all production masses;
% inset: position of the entanglement maximum, z^max*g vs t*g (t > 0 branch).
% EVEN startsites only; parameters (x, L, D=chi): mt=1: (20,90,200);
% mt=10: (100,100,120); all: Lcut=10, Ntau=100. Dashed line: t*g
% beyond which the matels are compatible with 0 (mt=10: 5.5; mt=1: never).
\begin{tikzpicture}
\begin{axis}[%
name=main,
width=\figWidth,
height=\figHeight,
scale only axis,
xmin=0,
xmax=12,
xlabel style={font=\color{white!15!black}},
xlabel={$t g$},
ymin=0.962,
ymax=1.672,
ylabel style={font=\color{white!15!black}},
ylabel={$S_\text{max}$},
axis background/.style={fill=white},
xmajorgrids,
ymajorgrids,
legend style={at={(0.5,-0.22)}, anchor=north, legend cell align=left, align=left, draw=white!15!black, font=\tiny},
legend columns=3,
ylabel style={{rotate=-90}}, scaled ticks=false, tick label style={/pgf/number format/fixed},
xticklabel style={yshift=-.5ex},
yticklabel style={xshift=-.5ex}
]

\addplot [myline5, only marks]
table[row sep=crcr]{%
0	1.00207094426\\
0.4	1.14707351992\\
0.8	1.38573873806\\
1.2	1.5560940041\\
1.6	1.58409650404\\
2	1.58409650404\\
2.4	1.5866032201\\
2.8	1.5866032201\\
3.2	1.61275350686\\
3.6	1.61275350686\\
4	1.61951726834\\
4.4	1.61951726834\\
4.8	1.62046237194\\
5.2	1.62046237194\\
5.6	1.62046237194\\
6	1.62233416787\\
6.4	1.62233416787\\
6.8	1.62291543242\\
7.2	1.62685234258\\
7.6	1.62685234258\\
8	1.6285839031\\
8.4	1.6285839031\\
8.8	1.6285839031\\
9.2	1.63325445396\\
9.6	1.63325445396\\
10	1.63325445396\\
10.4	1.63397249333\\
10.8	1.63397249333\\
11.2	1.63397249333\\
11.6	1.62755247918\\
12	1.63103035806\\
12.4	1.63389988985\\
12.8	1.63389988985\\
13.2	1.63389988985\\
13.6	1.63447152303\\
14	1.63447152303\\
14.4	1.63447152303\\
14.8	1.63447152303\\
15.2	1.63447152303\\
15.6	1.63661611957\\
16	1.63661611957\\
16.4	1.63661611957\\
16.8	1.63661611957\\
17.2	1.63698979959\\
17.6	1.63698979959\\
18	1.63698979959\\
18.4	1.64422597501\\
18.8	1.64422597501\\
19.2	1.64494818662\\
19.6	1.64494818662\\
};
\addlegendentry{$\mass = 2.5$}

\addplot [myline1, only marks]
table[row sep=crcr]{%
0	1.00018951187\\
0.2	1.06209122213\\
0.4	1.1745472581\\
0.6	1.27748795023\\
0.8	1.3403752483\\
1	1.36921134167\\
1.2	1.37437859158\\
1.4	1.37437859158\\
1.6	1.37688761179\\
1.8	1.37967701946\\
2	1.38906429491\\
2.2	1.39592942751\\
2.4	1.39592942751\\
2.6	1.39592942751\\
2.8	1.39592942751\\
3	1.39725839275\\
3.2	1.40219482371\\
3.4	1.40219482371\\
3.6	1.40219482371\\
3.8	1.40653044924\\
4	1.40653044924\\
4.2	1.40653044924\\
4.4	1.40653044924\\
4.6	1.40653044924\\
4.8	1.40667315337\\
};
\addlegendentry{$\mass = 10$}

\addplot [myline2, only marks]
table[row sep=crcr]{%
0	1.00005844291\\
0.0632456	1.0405590983\\
0.126491	1.1175224206\\
0.189737	1.19587884009\\
0.252982	1.25645904318\\
0.316228	1.29220294225\\
0.379473	1.30667229646\\
0.442719	1.30846034542\\
0.505964	1.30846034542\\
0.56921	1.30912208176\\
0.632456	1.31402965227\\
0.695701	1.31979577412\\
0.758947	1.32383837102\\
0.822192	1.32446753301\\
0.885438	1.32446753301\\
0.948683	1.32446753301\\
1.01193	1.32619698679\\
1.07517	1.32828733849\\
1.13842	1.33012726276\\
1.20167	1.33060709623\\
1.26491	1.33060709623\\
1.32816	1.33064255524\\
1.3914	1.33125009139\\
1.45465	1.33215326582\\
1.51789	1.33349392355\\
1.58114	1.33381963782\\
1.64438	1.33381963782\\
1.70763	1.3339566555\\
1.77088	1.3342032167\\
1.83412	1.33454035389\\
1.89737	1.33554234666\\
1.96061	1.33575586612\\
2.02386	1.33575586612\\
2.0871	1.33616597321\\
2.15035	1.33633940258\\
2.21359	1.33635934344\\
2.27684	1.33701452295\\
2.34009	1.33737348855\\
2.40333	1.33737348855\\
2.46658	1.33746464755\\
2.52982	1.33775013215\\
2.59307	1.33775013215\\
2.65631	1.33818101398\\
2.71956	1.33839869872\\
2.7828	1.33839869872\\
2.84605	1.33842637397\\
2.9093	1.33874800411\\
2.97254	1.33874800411\\
3.03579	1.33885984457\\
3.09903	1.33914308348\\
3.16228	1.33914308348\\
3.22552	1.33915047765\\
3.28877	1.33939564062\\
3.35201	1.33939564062\\
3.41526	1.33941137187\\
3.47851	1.33972768693\\
3.54175	1.33972768693\\
3.605	1.33972768693\\
3.66824	1.3401113375\\
3.73149	1.3401113375\\
3.79473	1.3401113375\\
3.85798	1.3401113375\\
3.92122	1.3401113375\\
3.98447	1.3401958796\\
};
\addlegendentry{$\mass = 40$}

% matels compatible with 0 beyond t*g = 10.8 (zlight*g = 15.2735, 2_5)
\addplot [myline5, dashed, mark=none, forget plot]
table[row sep=crcr]{%
10.8	0.962\\
10.8	1.672\\
};

% matels compatible with 0 beyond t*g = 3.4 (zlight*g = 4.80833, 10)
\addplot [myline1, dashed, mark=none, forget plot]
table[row sep=crcr]{%
3.4	0.962\\
3.4	1.672\\
};

% matels compatible with 0 beyond t*g = 2.27684 (zlight*g = 3.21994, 40)
\addplot [myline2, dashed, mark=none, forget plot]
table[row sep=crcr]{%
2.27684	0.962\\
2.27684	1.672\\
};

\end{axis}
% ---- inset: position of the entanglement maximum ----
\begin{axis}[%
at={(main.south east)}, anchor=south east, xshift=-10pt, yshift=14pt,
width=0.38*\figWidth,
height=0.4*\figHeight,
scale only axis,
xmin=0,
xmax=20.38,
ymin=-3.484,
ymax=0.484,
xlabel={$t g$},
ylabel={$\zmax g$},
xlabel style={font=\tiny},
ylabel style={font=\tiny, rotate=-90},
tick label style={font=\tiny, /pgf/number format/fixed}, scaled ticks=false,
axis background/.style={fill=white},
xmajorgrids,
ymajorgrids
]
\addplot [myline5, only marks, mark size=1pt, forget plot]
table[row sep=crcr]{%
0	-0.1\\
0.4	0.1\\
0.8	0.1\\
1.2	0.1\\
1.6	-0.3\\
2	-0.3\\
2.4	-0.7\\
2.8	-0.3\\
3.2	-0.3\\
3.6	-0.3\\
4	-0.7\\
4.4	-0.7\\
4.8	-1.1\\
5.2	-1.1\\
5.6	-1.5\\
6	-0.7\\
6.4	-1.9\\
6.8	-1.1\\
7.2	-1.1\\
7.6	-2.3\\
8	-1.5\\
8.4	-1.5\\
8.8	-1.9\\
9.2	-1.9\\
9.6	-1.9\\
10	-2.3\\
10.4	-2.3\\
10.8	-2.3\\
11.2	-1.5\\
11.6	-2.7\\
12	-2.7\\
12.4	-2.7\\
12.8	-1.9\\
13.2	-1.9\\
13.6	-1.9\\
14	-1.9\\
14.4	-1.9\\
14.8	-3.1\\
15.2	-3.1\\
15.6	-3.1\\
16	-3.1\\
16.4	-3.1\\
16.8	-3.1\\
17.2	-3.1\\
17.6	-1.9\\
18	-1.9\\
18.4	-1.9\\
18.8	-1.9\\
19.2	-1.9\\
19.6	-2.7\\
};
\addplot [myline1, only marks, mark size=1pt, forget plot]
table[row sep=crcr]{%
0	-0.05\\
0.2	0.05\\
0.4	0.05\\
0.6	0.05\\
0.8	0.05\\
1	-0.05\\
1.2	-0.05\\
1.4	0.05\\
1.6	0.05\\
1.8	0.05\\
2	-0.15\\
2.2	-0.15\\
2.4	-0.15\\
2.6	-0.35\\
2.8	-0.25\\
3	-0.15\\
3.2	-0.15\\
3.4	-0.15\\
3.6	-0.35\\
3.8	-0.35\\
4	-0.35\\
4.2	-0.55\\
4.4	-0.55\\
4.6	-0.35\\
4.8	-0.35\\
};
\addplot [myline2, only marks, mark size=1pt, forget plot]
table[row sep=crcr]{%
0	-0.0158114\\
0.0632456	0.0158114\\
0.126491	0.0158114\\
0.189737	0.0158114\\
0.252982	0.0158114\\
0.316228	-0.0158114\\
0.379473	-0.0158114\\
0.442719	-0.0158114\\
0.505964	0.0158114\\
0.56921	0.0158114\\
0.632456	0.0158114\\
0.695701	-0.0158114\\
0.758947	-0.0158114\\
0.822192	-0.0158114\\
0.885438	0.0158114\\
0.948683	0.0158114\\
1.01193	0.0158114\\
1.07517	-0.0158114\\
1.13842	-0.0158114\\
1.20167	-0.0158114\\
1.26491	0.0158114\\
1.32816	0.0158114\\
1.3914	0.0158114\\
1.45465	-0.0158114\\
1.51789	-0.0158114\\
1.58114	-0.0158114\\
1.64438	0.0158114\\
1.70763	-0.0790569\\
1.77088	-0.0790569\\
1.83412	0.0158114\\
1.89737	-0.0474342\\
1.96061	-0.0474342\\
2.02386	-0.0474342\\
2.0871	-0.0790569\\
2.15035	-0.0790569\\
2.21359	-0.0474342\\
2.27684	-0.0474342\\
2.34009	-0.0474342\\
2.40333	-0.0474342\\
2.46658	-0.0790569\\
2.52982	-0.0790569\\
2.59307	-0.0474342\\
2.65631	-0.0474342\\
2.71956	-0.0474342\\
2.7828	-0.0474342\\
2.84605	-0.0790569\\
2.9093	-0.0790569\\
2.97254	-0.0474342\\
3.03579	-0.0474342\\
3.09903	-0.0474342\\
3.16228	-0.0474342\\
3.22552	-0.11068\\
3.28877	-0.11068\\
3.35201	-0.11068\\
3.41526	-0.0474342\\
3.47851	-0.0474342\\
3.54175	-0.0474342\\
3.605	-0.11068\\
3.66824	-0.11068\\
3.73149	-0.11068\\
3.79473	-0.0474342\\
3.85798	-0.142302\\
3.92122	-0.0474342\\
3.98447	-0.11068\\
};
\end{axis}
\end{tikzpicture}%
		\end{minipage}%
	}\hfill
	\subfloat[first excited state\label{sfig:entanglement_excited}]{%
		\begin{minipage}{\columnwidth}
			%!TEX root =  ../Schwinger_PDF_LCDA.tex
% written by Make_entropy_panels_tikz.m -- S_max vs t*g (positive
% branch, linear axes), first excited state, all production masses;
% inset: position of the entanglement maximum, z^max*g vs t*g (t > 0 branch).
% EVEN startsites only; parameters (x, L, D=chi): mt=1: (20,90,200);
% mt=10: (100,100,120); all: Lcut=10, Ntau=100. Dashed line: t*g
% beyond which the matels are compatible with 0 (mt=10: 5.5; mt=1: never).
\begin{tikzpicture}
\begin{axis}[%
name=main,
width=\figWidth,
height=\figHeight,
scale only axis,
xmin=0,
xmax=12,
xlabel style={font=\color{white!15!black}},
xlabel={$t g$},
ymin=1.871,
ymax=3.465,
ylabel style={font=\color{white!15!black}},
ylabel={$S_\text{max}$},
axis background/.style={fill=white},
xmajorgrids,
ymajorgrids,
ylabel style={{rotate=-90}}, scaled ticks=false, tick label style={/pgf/number format/fixed},
xticklabel style={yshift=-.5ex},
yticklabel style={xshift=-.5ex}
]

\addplot [myline5, only marks]
table[row sep=crcr]{%
0	1.95599958779\\
0.4	2.22988155727\\
0.8	2.43952173724\\
1.2	2.59312192092\\
1.6	2.65710938781\\
2	2.68189507372\\
2.4	2.71399470344\\
2.8	2.73006674041\\
3.2	2.77906382266\\
3.6	2.81513777325\\
4	2.84401017075\\
4.4	2.87786812112\\
4.8	2.91057602748\\
5.2	2.93045643351\\
5.6	2.95955585479\\
6	2.99953223173\\
6.4	3.00880160085\\
6.8	3.03879594127\\
7.2	3.07055318445\\
7.6	3.09466867356\\
8	3.12574797453\\
8.4	3.14122339785\\
8.8	3.18097634839\\
9.2	3.20651714485\\
9.6	3.22360405048\\
10	3.25981426202\\
10.4	3.27433414554\\
10.8	3.3030628904\\
11.2	3.32909555198\\
11.6	3.36881551093\\
12	3.37964929519\\
12.4	3.41620726501\\
12.8	3.43723378838\\
13.2	3.4666132555\\
13.6	3.49351499376\\
14	3.52020902136\\
14.4	3.55053719747\\
14.8	3.59326096832\\
15.2	3.61163882959\\
15.6	3.64809277365\\
16	3.66350121065\\
16.4	3.69779386888\\
16.8	3.7239756902\\
17.2	3.74796065478\\
17.6	3.77791476887\\
18	3.79577907139\\
18.4	3.83377350847\\
18.8	3.85412304957\\
19.2	3.87867740849\\
19.6	3.90427096004\\
};

\addplot [myline1, only marks]
table[row sep=crcr]{%
0	1.99257382393\\
0.2	2.2870925233\\
0.4	2.2870925233\\
0.6	2.51990059241\\
0.8	2.51990059241\\
1	2.59249211973\\
1.2	2.59249211973\\
1.4	2.59249211973\\
1.6	2.59249211973\\
1.8	2.59249211973\\
2	2.60480808505\\
2.2	2.60480808505\\
2.4	2.60733896683\\
2.6	2.61401972848\\
2.8	2.61918388423\\
3	2.61918388423\\
3.2	2.62766363553\\
3.4	2.62766363553\\
3.6	2.65110606951\\
3.8	2.65560630877\\
4	2.67193902251\\
4.2	2.68520321898\\
4.4	2.69338592963\\
4.6	2.69467008758\\
4.8	2.71250746777\\
};

\addplot [myline2, only marks]
table[row sep=crcr]{%
0	2.18275934435\\
0.0632456	2.33689637827\\
0.126491	2.34914662824\\
0.189737	2.46815575023\\
0.252982	2.54919566654\\
0.316228	2.54973488731\\
0.379473	2.59500939023\\
0.442719	2.59500939023\\
0.505964	2.59500939023\\
0.56921	2.59750584755\\
0.632456	2.59750584755\\
0.695701	2.60331509222\\
0.758947	2.60331509222\\
0.822192	2.60625278138\\
0.885438	2.60770891263\\
0.948683	2.60770891263\\
1.01193	2.60770891263\\
1.07517	2.61374696768\\
1.13842	2.61374696768\\
1.20167	2.61374696768\\
1.26491	2.61374696768\\
1.32816	2.61374696768\\
1.3914	2.61718288136\\
1.45465	2.61718288136\\
1.51789	2.61718288136\\
1.58114	2.61718288136\\
1.64438	2.61730505529\\
1.70763	2.61730505529\\
1.77088	2.61730505529\\
1.83412	2.61730505529\\
1.89737	2.62061135611\\
1.96061	2.62061135611\\
2.02386	2.62061135611\\
2.0871	2.62061135611\\
2.15035	2.62061135611\\
2.21359	2.62288364673\\
2.27684	2.62288364673\\
2.34009	2.62288364673\\
2.40333	2.62288364673\\
2.46658	2.62352456967\\
2.52982	2.62352456967\\
2.59307	2.62352456967\\
2.65631	2.62352456967\\
2.71956	2.62744437765\\
2.7828	2.62744437765\\
2.84605	2.62744437765\\
2.9093	2.62744437765\\
2.97254	2.62942229098\\
3.03579	2.6307460137\\
3.09903	2.6307460137\\
3.16228	2.6317267335\\
3.22552	2.63363269569\\
3.28877	2.6352303333\\
3.35201	2.6352303333\\
3.41526	2.6352303333\\
3.47851	2.63996404833\\
3.54175	2.64003397305\\
3.605	2.64003397305\\
3.66824	2.64184167544\\
3.73149	2.64576386056\\
3.79473	2.64726603129\\
3.85798	2.64726603129\\
3.92122	2.64841418515\\
3.98447	2.6546500848\\
};

% matels compatible with 0 beyond t*g = 10.8 (zlight*g = 15.2735, 2_5)
\addplot [myline5, dashed, mark=none, forget plot]
table[row sep=crcr]{%
10.8	1.871\\
10.8	3.465\\
};

% matels compatible with 0 beyond t*g = 3.4 (zlight*g = 4.80833, 10)
\addplot [myline1, dashed, mark=none, forget plot]
table[row sep=crcr]{%
3.4	1.871\\
3.4	3.465\\
};

% matels compatible with 0 beyond t*g = 2.27684 (zlight*g = 3.21994, 40)
\addplot [myline2, dashed, mark=none, forget plot]
table[row sep=crcr]{%
2.27684	1.871\\
2.27684	3.465\\
};

\end{axis}
% ---- inset: position of the entanglement maximum ----
\begin{axis}[%
at={(main.south east)}, anchor=south east, xshift=-10pt, yshift=14pt,
width=0.38*\figWidth,
height=0.4*\figHeight,
scale only axis,
xmin=0,
xmax=20.38,
ymin=-0.444,
ymax=1.044,
xlabel={$t g$},
ylabel={$\zmax g$},
xlabel style={font=\tiny},
ylabel style={font=\tiny, rotate=-90},
tick label style={font=\tiny, /pgf/number format/fixed}, scaled ticks=false,
axis background/.style={fill=white},
xmajorgrids,
ymajorgrids
]
\addplot [myline5, only marks, mark size=1pt, forget plot]
table[row sep=crcr]{%
0	-0.1\\
0.4	0.1\\
0.8	0.1\\
1.2	0.1\\
1.6	-0.1\\
2	-0.3\\
2.4	0.1\\
2.8	-0.3\\
3.2	0.1\\
3.6	-0.3\\
4	0.1\\
4.4	0.1\\
4.8	0.5\\
5.2	0.1\\
5.6	0.5\\
6	0.1\\
6.4	0.5\\
6.8	0.5\\
7.2	0.5\\
7.6	0.5\\
8	0.9\\
8.4	0.5\\
8.8	0.1\\
9.2	0.5\\
9.6	0.9\\
10	0.5\\
10.4	0.5\\
10.8	0.1\\
11.2	0.9\\
11.6	0.1\\
12	0.5\\
12.4	0.1\\
12.8	0.5\\
13.2	0.1\\
13.6	0.5\\
14	0.5\\
14.4	0.1\\
14.8	-0.3\\
15.2	0.1\\
15.6	-0.3\\
16	0.1\\
16.4	-0.3\\
16.8	-0.3\\
17.2	-0.3\\
17.6	-0.3\\
18	0.1\\
18.4	-0.3\\
18.8	0.1\\
19.2	-0.3\\
19.6	0.1\\
};
\addplot [myline1, only marks, mark size=1pt, forget plot]
table[row sep=crcr]{%
0	-0.05\\
0.2	-0.05\\
0.4	0.05\\
0.6	0.05\\
0.8	-0.05\\
1	-0.05\\
1.2	-0.15\\
1.4	0.05\\
1.6	0.05\\
1.8	-0.15\\
2	0.05\\
2.2	-0.15\\
2.4	0.05\\
2.6	-0.15\\
2.8	0.05\\
3	-0.15\\
3.2	0.05\\
3.4	0.15\\
3.6	0.25\\
3.8	0.05\\
4	0.25\\
4.2	0.05\\
4.4	0.25\\
4.6	0.35\\
4.8	0.25\\
};
\addplot [myline2, only marks, mark size=1pt, forget plot]
table[row sep=crcr]{%
0	-0.0158114\\
0.0632456	0.0158114\\
0.126491	0.0158114\\
0.189737	0.0158114\\
0.252982	0.0158114\\
0.316228	-0.0158114\\
0.379473	0.0158114\\
0.442719	-0.0158114\\
0.505964	-0.0474342\\
0.56921	0.0158114\\
0.632456	-0.0474342\\
0.695701	0.0158114\\
0.758947	-0.0158114\\
0.822192	-0.0474342\\
0.885438	0.0158114\\
0.948683	-0.0474342\\
1.01193	0.0158114\\
1.07517	0.0158114\\
1.13842	-0.0474342\\
1.20167	0.0158114\\
1.26491	0.0158114\\
1.32816	-0.0474342\\
1.3914	0.0158114\\
1.45465	-0.0474342\\
1.51789	0.0158114\\
1.58114	0.0158114\\
1.64438	-0.0474342\\
1.70763	0.0158114\\
1.77088	-0.0474342\\
1.83412	0.0790569\\
1.89737	0.0158114\\
1.96061	-0.0474342\\
2.02386	0.0158114\\
2.0871	0.0158114\\
2.15035	0.0790569\\
2.21359	0.0158114\\
2.27684	-0.0474342\\
2.34009	0.0790569\\
2.40333	0.0158114\\
2.46658	0.0790569\\
2.52982	0.0158114\\
2.59307	0.0158114\\
2.65631	0.0790569\\
2.71956	0.0158114\\
2.7828	0.0790569\\
2.84605	0.0790569\\
2.9093	0.0158114\\
2.97254	0.0790569\\
3.03579	0.0158114\\
3.09903	0.0790569\\
3.16228	0.0790569\\
3.22552	0.0158114\\
3.28877	0.0790569\\
3.35201	0.0158114\\
3.41526	0.0158114\\
3.47851	0.0790569\\
3.54175	0.0158114\\
3.605	0.0790569\\
3.66824	0.0790569\\
3.73149	0.142302\\
3.79473	0.0790569\\
3.85798	0.0158114\\
3.92122	0.142302\\
3.98447	0.0790569\\
};
\end{axis}
\end{tikzpicture}%
		\end{minipage}%
	}
	\caption{Maximum entanglement entropy during the time evolution at times $t$ when computing matrix elements connecting an even site at the origin with an even site at a different light-front distance $\zlight = \sqrt{2} t$. The initial state is \protect\subref{sfig:entanglement_gs} the ground state (vacuum) or \protect\subref{sfig:entanglement_excited} the first excited state (vector meson); dashed lines indicate the times $t_0$ where LCDA matrix elements vanish: $\left| \frac{\mel_{\text{LCDA}}(t \ge t_0)}{\mel_{\text{LCDA}}(0)} \right| \le 10^{-3}$. The lattice parameters are the base parameters shown in \cref{tab:sim_params} of \cref{app:extrapolations}, but at the largest available bond dimension. Insets: spatial coordinate \zmax where the entanglement entropy is maximal.}
	\label{fig:entanglement}
\end{figure}
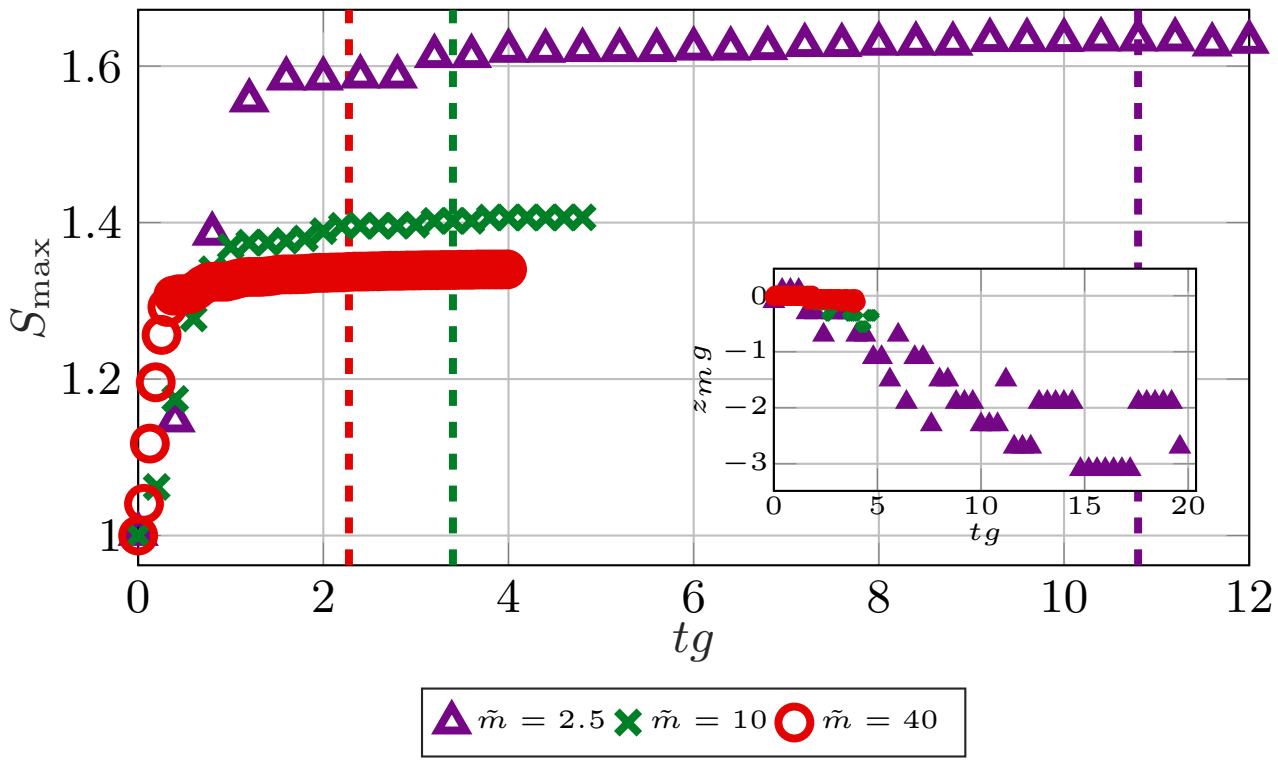
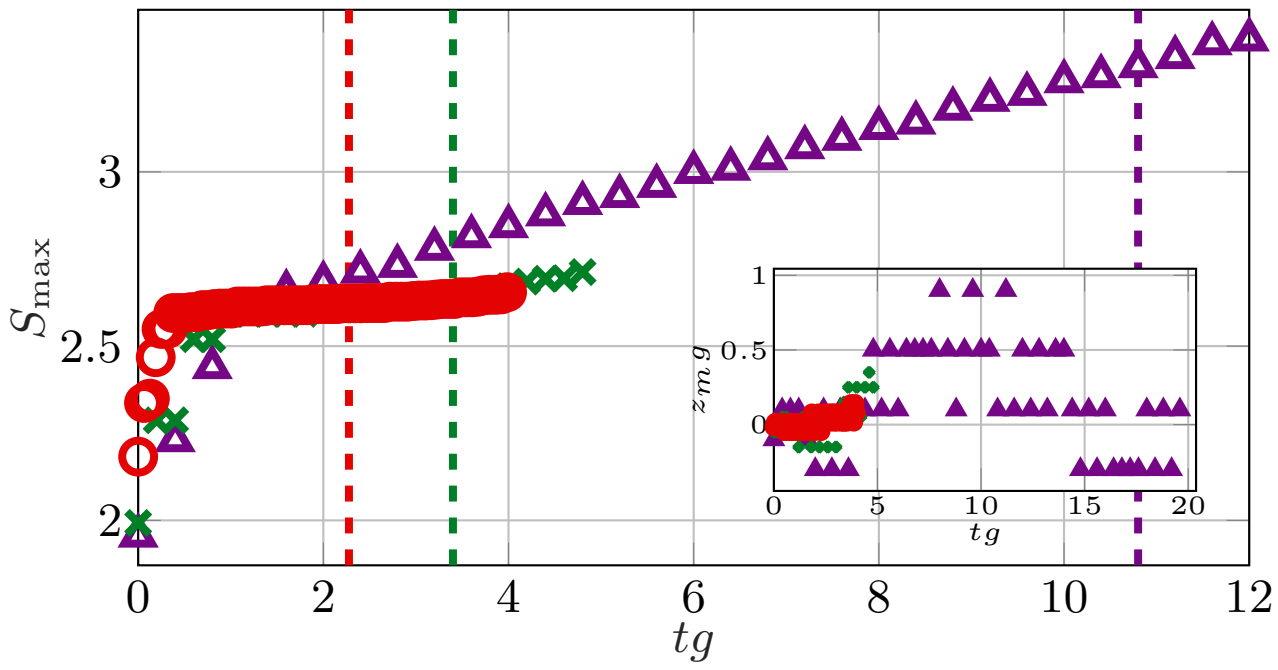

In general, the bond dimension required to describe the time-evolved states grows fast with time, and limits the reach of a faithful MPS description to only short times. Even though the entanglement and subsequent bond dimension growth are expected to be slower in the case of local perturbations onto an equilibrium state~\cite{Calabrese_2007,Vasseur2017local}, we need to systematically control the convergence of the simulations. To monitor the entanglement growth, we compute the von Neumann entanglement entropy $S_A = - \trace{\rho_A \log_2(\rho_A)}$ in the evolution, where $\rho_A$ is the reduced density matrix of a subsystem $A$. We calculate $S_A$ for all possible cuts of the linear chain and identify the maximum $S_\text{max}$ at each evolution time $t = \frac{\zlight}{\sqrt{2}}$ as shown in \cref{fig:entanglement}. We observe that the location \zmax of the maximum remains close to the center of the chain, where the perturbation by the fermion operator was applied [see insets in \cref{fig:entanglement}]. Starting in the ground state, we observe that the maximal entanglement grows fast at the beginning of the evolution, and then becomes almost constant. In the case of the meson state, the entanglement entropy keeps increasing slowly, faster for smaller fermion masses, but it stays far below the maximal value allowed by the system size. We vary the bond dimension $\chi$ of the time-evolved MPS between runs and check that the entanglement growth is not limited by the values of $\chi$ used in the calculations.

Even though a small entanglement entropy does not guarantee that the MPS description is accurate, this check provides a picture of how the states' complexity evolves after perturbing them along the light front, and further supports the idea that tensor networks can reliably represent the time-evolved states. Furthermore, the decay of matrix elements to zero for large light-front distances [see, e.g., \cref{fig:lcda_matrix_elements}] implies that the evolution times that we need to simulate are limited. For example, for the case of mass $\mass = 2.5$, for which we observe the fastest entanglement growth in \cref{fig:entanglement}, the LCDA matrix elements vanish for $|t| \ge 10.8 / g$. This is indicated by dashed lines in the figure, and the maximal entanglement entropy is still modest ($S_\text{max} < 3.31$) there. A further evolution all the way to the boundary at $t = 19.6 / g$ is therefore, in principle, not necessary.

In our error analysis, we check that the necessary bond dimensions remain moderate in our calculations ($D = \chi \in [80, 160]$, with larger bond dimensions required for smaller masses in agreement with the faster entanglement growth), while attaining convergence of the computed quantities. We conclude that the entanglement growth is not a major obstacle for the simulations [see also \cref{fig:PDF_errors,fig:LCDA_errors} in \cref{app:extrapolations} for contributions to the uncertainties].

\section{Conclusion and outlook}
\label{sec:conclusion}
%!TEX root = ./Schwinger_PDF_LCDA.tex

We demonstrated that parton functions can be directly calculated in Minkowski space from light-front matrix elements with controlled uncertainties, using tensor network methods to efficiently represent and time-evolve states. In particular, we calculated fermion and antifermion PDFs as well as LCDAs and decay constants of the vector meson in the Schwinger model for several fermion masses. Systematic uncertainties were quantified by varying lattice parameters, and we provide error bands for all calculated quantities. We discussed symmetries of the matrix elements and parton functions and found good agreement of our data with the expectations and known analytical limits of vanishing or infinite fermion masses.

This direct Hamiltonian approach does not require an expansion in Mellin moments or a perturbative matching in the UV regime and therefore provides a complementary approach to Euclidean methods. Boosts to large momenta of the hadrons are not required, and the meson can be studied at rest or, as in this work, with open boundary conditions.

We implemented a stepwise evolution of the light-front Wilson line in which spatial shifts of an external charge alternate with time-evolution steps. The numerically most demanding part of the algorithm is the time evolution with the electric term. At fixed bond dimension, the algorithm scales quadratically in the number of lattice sites and therefore allows for efficient time evolution. We found that the time-evolved states can be represented with moderate bond dimensions $\chi$. Generally, the entanglement and therefore the required $\chi$ grow with the evolution time. We observed that this growth is mild in the Schwinger model and bond dimensions of $\chi \in [80, 160]$ are sufficient, with slightly larger $\chi$ for smaller fermion masses. We attribute the slow entanglement growth to the fact that the matrix elements require only localized perturbations instead of global quenches. Additionally, the decay of the matrix elements for large light-front separations limits the necessary evolution times. These findings support the use of tensor network states, which represent states with limited entanglement entropy, for light-front matrix elements.

With our parameter choices, the uncertainties are mostly dominated by finite-volume effects and lattice artifacts. Further improvements can thus be achieved by increasing the volume and the inverse lattice spacing, with the number of lattice sites being proportional to both of them.

Looking ahead, the success of the PDF and LCDA calculations in the Schwinger model further encourages the application of TNS to gauge theory problems, with the outlook of approaching QCD calculations.  Moving in this direction requires, on the one hand, dealing with more complex gauge theories.  In this sense, TNS have already been applied to some extent to low-dimensional multi-flavor and non-Abelian models [see Refs.~\cite{banuls2020ropp,banuls2020epjd,meurice2022trg} for reviews]. On the other hand, a crucial step is simulating lattice gauge theories in higher dimensions, where transverse directions and spin exist, and gauge bosons become truly dynamical degrees of freedom.  This will open the possibility to study gauge-boson distributions, generalized parton distributions, and transverse-momentum-dependent parton distributions. Progress of TNS in higher dimensions has been impressive, allowing for the first simulations of a 3+1 dimensional lattice gauge theory~\cite{magnifico2021qed3d}, but important challenges remain, including the increased algorithmic complexity and numerical costs, and the fact that gauge degrees of freedom need to be discretized and truncated in higher dimensions~\cite{Magnifico:2024eiy}. It is worth noting that many of the challenges faced by TNS methods in their potential application to high-energy problems are shared by quantum simulation and quantum computing proposals [see Refs.~\cite{Zohar2015,halimeh2023coldatom,Bauer:2022hpo,dimeglio2023quantum} for reviews], so that advances in one area may help developments in the other.

Another difficulty in higher-dimensional simulations is renormalization. Because of the presence of UV divergences, the lattice-regularized matrix elements are often not well defined at vanishing lattice spacing. Before performing the continuum extrapolation, it is therefore necessary to subtract these divergences and match the lattice results to a scheme that is suitable for investigating continuum physics. Such a renormalization may have to be implemented non-perturbatively due to the appearance of power-law and cusp UV divergences in a stepwise Wilson line on the lattice. Even though TNS calculations have been able to identify and to subtract UV divergences in the Schwinger model~\cite{Buyens2015pos,buyens2016confinement}, the task may be more challenging for higher dimensions, and a dedicated study of the problem is necessary.

Quantum hardware could open up another route to computing parton functions, exploiting its natural capability to efficiently simulate quantum dynamics. Recent works have indeed started exploring the possibility of addressing hadron physics with quantum devices~\cite{Lamm2020,Echevarria2021,Kreshchuk2021,Kreshchuk2021Entropy,Kreshchuk2022,Li2022,Qian2022,Chen2025,GalvezViruet2025,GalvezViruet2026,Zou2026}. The Hamiltonian formalism presented in this paper could in principle also be implemented on quantum computers or quantum simulators, even though a large number of Trotter steps, such as we used in our calculations, is difficult to achieve in current quantum hardware, limited by noise. Nevertheless, studying the different approaches for further gauge theories towards QCD and identifying the challenges and opportunities for parton physics from TNS and quantum computers remain relevant and exciting research directions.

\begin{acknowledgments}
	We thank Hsiang-nan Li, Enrique Rico, and Torsten Zache for very helpful discussions. This work was partly supported by the DFG (German Research Foundation) under Germany's Excellence Strategy -- EXC-2111 -- 390814868, and Research Unit FOR 5522 (Grant No. 499180199), and by the EU-QUANTERA project TNiSQ (BA 6059/1-1), as well as Taiwanese NSTC grants 112-2112-M-A49-021-MY3, 112-2811-M-A49-543-MY2, 113-2119-M-007-013, 113-2123-M-A49-001, 114-2119-M-007-013, 114-2123-M-A49-001, and 115-2811-M-A49-501. K.C.\ is supported by the National Science Centre (Poland) grant OPUS No.\ 2021/43/B/ST2/00497.
\end{acknowledgments}

%%%%%%%%%%%%%%%%%%%%%%%%%%%%%%%%%%%%%%%%%%%%%%%%%%%%%%%%%
%                           BIBLIOGRAPHY
%%%%%%%%%%%%%%%%%%%%%%%%%%%%%%%%%%%%%%%%%%%%%%%%%%%%%%%%%
\bibliography{references} 

%%%%%%%%%%%%%%%%%%%%%%%%%%%%%%%%%%%%%%%%%%%%%%%%%%%%%%%%%
%                           APPENDIX
%%%%%%%%%%%%%%%%%%%%%%%%%%%%%%%%%%%%%%%%%%%%%%%%%%%%%%%%%
%\clearpage
\appendix

%!TEX root = ./Schwinger_PDF_LCDA.tex

\section{The Schwinger model: from Lagrangian density to spin Hamiltonian}
\label{app:SchwingerModel}

\subsection{Lagrangian and Hamiltonian}

The Lagrangian density of the massive Schwinger model is~\cite{SchwingerModelMassive}
\begin{equation}
	{\mathcal{L}} = \overline{\psi} (i \slashed{\partial}- g \slashed{A }-m )\psi-\frac{1}{4}F_{\mu\nu}F^{\mu \nu} - A_{0} \rho .
\end{equation}
Here, $\psi$ are Dirac fermions, $A_\mu$ is the $U(1)$ gauge field, $g$ is the coupling strength, $m$ is the bare fermion mass, and $F_{\mu\nu} = \partial_{\mu}A_{\nu}-\partial_{\nu}A_{\mu}$ is the field strength tensor. We include a coupling of the gauge field to an external charge density $\rho$.

The Hamiltonian density obtained by a Legendre transformation is
\begin{align}
	{\mathcal{H}} =& -i \overline{\psi} \gamma^1 (\partial_1 + i g A_1) \psi + m \overline{\psi} \psi + \frac{1}{2} E^2 \nonumber\\
	&+ (\partial_1 A_0) E	+ A_0 \left(g \psi^\dagger \psi + \rho \right) ,
\end{align}
with electric fields $E = F_{0 1} = \partial_0 A_1 - \partial_1 A_0$. The Hamiltonian remains unchanged if the term $(\partial_1 A_0) E$ is integrated by parts, which leads to the equivalent Hamiltonian density
\begin{align}
	{\mathcal{H}} =& -i \overline{\psi} \gamma^1 (\partial_1 + i g A_1) \psi + m \overline{\psi} \psi + \frac{1}{2} E^2 \nonumber\\
	&+ A_0 \underbrace{\left(g \psi^\dagger \psi + \rho - \partial_1 E \right)}_{G} .
\end{align}

Note that $A_0$ is not a dynamical degree of freedom because its canonical momentum $\frac{\partial {\mathcal{L}}}{\partial (\partial_0 A_0)} \equiv 0$ vanishes. In the Lagrangian formalism, it acts as a Lagrange multiplier ensuring that Gauss's law $G = 0$ is fulfilled. In the Hamiltonian framework, we partially fix the gauge to remove one (non-dynamical) bosonic degree of freedom. Gauss's law is ensured by restricting physical states to those fulfilling $G \Ket{\Phi}_\text{phys} = 0$ [see \cref{eq:Gauss_law}]. In this work, we choose the temporal gauge $A^0 \equiv 0$, which leads to the Hamiltonian density in \cref{eq:HamiltonDensity}. The electric field reduces to $E = \partial_0 A_1$.

This approach is similar to the common procedure in QED in 3+1 dimensions, where the initial four degrees of freedom of the gauge field are reduced to two dynamical degrees of freedom, corresponding to the two transverse photon polarizations. This is achieved in a similar way by partially fixing the gauge (such as the Coulomb, Lorenz, or temporal gauge) and by imposing a constraint on physical states.

\subsection{Naive fermion discretization}

A naive lattice discretization with lattice spacing $a$ is
\begin{align}
	H_\text{naive} =& -\frac{i}{2} \sum_{\site} \left( \overline{\psi}_{\site} \gamma^1 e^{i\theta_\site} \psi_{\site+1} - \overline{\psi}_{\site+1} \gamma^1 e^{-i\theta_\site} \psi_{\site} \right) \nonumber\\
	&+ m_{\text{lat}} a \sum_{\site} \overline{\psi}_{\site} \psi_{\site} +\frac{a g^2}{2} \sum_{\site} L_{\site}^2 .
\end{align}
We introduced the new lattice variables $\theta$ and $L$ that correspond to the gauge field and the electric field in the continuum limit:
\begin{align}
	\frac{1}{ag} \theta_{\site} &\xrightarrow[z = \site a]{a \rightarrow 0} A_{1}(z) , \\
	g L_{\site} &\xrightarrow[z = \site a]{a \rightarrow 0} E(z) .
\end{align}

A variable transformation to new Dirac fermions $\chi$ of the form
\begin{equation}
	\psi_{\site} \rightarrow \left(\gamma^0 \gamma^1 \right)^{\site} \chi_\site
\end{equation}
diagonalizes the Dirac structure of the Hamiltonian:
\begin{align}
	H_{\chi}=&-\frac{i}{2}\sum_{\site} \left( \chi^{\dagger}_{\site} e^{i\theta_\site} \chi_{\site+1} - \chi^{\dagger}_{\site+1} e^{-i\theta_\site} \chi_{\site}\right) \nonumber\\
	&+m_\text{lat} a \sum_{\site} (-1)^{\site} \overline{\chi}_{\site} \chi_\site +\frac{a g^2}{2} \sum_{\site} L_{\site}^2 . \label{appeq:naive_fermions}
\end{align}
While the continuum limit $a \rightarrow 0$ formally reproduces the continuum Hamiltonian in \cref{eq:HamiltonDensity}, the spectrum of the naive fermion discretization is spoiled by doublers. These states occur at the boundaries of the Brillouin zone and have the same energy as the low-momentum states.

\subsection{Staggered fermions}

By keeping only one component of the Dirac spinors, the doublers can be completely removed in 1+1 dimensions. This leads to Kogut-Susskind staggered fermions~\cite{staggeredFermions}. We keep only the upper component of the $\chi$ fields in \cref{appeq:naive_fermions}, or, equivalently, the upper (lower) component of the two-component Dirac fermions $\psi$ on even (odd) sites:
\begin{equation}
	\frac{1}{\sqrt{a}} \psilat_{\site} \leftarrow
	\begin{cases*}
		\psi_\text{upper}(\site a) & \text{if}~\site~\text{even} \\
		\psi_\text{lower}(\site a) & \text{if}~\site~\text{odd}.
	\end{cases*} \label{appeq:Schwinger_phi_def}
\end{equation}
The corresponding staggered Hamiltonian
\begin{align}
	H_\text{stag}=&-\frac{i}{2 a}\sum_{\site} \left( \psilat^{\dagger}_{\site} e^{i\theta_\site} \psilat_{\site+1} - \psilat^{\dagger}_{\site+1} e^{-i\theta_\site} \psilat_{\site}\right) \nonumber\\
	&+m_\text{lat}\sum_\site(-1)^{\site} \psilat_{\site}^{\dagger} \psilat_\site
	+\frac{a g^2}{2} \sum_{\site} L_{\site}^2
	\label{appeq:Hferm}
\end{align}
converges in the naive $a \rightarrow 0$ limit to the continuum theory without doublers. The commutation relation $[\theta_\site, L_\sitesum] = i \delta_{\site,\sitesum}$ on the lattice restores the equal-time quantization relation $[A_1(z), E(z')] = i \delta(z-z')$ in the continuum limit.

The physical, two-component fermionic operators can be obtained from the staggered fermions in the continuum as
\begin{equation}
	\psi_{\spinind} (\siteeven a) = \frac{1}{\sqrt{2a}} \psilat_{\siteeven + \spinind} ,
\end{equation}
where $\spinind$ is the spinor index and $\spinind=0$ corresponds to the upper, $\spinind=1$ to the lower component, and $\siteeven$ numbers the even sites on the lattice. The combination of an even site and the following odd site thus forms a two-component Dirac spinor. The normalization is chosen such that the continuum anti-commutation relation $\left\{ \psi_{\spinind}(z), \psi^\dagger_{\spinind'}(z') \right\} = \delta(z - z') \delta_{\spinind,\spinind'}$ corresponds to $\left\{ \psilat_\site, \psilat^\dagger_\sitesum \right\} = \delta_{\site,\sitesum}$ on the lattice. Note that the reconstruction of the two-component Dirac spinors from neighboring $\psilat$ fields leads to an effective doubling of the lattice spacing compared to \cref{appeq:Schwinger_phi_def}.

The lattice Hamiltonian simplifies with a $U(1)$ gauge transformation $\psilat_{\site} \rightarrow \prod_{\sitesum \ge \site}\left(e^{-i \theta_\sitesum}\right) \psilat_\site$ to
\begin{align}
	H_\text{dec}=&-\frac{i}{2 a}\sum_{\site} \left( \psilat^{\dagger}_{\site} \psilat_{\site+1} - \psilat^{\dagger}_{\site+1} \psilat_{\site}\right) \nonumber\\
	&+m_\text{lat}\sum_\site(-1)^{\site} \psilat_{\site}^{\dagger} \psilat_\site +\frac{a g^2}{2} \sum_{\site} L_{\site}^2 .
\end{align}

The Gauss operator on the lattice is
\begin{equation}
	G_{\site} = \psilat_{\site}^\dagger \psilat_{\site} + Q_{\site} -\frac{1}{2}\left[1-(-1)^\site\right] - L_{\site} + L_{\site-1} ,
\end{equation}
with the external charge $Q_\site$ on the lattice reproducing the charge density in the continuum limit as $\frac{g}{a} Q_\site \xrightarrow[z = \site a]{a \rightarrow 0} \rho(z)$.

\subsection{From fermionic degrees of freedom to a spin model}

A Jordan-Wigner transformation allows for a replacement of fermionic operators by Pauli matrices:\footnote{The sigma matrices are
$
	\sigma^{\pm}=\frac{1}{2}\left(\sigma^{x}\pm i\sigma^{y}\right)
	= \frac{1}{2} \begin{pmatrix}
		0	&	1 \pm 1\\
		1 \mp 1	&	0
	\end{pmatrix}
$, with the Pauli matrices
$
	\sigma^{x} = \begin{pmatrix}
		0	&	1 \\
		1	&	0
	\end{pmatrix}
$ and
$
\sigma^{y} =
\begin{pmatrix}
	0	&	-i \\
	i	&	0
\end{pmatrix}
$.}
\begin{align}
	\psilat_\site&=\prod_{\sitesum<\site}\left(-i\sigma_\sitesum^{z}\right)\sigma_{\site}^{-} , \nonumber\\
	\psilat^{\dagger}_\site&=\prod_{\sitesum<\site}\left(i\sigma_\sitesum^{z}\right)\sigma_{\site}^{+} .
\end{align}

Additionally, we subtract the ground state energy $E_0 = - \sum_{\site~\text{odd}} m_\text{lat}$ in the strong-coupling limit $\frac{m_\text{lat}}{ag^2} \rightarrow 0, ag \rightarrow \infty$~\cite{Hamer1982}. The spin Hamiltonian then reads
\begin{align}
	H_\text{spin} =& H_\text{dec} - E_0 \nonumber\\
	=&\frac{1}{2 a}\sum_{\site} \left( \sigma^{+}_{\site} \sigma^{-}_{\site+1} + \sigma^{-}_{\site} \sigma^{+}_{\site+1}\right) \nonumber\\
	&+\frac{m_\text{lat}}{2}\sum_{\site=0}^{N-1} \left [ 1 + (-1)^{\site} \sigma_{\site}^z \right ] \nonumber\\
	&+\frac{a g^2}{2} \sum_{\site} L_{\site}^2 .
\end{align}

Using Gauss's law,
\begin{align}
	L_\site-L_{\site-1} =& \psilat_{\site}^{\dagger} \psilat_{\site} -\frac{1}{2}\left[1-(-1)^\site\right] + Q_{\site} \nonumber\\
	=& \sigma^{+}_\site\sigma^{-}_{\site} -\frac{1}{2}\left[1-(-1)^\site\right] + Q_{\site} \nonumber\\
	=& \frac{1}{2}\left[(-1)^{\site} + \sigma^z_{\site} \right] + Q_{\site} ,
\end{align}
the gauge degrees of freedom can be removed, and the Hamiltonian can be formulated within the Hilbert space restricted to physical states only. For open boundary conditions (obc), the Hamiltonian becomes
\begin{align}
	H_\text{spin} =& \frac{1}{2 a}\sum_{\site=0}^{N-2} \left [ \sigma_\site^+\sigma_{\site+1}^- + \sigma_\site^-\sigma_{\site+1}^+ \right ] \nonumber\\
	&+ \frac{m_\text{lat}}{2}\sum_{\site=0}^{N-1} \left [ 1 + (-1)^\site \sigma_\site^z \right ] \\
	&+ \frac{a g^2}{2} \sum_{\site=0}^{N-2} \left [ \ell +\frac{1}{2}\sum_{\sitesum=0}^\site \left((-1)^\sitesum+\sigma_\sitesum^{z} + 2 Q_{\sitesum}\right) \right ]^2 .\nonumber
\end{align}
A similar procedure can be applied for periodic or infinite boundary conditions~\cite{SchwingerMPO_infinite}. The background electric field $\ell$ is set to zero in this work.

Finally, we rescale the Hamiltonian by a factor $\frac{2}{ag^2}$ to obtain the dimensionless form
\begin{align}
	H =& \frac{2}{ag^2} H_\text{spin} \nonumber\\
	=& x \sum_{\site=0}^{N-2} \left [ \sigma_{\site}^+\sigma_{\site+1}^- + \sigma_{\site}^-\sigma_{\site+1}^+ \right ] \nonumber\\
	&+ \frac{\mu}{2} \sum_{\site=0}^{N-1} \left [ 1 + (-1)^{\site} \sigma_{\site}^z \right ] \nonumber\\
	&+ \sum_{\site=0}^{N-2} \left [ \ell +\frac{1}{2}\sum_{\sitesum=0}^{\site} \left((-1)^\sitesum+\sigma_\sitesum^{z} + 2 Q_\sitesum\right)\right ]^2 \label{appeq:Hdimless}
\end{align}
with the dimensionless variables $x = \frac{1}{a^2 g^2}$ and $\mu = 2 \sqrt{x} \frac{m_\text{lat}}{g}$. Using the additive mass renormalization $m_\text{lat} = m - \frac{a g^2}{8}$~\cite{Schwinger_latticeMass,MassRenormalizationNumerical,MassRenormalizationNumerical_PoS} and $\ell = 0$ leads to \cref{eq:Hdimless}.

%!TEX root = ./Schwinger_PDF_LCDA.tex

\section{Matrix product operator formulation of the Schwinger model}
\label{app:mpo}

We make use of the MPO form in the variational optimization of states, when applying Trotterized time-evolution operators, and in the calculation of expectation values such as the energy. The Hamiltonian in \cref{appeq:Hdimless} of \cref{app:SchwingerModel} can be represented by local MPO tensors $W_\site$. The tensors have two internal indices, each connecting to one nearest neighbor. Taking the ordered product of all MPO tensors, where a summation is assumed over these internal indices of dimension 5, recovers the original Hamiltonian: $H = \prod_{\site=0}^{N-1} W_\site$. The MPO tensors act as operators on the physical degrees of freedom and, thus, have two physical indices each. We use open boundary conditions with the number of lattice sites $N$ assumed to be even~\cite{Schwinger_massSpectrum} [see Ref.~\cite{SchwingerMPO_infinite} for infinite systems]. Explicitly, the Hamiltonian is expressed by the MPO tensors:
\begin{widetext}
\begin{align}
	W_\site =& \begin{pmatrix}
		\eye_\site	&	\sqrt{\frac{x}{2}} \sigma^x_\site	&	\sqrt{\frac{x}{2}} \sigma^y_\site	&	\sigma^z_\site	&	\left( \alpha_\site^2 + \frac{\site+1}{4} + \frac{\mu}{2} \right) \eye_\site + \left( (-1)^{\site} \frac{\mu}{2} + S_\site \right) \sigma^z_\site\\
		0	&	0	&	0	&	0	&	\sqrt{\frac{x}{2}} \sigma^x_\site \\
		0	&	0	&	0	&	0	&	\sqrt{\frac{x}{2}} \sigma^y_\site \\
		0	&	0	&	0	&	\eye_\site	&	\frac{N-\site-1}{2} \sigma^z_\site \\
		0	&	0	&	0	&	0	&	\eye_\site
	\end{pmatrix} , \\
	W_0 =& \begin{pmatrix}
		\eye_0	&	\sqrt{\frac{x}{2}} \sigma^x_0	&	\sqrt{\frac{x}{2}} \sigma^y_0	&	\sigma^z_0	&	\left( \alpha_0^2 + \frac{1}{4} + \frac{\mu}{2} \right) \eye_0 + \left( \frac{\mu}{2} + S_0 \right) \sigma^z_0
	\end{pmatrix} , \\
	W_{N-1} =& \begin{pmatrix}
		\frac{\mu}{2} \eye_{N-1} -\frac{\mu}{2} \sigma^z_{N-1}\\
		\sqrt{\frac{x}{2}} \sigma^x_{N-1} \\
		\sqrt{\frac{x}{2}} \sigma^y_{N-1} \\
		0 \\
		\eye_{N-1}
	\end{pmatrix} ,
\end{align}
with
\begin{align}
	\alpha_\site =& \ell + \sum_{k=0}^\site \left( \frac{1}{2} (-1)^k + Q_k \right)
		= \ell + \frac{1}{4} \left( 1 + (-1)^\site \right) + \sum_{k=0}^\site Q_k , \\
	S_\site =& \sum_{k=\site}^{N-2} \alpha_k.
\end{align}
\end{widetext}

%!TEX root = ./Schwinger_PDF_LCDA.tex

\section{Symmetry properties}
\label{app:symmetries}

The matrix elements in \cref{eq:matel_PDF_twoargument,eq:matel_LCDA_twoargument} fulfill symmetry properties that can be derived from transformations under charge conjugation \Charge, parity transformation \Parity, and time reversal \Time.

\subsection{Operators under C, P, T transformations}

We first consider how \Charge, \Parity, and \Time act on the quantum fields. Note that \Time is an anti-unitary operator in the Hilbert space.

The Abelian gauge field $A_\mu$ is real-valued and transforms as
\begin{align}
	\Charge g A_\mu(t,z) \Charge^{-1} &= -g A_\mu(t,z) , \nonumber\\
	\Parity A_0(t,z) \Parity^{-1} &= A_0(t,-z) , \nonumber\\
	\Time A_0(t,z) \Time^{-1} &= A_0(-t,z) , \nonumber\\
	\Parity A_i(t,z) \Parity^{-1} &= -A_i(t,-z) , \nonumber\\
	\Time A_i(t,z) \Time^{-1} &= -A_i(-t,z) ,
\end{align}
where $g$ is the charge. This leads to the transformation rules of the Wilson line:
\begin{align}
	\wilson{b}{a} &= e^{-i g \int_a^b dz^\mu A_\mu(z^\mu)} , \\
	\Charge \wilson{b}{a} \Charge^{-1} &= e^{-i (-g) \int_a^b dz^\mu A_\mu(z^\mu)} \nonumber\\
	&= \wilsondagger{b}{a} = \wilson{a}{b} , \\
	\PT \wilson{b}{a} \PT^{-1} &= e^{i g \int_{a}^{b} dz^\mu A_\mu(-z^\mu)} \nonumber\\
	&= e^{-i g \int_{-a}^{-b} dz^\mu A_\mu(z^\mu)} = \wilson{-b}{-a} .
\end{align}
Note that the parity-time transformation $\PT = \Parity \Time$ is anti-unitary and conjugates complex numbers.

The fermionic operators transform as:\footnote{The normalization of the $\gamma_{C/P/T}$ matrices is constrained by the (anti)unitarity properties of \Charge, \Parity, \Time. The phase can be chosen arbitrarily and cancels in bilinears.}
\begin{align}
	\Charge \psi(t,z) \Charge^{-1} &= \gamma_C \psi^*(t,z) = \gamma^0 \gamma^1 \psi^*(t,z) , \nonumber\\
	\Parity \psi(t,z) \Parity^{-1} &= \gamma_P \psi(t,-z) = \gamma^0 \psi(t,-z) , \label{appeq:gamma_CPT} \nonumber\\
	\Time \psi(t,z) \Time^{-1} &= \gamma_T \psi(-t,z) = \gamma^0 \psi(-t,z) , \nonumber\\
	\PT \psi(t,z) \PT^{-1} &= \gamma_P \gamma_T \psi(-t,-z) = \psi(-t,-z) .
\end{align}
This can be derived from the requirement that the transformed operators need to fulfill the same Dirac equation
\begin{equation}
	\left( i \gamma^\mu (\partial_\mu + i g A_\mu) -m \right) \psi = 0 ,
\end{equation}
which leads to the requirements
\begin{equation}
	\begin{aligned}[c]
		\gamma_C^{-1} \gamma^\mu \gamma_C &= - \left(\gamma^\mu\right)^* , \\
		\gamma_P^{-1} \gamma^0 \gamma_P &= \gamma^0 , \\
		\gamma_T^{-1} \left(\gamma^0\right)^* \gamma_T &= \gamma^0 , 
	\end{aligned}
	\qquad
	\begin{aligned}[c]
		\\
		\gamma_P^{-1} \gamma^i \gamma_P &= -\gamma^i , \\
		\gamma_T^{-1} \left(\gamma^i\right)^* \gamma_T &= -\gamma^i .
	\end{aligned}
\end{equation}
For our representation of the $\gamma$ matrices in 1+1 dimensions, these conditions are fulfilled by the right-hand sides of \cref{appeq:gamma_CPT}.\footnote{Note that for classical fields, \Time is unitary instead of anti-unitary. In this case, the classical spinors transform as $\psi(t,z) \xrightarrow{\Time} \gamma_T^* \psi^*(-t,z)$, while the classical Wilson line transforms according to $\wilson{t,z}{t',z'} \xrightarrow{\Time} \wilsondagger{-t,z}{-t',z'}$. Both differ from the operator-valued fields by a conjugation.}

\subsection{Bilinear operators}
We consider the transformation behavior of bilinear operators as appearing in \cref{eq:operator}. Since $\Gamma$ is real, $\Gamma^* = \Gamma$ holds. We define $\tilde{\Gamma} = \gamma_C^\dagger \Gamma \gamma_C$ and use $\left(\gamma_C^\dagger \Gamma \gamma_C \right)^T = \Gamma$ as well as $\gamma_T^\dagger \gamma_P^\dagger \Gamma \gamma_P \gamma_T = \Gamma$ to derive the symmetry properties of bilinears:
\begin{align}
	\op(z_f, z_i) &= \psi^\dagger(z_f) \wilson{z_f}{z_i} \Gamma \psi(z_i) , \\
	\Charge \op(z_f, z_i) \Charge^{-1} &= \left( \gamma_C \psi^*(z_f) \right)^\dagger \wilson{z_i}{z_f} \Gamma \gamma_C \psi^*(z_i) \nonumber\\
	&= \sum_{a,b} \psi_a(z_f) \wilson{z_i}{z_f} \tilde{\Gamma}_{a,b} \psi_b^*(z_i) \nonumber\\
	&= - \sum_{a,b} \psi_b^*(z_i) \wilson{z_i}{z_f} \tilde{\Gamma}^T_{b,a} \psi_a(z_f) \nonumber\\
	&= - \psi^\dagger(z_i) \wilson{z_i}{z_f} \left(\gamma_C^\dagger \Gamma \gamma_C \right)^T \psi(z_f) \nonumber\\
	&= - \psi^\dagger(z_i) \wilson{z_i}{z_f} \Gamma \psi(z_f) \nonumber\\
	&= -\op(z_i, z_f) , \\
	\PT \op(z_f, z_i) \PT^{-1} &= \psi^\dagger(-z_f) \wilson{-z_f}{-z_i} \Gamma \psi(-z_i) \nonumber\\
	&= \op(-z_f, -z_i) .
\end{align}

\subsection{Matrix elements}

The bilinear operator acts on the vacuum or hadrons as external states. For hadrons $h$ that are eigenstates of \PT with eigenvalue $\eta_{PT}(h)$, the matrix elements in \cref{eq:matel} are symmetric with respect to the origin, up to a phase:\footnote{We do not consider degenerate states here and restrict ourselves to 1+1 dimensions, where no spin exists.}
\begin{align}
	\matel(z_f, z_i) =& \Braket{F| \op(z_f, z_i) |I} \nonumber\\
	=& \Braket{\PT F| \PT \op(z_f, z_i) \PT^{-1} \PT |I}^* \nonumber\\
	=& \eta_{PT}(F) \eta^*_{PT}(I) \Braket{F| \op(-z_f, -z_i) |I}^* \nonumber\\
	=& \eta_{PT}(F) \eta^*_{PT}(I) \matel^*(-z_f, -z_i) .
\end{align}
Note again that \Time is anti-unitary and matrix elements have to be conjugated.

For PDF matrix elements with $\Ket{I} = \Ket{F}$, this enforces that
\begin{align}
	\matelpdf(\zlight) &\equiv \matelpdf\left(\frac{\zlight}{2}, -\frac{\zlight}{2}\right) \nonumber\\
	&= \matel^*_{\text{PDF}}\left(-\frac{\zlight}{2}, \frac{\zlight}{2}\right) \equiv \matelpdf^*(-\zlight). \label{appeq:symmetry_PDF}
\end{align}
The real part is thus symmetric and the imaginary part antisymmetric, as was shown in the supplemental material of Ref.~\cite{SchwingerPDFletter2025} before.

For LCDA matrix elements, we can conclude that
\begin{equation}
	\matellcda\left(\frac{\zlight}{2}, -\frac{\zlight}{2}\right) = \eta_{PT} \matellcda^*\left(-\frac{\zlight}{2}, \frac{\zlight}{2}\right) ,
\end{equation}
with a phase $\eta_{PT} = \eta_{PT}(F) \eta^*_{PT}(I) = \frac{\matellcda(0,0)}{\matellcda^*(0,0)}$. Note that $\PT$ is anti-unitary, and the phase $\eta_{PT}$ can be chosen arbitrarily by the phase factor of the states, which does not change the physics. We set $\eta_{PT}=1$ by multiplying the matrix elements by a phase factor, such that $\matellcda(0,0)$ is real (and positive). In this case, we can draw the same conclusions as for the PDF matrix elements: the real part of $\matellcda$ is symmetric, while the imaginary part is antisymmetric.

For the vacuum and for hadrons that are their own antiparticles, the initial and final states are furthermore eigenstates of \Charge with eigenvalues $\eta_C(I)$ and $\eta_C(F)$. It follows:
\begin{align}
	\matel(z_f, z_i) =& \Braket{F| \Charge^{-1} \Charge \op(z_f, z_i) \Charge^{-1} \Charge |I} \nonumber\\
	=& - \eta^*_C(F) \eta_C(I) \Braket{F| \op(z_i, z_f) |I} \nonumber\\
	=& - \eta^*_C(F) \eta_C(I) \matel(z_i, z_f) . \label{appeq:matel_C}
\end{align}
The PDF matrix elements, where $F = I$, are therefore antisymmetric. Together with the previous general symmetry properties in \cref{appeq:symmetry_PDF}, this means that the real part of the PDF matrix elements has to vanish for hadrons that are their own antiparticles~\cite{SchwingerPDFletter2025}.

The LCDA matrix elements depend on the charge parity of the state. Because the vacuum fulfills $\eta_C(0) = 1$, we find that
\begin{equation}
	\matellcda\left(\frac{\zlight}{2}, -\frac{\zlight}{2}\right) = -\eta_C(h) \matellcda\left(-\frac{\zlight}{2}, \frac{\zlight}{2}\right).
\end{equation}
For the vector meson, the LCDA matrix elements are thus symmetric, and the imaginary part has to vanish (with the choice $\eta_{PT}=1$). For the scalar meson, the LCDA matrix elements would be antisymmetric and purely imaginary.

\subsection{LCDA and PDF}

We detail the relation between $\lcda(\frac{1}{2} + u)$ and $\lcda(\frac{1}{2} - u)$ in the following. A translation in the LCDA variable $\uscale$ corresponds to a translation in the $\zlight$ direction, as shown in \cref{eq:matel_LCDA_twoargument}. We can thus derive:
\begin{align}
	\lcda&\left(\frac{1}{2} \pm u \right) \nonumber\\
	&= \frac{1}{\deconst} \int_{-\infty}^{\infty} \frac{d\zlight}{2\pi} e^{i (\frac{1}{2} \mp \uscale) \zlight \Plight} \matellcda(\zlight, 0) \nonumber\\
	&= \frac{1}{\deconst} \int_{-\infty}^{\infty} \frac{d\zlight}{2\pi} e^{\mp i \uscale \zlight \Plight} \matellcda\left(\frac{\zlight}{2}, -\frac{\zlight}{2} \right) .
\end{align}

We can relate $\lcda(\frac{1}{2} + u)$ and $\lcda(\frac{1}{2} - u)$ by the substitution $\zlight \rightarrow -\zlight$ and using \cref{appeq:matel_C}:
\begin{align}
	\lcda&\left(\frac{1}{2} - u \right) \nonumber\\
	&= \frac{1}{\deconst} \int_{-\infty}^{\infty} \frac{d\zlight}{2\pi} e^{i \uscale \zlight \Plight} \matellcda\left(\frac{\zlight}{2}, -\frac{\zlight}{2} \right) \nonumber\\
	&= -\frac{1}{\deconst} \int_{\infty}^{-\infty} \frac{d\zlight}{2\pi} e^{-i \uscale \zlight \Plight} \matellcda\left(-\frac{\zlight}{2}, \frac{\zlight}{2} \right) \nonumber\\
	&= \frac{1}{\deconst} \int_{-\infty}^{\infty} \frac{d\zlight}{2\pi} e^{-i \uscale \zlight \Plight} \nonumber\\
	&\phantom{= \frac{1}{\deconst} \int_{-\infty}^{\infty} \frac{d\zlight}{2\pi}} \left( - \eta_C(h) \matellcda\left(\frac{\zlight}{2}, -\frac{\zlight}{2} \right) \right) \nonumber\\
	&= - \eta_C(h) \lcda\left(\frac{1}{2} + u \right) .
\end{align}
For hadrons with charge conjugation number $\eta_C(h) = -1$ ($\eta_C(h) = +1$), the LCDA is thus symmetric (antisymmetric) with respect to $\uscale = \frac{1}{2}$.

Similarly, one can derive for eigenstates of the charge conjugation operator that
\begin{align}
	\pdf = - \pdfbare(-\bjorx).
\end{align}
Together with \cref{eq:antifermion_from_fermion_PDF}, it follows that the fermion PDF and the antifermion PDF are equal for hadrons that are their own antiparticles, as expected.

%!TEX root = ./Schwinger_PDF_LCDA.tex

\section{Data analysis, extrapolations and systematic errors}
\label{app:extrapolations}

To study systematic errors, we follow a similar procedure to the one outlined for PDFs in the supplemental material of Ref.~\cite{SchwingerPDFletter2025} and also apply it to LCDAs. All plots shown in this appendix are representative examples at $\mass=10$ where not otherwise stated.

The obtained lattice data points for matrix elements, PDF, and LCDA have systematic errors that vanish when the limits $D \rightarrow \infty$, $\Lcut \rightarrow \infty$, $\Nt \rightarrow \infty$, $x \rightarrow \infty$, and $\vol \rightarrow \infty$ are taken. We quantify these uncertainties by the following procedure:

\begin{table*}[htb]
	\centering
	%!TEX root =  ../Schwinger_PDF_LCDA.tex
\footnotesize 
\begin{tabular}{|c|c|c|c|c|c|}
	\hline
	\normalsize $\mass$ & \normalsize $D$ & \normalsize $\Lcut$ & \normalsize $\Nt$ & \normalsize $x$ & \normalsize $\vol$ \\
	\hline
	1 & 80, 100, 120, \marktext{160}, 200 & 4, 6, \marktext{10}, 12 & 20, 40, 60, 80, \marktext{100}, 120, 200 & 6, 10, 16, \marktext{20}, 40 & 10, 16, \marktext{20}, 40 \\
	2.5 & 60, 80, \marktext{120}, 160 & 4, 6, \marktext{10}, 12 & 20, 40, 60, 80, \marktext{100}, 120 & 6, 10, 16, \marktext{25}, 40, 100 & 50, 75, \marktext{100}, 125 \\
	5 & 40, 60, \marktext{80}, 100, 120 & 4, 6, \marktext{10}, 12 & 20, 40, 60, 80, \marktext{100}, 120 & 20, \marktext{25}, 50, 100, 200, 400 & 75, \marktext{100}, 125, 150 \\
	10 & 40, 60, \marktext{80}, 120 & 4, 8, \marktext{10}, 12, 20 & 20, 50, \marktext{100}, 200 & 80, \marktext{100}, 200, 300, 500, 600, 1000 & 80, \marktext{100}, 120, 140, 160 \\
	20 & 40, 60, \marktext{80}, 120 & 4, 6, \marktext{10}, 12 & 20, 50, 80, \marktext{100}, 200 & 300, \marktext{400}, 600, 1000 & 140, \marktext{200}, 240, 280, 400 \\
	40 & 40, 60, \marktext{80}, 120 & 4, 6, \marktext{10}, 12 & 50, 80, \marktext{100}, 200 & \marktext{1000}, 1200, 1400, 1600, 2000 & 240, \marktext{320}, 400, 480 \\
	\hline
\end{tabular}
	\caption{Simulation parameters used in the error analysis. Base parameters in bold and underlined. The simulations are performed by keeping all parameters but one fixed at their base value.}
	\label{tab:sim_params}
\end{table*}

First, a set of base parameters $D^b, \Lcut^b, \Nt^b, x^b, \vol^b$ is chosen, with each value large enough such that results are close to convergence to the relevant limit. These base parameters are listed as bold, underlined values in \cref{tab:sim_params} for each fermion mass, together with all other parameter values that enter the data analysis. The parameters $D$, \Lcut, \Nt, \vol and $ma \propto \frac{m}{\sqrt{x}}$ are kept approximately mass-independent, with adaptations to account for the larger bond dimensions needed at small masses and, conversely, finer lattices and increased volumes at large masses to account for enhanced discretization and volume effects.

We then estimate the uncertainties for this set by varying each parameter in subsequent runs independently, keeping all others fixed at their base values. For each parameter, we extrapolate to the relevant limit using several fit ansätze. To estimate the uncertainties due to finite $D, \Lcut$, and $\Nt$, we extrapolate the matrix elements at each light-front distance and obtain the extrapolated PDF and LCDA via a Fourier transform. For the errors due to finite $x$ and $\vol$, we first take a Fourier transform of the data and then extrapolate the PDFs and LCDAs for each value of $\bjorx$ and $\uscale$, respectively. The reason for this treatment is that matrix elements are calculated at different light-front distances when $x$ is varied, such that the matrix elements cannot directly be extrapolated for each fixed $\zlight$. When varying \vol, matrix elements at large separations exist only for the largest volumes, so an extrapolation at fixed $z^-$ would lack data for the smaller volumes. We therefore extrapolate the parton functions instead.

For each parameter that we vary, we assign symmetric error bars that include all simulation results with the base parameter or larger (for the $D$ uncertainty: also values smaller than the base parameter), as well as all extrapolated values. We treat the $\vol$-dependence of PDFs and the $x$-dependence of LCDAs and decay constants differently, as explained in the corresponding subsections below. With the complete procedure, we independently estimate the uncertainties due to each finite simulation parameter. Finally, we combine the uncertainties in quadrature. The procedure applied in this work corresponds to a conservative uncertainty estimation. The size of the error bands is driven by the points that deviate the most from the base value. For example, many uncertainties caused by the bond dimension are driven by simulations with small bond dimensions that are only half of the base value.

\subsection{Bond dimension}

\begin{figure*}[htbp]
	\centering
	\def\figWidth{\textwidth/2.6}
	\def\figHeight{\textheight/7}
	\subfloat[Real part of matrix elements for $\zlight g = 14 \sqrt{\frac{2}{x}} \approx 1.980$\label{sfig:LCDA_M_real_D_d14}]{%
		%!TEX root =  ../Schwinger_PDF_LCDA.tex
% Generated by writeTikzPlot.m -- do not edit data by hand.
\begin{tikzpicture}
% y values written with offset y0 = 0.6364 subtracted (tick labels are absolute)
\begin{axis}[%
width=\figWidth,
height=\figHeight,
at={(0\figWidth,0\figHeight)},
scale only axis,
xmin=0,
xmax=0.03,
xlabel style={font=\color{white!15!black}},
xlabel={$1 \big/ D$},
ylabel={$\Re\left(\mathcal{M}_\mathrm{LCDA}\right)$},
ymin=-0.000144088787046,
ymax=0.000102459135981,
ytick={-0.0001, 0, 0.0001},
yticklabels={0.6363, 0.6364, 0.6365},
ylabel style={font=\color{white!15!black}},
ylabel style={{rotate=-90}}, scaled ticks=false, tick label style={/pgf/number format/fixed},
axis background/.style={fill=white},
xmajorgrids,
ymajorgrids,
legend pos=south west,
legend style={legend cell align=left, cells={align=left}, align=left, draw=white!15!black},
xticklabel style={yshift=-.5ex, /pgf/number format/fixed, /pgf/number format/precision=4},
yticklabel style={xshift=-.5ex, /pgf/number format/fixed, /pgf/number format/precision=6}
]
\addplot [myline2, only marks, forget plot]
  table[row sep=crcr]{%
0.025	-0.000119220375275453\\
};
\addplot [myline3, only marks, forget plot]
  table[row sep=crcr]{%
0.0166666666666667	1.57399492192178e-05\\
};
\addplot [myline1, only marks, forget plot]
  table[row sep=crcr]{%
0.0125	-5.37410102385838e-06\\
};
\addplot [myline4, only marks, forget plot]
  table[row sep=crcr]{%
0.00833333333333333	6.91735199243038e-06\\
};
\addplot [myline8, mark=none]
  table[row sep=crcr]{%
0	9.12524122068659e-05\\
0.000303030303030303	8.89884276060737e-05\\
0.000606060606060606	8.67244430052816e-05\\
0.000909090909090909	8.44604584044895e-05\\
0.00121212121212121	8.21964738036973e-05\\
0.00151515151515152	7.99324892029052e-05\\
0.00181818181818182	7.76685046021131e-05\\
0.00212121212121212	7.54045200013209e-05\\
0.00242424242424242	7.31405354004178e-05\\
0.00272727272727273	7.08765507996256e-05\\
0.00303030303030303	6.86125661988335e-05\\
0.00333333333333333	6.63485815980414e-05\\
0.00363636363636364	6.40845969972492e-05\\
0.00393939393939394	6.18206123964571e-05\\
0.00424242424242424	5.9556627795665e-05\\
0.00454545454545455	5.72926431948728e-05\\
0.00484848484848485	5.50286585940807e-05\\
0.00515151515151515	5.27646739932885e-05\\
0.00545454545454545	5.05006893924964e-05\\
0.00575757575757576	4.82367047917043e-05\\
0.00606060606060606	4.59727201909121e-05\\
0.00636363636363636	4.370873559012e-05\\
0.00666666666666667	4.14447509893279e-05\\
0.00696969696969697	3.91807663884247e-05\\
0.00727272727272727	3.69167817876326e-05\\
0.00757575757575758	3.46527971868404e-05\\
0.00787878787878788	3.23888125860483e-05\\
0.00818181818181818	3.01248279852562e-05\\
0.00848484848484848	2.7860843384464e-05\\
0.00878787878787879	2.55968587836719e-05\\
0.00909090909090909	2.33328741828798e-05\\
0.00939393939393939	2.10688895820876e-05\\
0.0096969696969697	1.88049049812955e-05\\
0.01	1.65409203805034e-05\\
0.0103030303030303	1.42769357797112e-05\\
0.0106060606060606	1.20129511789191e-05\\
0.0109090909090909	9.74896657812696e-06\\
0.0112121212121212	7.48498197733483e-06\\
0.0115151515151515	5.22099737643167e-06\\
0.0118181818181818	2.95701277563953e-06\\
0.0121212121212121	6.93028174847399e-07\\
0.0124242424242424	-1.57095642594474e-06\\
0.0127272727272727	-3.83494102673687e-06\\
0.013030303030303	-6.09892562752901e-06\\
0.0133333333333333	-8.36291022832114e-06\\
0.0136363636363636	-1.06268948291133e-05\\
0.0139393939393939	-1.28908794299054e-05\\
0.0142424242424242	-1.51548640306975e-05\\
0.0145454545454545	-1.74188486314897e-05\\
0.0148484848484848	-1.96828332322818e-05\\
0.0151515151515152	-2.1946817833074e-05\\
0.0154545454545455	-2.42108024338661e-05\\
0.0157575757575758	-2.64747870346582e-05\\
0.0160606060606061	-2.87387716355614e-05\\
0.0163636363636364	-3.10027562363535e-05\\
0.0166666666666667	-3.32667408371456e-05\\
0.016969696969697	-3.55307254379378e-05\\
0.0172727272727273	-3.77947100387299e-05\\
0.0175757575757576	-4.00586946395221e-05\\
0.0178787878787879	-4.23226792403142e-05\\
0.0181818181818182	-4.45866638411063e-05\\
0.0184848484848485	-4.68506484418985e-05\\
0.0187878787878788	-4.91146330426906e-05\\
0.0190909090909091	-5.13786176434827e-05\\
0.0193939393939394	-5.36426022442749e-05\\
0.0196969696969697	-5.5906586845067e-05\\
0.02	-5.81705714458591e-05\\
0.0203030303030303	-6.04345560466513e-05\\
0.0206060606060606	-6.26985406475544e-05\\
0.0209090909090909	-6.49625252483466e-05\\
0.0212121212121212	-6.72265098491387e-05\\
0.0215151515151515	-6.94904944499308e-05\\
0.0218181818181818	-7.1754479050723e-05\\
0.0221212121212121	-7.40184636515151e-05\\
0.0224242424242424	-7.62824482523072e-05\\
0.0227272727272727	-7.85464328530994e-05\\
0.023030303030303	-8.08104174538915e-05\\
0.0233333333333333	-8.30744020546836e-05\\
0.0236363636363636	-8.53383866554758e-05\\
0.0239393939393939	-8.76023712562679e-05\\
0.0242424242424242	-8.986635585706e-05\\
0.0245454545454545	-9.21303404578522e-05\\
0.0248484848484848	-9.43943250586443e-05\\
0.0251515151515152	-9.66583096595475e-05\\
0.0254545454545455	-9.89222942603396e-05\\
0.0257575757575758	-0.000101186278861132\\
0.0260606060606061	-0.000103450263461924\\
0.0263636363636364	-0.000105714248062716\\
0.0266666666666667	-0.000107978232663508\\
0.026969696969697	-0.0001102422172643\\
0.0272727272727273	-0.000112506201865092\\
0.0275757575757576	-0.000114770186465885\\
0.0278787878787879	-0.000117034171066677\\
0.0281818181818182	-0.000119298155667469\\
0.0284848484848485	-0.000121562140268261\\
0.0287878787878788	-0.000123826124869053\\
0.0290909090909091	-0.000126090109469845\\
0.0293939393939394	-0.000128354094070637\\
0.0296969696969697	-0.000130618078671541\\
0.03	-0.000132882063272333\\
};
\addlegendentry{$0.636-0.00747\,(1/D)$}
\end{axis}
\end{tikzpicture}%
	}
	\hfill
	\subfloat[Imaginary part of matrix elements for $\zlight g = 14 \sqrt{\frac{2}{x}} \approx 1.980$\label{sfig:LCDA_M_imag_D_d14}]{%
		%!TEX root =  ../Schwinger_PDF_LCDA.tex
% Generated by writeTikzPlot.m -- do not edit data by hand.
\begin{tikzpicture}
\begin{axis}[%
width=\figWidth,
height=\figHeight,
at={(0\figWidth,0\figHeight)},
scale only axis,
xmin=0,
xmax=0.03,
xlabel style={font=\color{white!15!black}},
xlabel={$1 \big/ D$},
ylabel={$\Im\left(\mathcal{M}_\mathrm{LCDA}\right)$},
ymin=-0.0195815495683,
ymax=-0.0188024005383,
ylabel style={font=\color{white!15!black}},
ylabel style={{rotate=-90}}, scaled ticks=false, tick label style={/pgf/number format/fixed},
axis background/.style={fill=white},
xmajorgrids,
ymajorgrids,
legend pos=south west,
legend style={legend cell align=left, cells={align=left}, align=left, draw=white!15!black},
xticklabel style={yshift=-.5ex, /pgf/number format/fixed, /pgf/number format/precision=4},
yticklabel style={xshift=-.5ex, /pgf/number format/fixed, /pgf/number format/precision=6}
]
\addplot [myline2, only marks, forget plot]
  table[row sep=crcr]{%
0.025	-0.0194988308551663\\
};
\addplot [myline3, only marks, forget plot]
  table[row sep=crcr]{%
0.0166666666666667	-0.0191014108624684\\
};
\addplot [myline1, only marks, forget plot]
  table[row sep=crcr]{%
0.0125	-0.0191097787745973\\
};
\addplot [myline4, only marks, forget plot]
  table[row sep=crcr]{%
0.00833333333333333	-0.0191169061625523\\
};
\addplot [myline8, mark=none]
  table[row sep=crcr]{%
0	-0.0188378164032703\\
0.000303030303030303	-0.0188449711234725\\
0.000606060606060606	-0.0188521258436747\\
0.000909090909090909	-0.0188592805638769\\
0.00121212121212121	-0.0188664352840791\\
0.00151515151515152	-0.0188735900042813\\
0.00181818181818182	-0.0188807447244835\\
0.00212121212121212	-0.0188878994446857\\
0.00242424242424242	-0.0188950541648879\\
0.00272727272727273	-0.0189022088850901\\
0.00303030303030303	-0.0189093636052923\\
0.00333333333333333	-0.0189165183254945\\
0.00363636363636364	-0.0189236730456967\\
0.00393939393939394	-0.0189308277658989\\
0.00424242424242424	-0.0189379824861011\\
0.00454545454545455	-0.0189451372063033\\
0.00484848484848485	-0.0189522919265055\\
0.00515151515151515	-0.0189594466467077\\
0.00545454545454545	-0.0189666013669099\\
0.00575757575757576	-0.0189737560871121\\
0.00606060606060606	-0.0189809108073143\\
0.00636363636363636	-0.0189880655275165\\
0.00666666666666667	-0.0189952202477186\\
0.00696969696969697	-0.0190023749679208\\
0.00727272727272727	-0.019009529688123\\
0.00757575757575758	-0.0190166844083252\\
0.00787878787878788	-0.0190238391285274\\
0.00818181818181818	-0.0190309938487296\\
0.00848484848484848	-0.0190381485689318\\
0.00878787878787879	-0.019045303289134\\
0.00909090909090909	-0.0190524580093362\\
0.00939393939393939	-0.0190596127295384\\
0.0096969696969697	-0.0190667674497406\\
0.01	-0.0190739221699428\\
0.0103030303030303	-0.019081076890145\\
0.0106060606060606	-0.0190882316103472\\
0.0109090909090909	-0.0190953863305494\\
0.0112121212121212	-0.0191025410507516\\
0.0115151515151515	-0.0191096957709538\\
0.0118181818181818	-0.019116850491156\\
0.0121212121212121	-0.0191240052113582\\
0.0124242424242424	-0.0191311599315604\\
0.0127272727272727	-0.0191383146517626\\
0.013030303030303	-0.0191454693719648\\
0.0133333333333333	-0.019152624092167\\
0.0136363636363636	-0.0191597788123692\\
0.0139393939393939	-0.0191669335325714\\
0.0142424242424242	-0.0191740882527736\\
0.0145454545454545	-0.0191812429729758\\
0.0148484848484848	-0.019188397693178\\
0.0151515151515152	-0.0191955524133802\\
0.0154545454545455	-0.0192027071335823\\
0.0157575757575758	-0.0192098618537845\\
0.0160606060606061	-0.0192170165739867\\
0.0163636363636364	-0.0192241712941889\\
0.0166666666666667	-0.0192313260143911\\
0.016969696969697	-0.0192384807345933\\
0.0172727272727273	-0.0192456354547955\\
0.0175757575757576	-0.0192527901749977\\
0.0178787878787879	-0.0192599448951999\\
0.0181818181818182	-0.0192670996154021\\
0.0184848484848485	-0.0192742543356043\\
0.0187878787878788	-0.0192814090558065\\
0.0190909090909091	-0.0192885637760087\\
0.0193939393939394	-0.0192957184962109\\
0.0196969696969697	-0.0193028732164131\\
0.02	-0.0193100279366153\\
0.0203030303030303	-0.0193171826568175\\
0.0206060606060606	-0.0193243373770197\\
0.0209090909090909	-0.0193314920972219\\
0.0212121212121212	-0.0193386468174241\\
0.0215151515151515	-0.0193458015376263\\
0.0218181818181818	-0.0193529562578285\\
0.0221212121212121	-0.0193601109780307\\
0.0224242424242424	-0.0193672656982329\\
0.0227272727272727	-0.0193744204184351\\
0.023030303030303	-0.0193815751386373\\
0.0233333333333333	-0.0193887298588395\\
0.0236363636363636	-0.0193958845790417\\
0.0239393939393939	-0.0194030392992438\\
0.0242424242424242	-0.019410194019446\\
0.0245454545454545	-0.0194173487396482\\
0.0248484848484848	-0.0194245034598504\\
0.0251515151515152	-0.0194316581800526\\
0.0254545454545455	-0.0194388129002548\\
0.0257575757575758	-0.019445967620457\\
0.0260606060606061	-0.0194531223406592\\
0.0263636363636364	-0.0194602770608614\\
0.0266666666666667	-0.0194674317810636\\
0.026969696969697	-0.0194745865012658\\
0.0272727272727273	-0.019481741221468\\
0.0275757575757576	-0.0194888959416702\\
0.0278787878787879	-0.0194960506618724\\
0.0281818181818182	-0.0195032053820746\\
0.0284848484848485	-0.0195103601022768\\
0.0287878787878788	-0.019517514822479\\
0.0290909090909091	-0.0195246695426812\\
0.0293939393939394	-0.0195318242628834\\
0.0296969696969697	-0.0195389789830856\\
0.03	-0.0195461337032878\\
};
\addlegendentry{$-0.0188-0.0236\,(1/D)$}
\end{axis}
\end{tikzpicture}%
	}
	\\
	\subfloat[LCDA for $\uscale = 0.5$\label{sfig:LCDA_real_D_u5}]{%
		%!TEX root =  ../Schwinger_PDF_LCDA.tex
% Generated by writeTikzPlot.m -- do not edit data by hand.
\begin{tikzpicture}
% y values written with offset y0 = 4.0532 subtracted (tick labels are absolute)
\begin{axis}[%
width=\figWidth,
height=\figHeight,
at={(0\figWidth,0\figHeight)},
scale only axis,
xmin=-0.00075,
xmax=0.02575,
xlabel style={font=\color{white!15!black}},
xlabel={$1 \big/ D$},
ylabel={$\Re\left(\lcda\right)$},
ymin=-0.000382898670536,
ymax=0.000392923864436,
ytick={-0.0002, 0, 0.0002},
yticklabels={4.0530, 4.0532, 4.0534},
ylabel style={font=\color{white!15!black}},
ylabel style={{rotate=-90}}, scaled ticks=false, tick label style={/pgf/number format/fixed},
axis background/.style={fill=white},
xmajorgrids,
ymajorgrids,
legend pos=south west,
legend style={legend cell align=left, cells={align=left}, align=left, draw=white!15!black},
xticklabel style={yshift=-.5ex, /pgf/number format/fixed, /pgf/number format/precision=4},
yticklabel style={xshift=-.5ex, /pgf/number format/fixed, /pgf/number format/precision=6}
]
\addplot [myline2, only marks, forget plot]
  table[row sep=crcr]{%
0.025	-0.000347634009855824\\
};
\addplot [myline3, only marks, forget plot]
  table[row sep=crcr]{%
0.0166666666666667	9.66003481748245e-05\\
};
\addplot [myline1, only marks, forget plot, error bar legend]
 plot [error bars/.cd, y dir=both, y explicit, error bar style={line width=\tikzlinewidth}, error mark options={line width=\tikzlinewidth, mark size=\tikzmarksize, rotate=90, mycolor1}]
 table[row sep=crcr, y error plus index=2, y error minus index=3]{%
0.0125	5.01259694996037e-06	0.000352646606805784	0.000352646606805784\\
};
\addplot [myline4, only marks, forget plot]
  table[row sep=crcr]{%
0.00833333333333333	2.50245488082612e-05\\
};
\addplot [myline8, only marks]
  table[row sep=crcr]{%
0	0.000290015226457108\\
};
\addlegendentry{$D \rightarrow \infty$ fit}
\end{axis}
\end{tikzpicture}%
	}
	\hfill
	\subfloat[Imaginary part of LCDA for $\uscale = 0.5$\label{sfig:LCDA_imag_D_u5}]{%
		%!TEX root =  ../Schwinger_PDF_LCDA.tex
% Generated by writeTikzPlot.m -- do not edit data by hand.
\begin{tikzpicture}
% y values written with offset y0 = -0.06682 subtracted (tick labels are absolute)
\begin{axis}[%
width=\figWidth,
height=\figHeight,
at={(0\figWidth,0\figHeight)},
scale only axis,
xmin=-0.00075,
xmax=0.02575,
xlabel style={font=\color{white!15!black}},
xlabel={$1 \big/ D$},
ylabel={$\Im\left(\lcda\right)$},
ymin=-6.34269380085e-05,
ymax=5.9579882231e-05,
ytick={-6e-05, -4e-05, -2e-05, 0, 2e-05, 4e-05},
yticklabels={-0.06688, -0.06686, -0.06684, -0.06682, -0.06680, -0.06678},
ylabel style={font=\color{white!15!black}},
ylabel style={{rotate=-90}}, scaled ticks=false, tick label style={/pgf/number format/fixed},
axis background/.style={fill=white},
xmajorgrids,
ymajorgrids,
legend style={at={(0,-0.2)}, anchor=north west, legend cell align=left, cells={align=left}, align=left, draw=white!15!black},
xticklabel style={yshift=-.5ex, /pgf/number format/fixed, /pgf/number format/precision=4},
yticklabel style={xshift=-.5ex, /pgf/number format/fixed, /pgf/number format/precision=6}
]
\addplot [myline2, only marks, forget plot]
  table[row sep=crcr]{%
0.025	-5.78357189066625e-05\\
};
\addplot [myline3, only marks, forget plot]
  table[row sep=crcr]{%
0.0166666666666667	-1.20013732629598e-06\\
};
\addplot [myline1, only marks, forget plot, error bar legend]
 plot [error bars/.cd, y dir=both, y explicit, error bar style={line width=\tikzlinewidth}, error mark options={line width=\tikzlinewidth, mark size=\tikzmarksize, rotate=90, mycolor1}]
 table[row sep=crcr, y error plus index=2, y error minus index=3]{%
0.0125	-1.92352788871974e-06	5.59121910179428e-05	5.59121910179428e-05\\
};
\addplot [myline4, only marks, forget plot]
  table[row sep=crcr]{%
0.00833333333333333	-2.93525534363903e-05\\
};
\addplot [myline8, only marks, forget plot]
  table[row sep=crcr]{%
0	1.06904953404707e-05\\
};
\end{axis}
\end{tikzpicture}%
	}
	\\
	\subfloat[LCDA for $\uscale = 0.3$\label{sfig:LCDA_real_D_u3}]{%
		%!TEX root =  ../Schwinger_PDF_LCDA.tex
% Generated by writeTikzPlot.m -- do not edit data by hand.
\begin{tikzpicture}
% y values written with offset y0 = 0.5508 subtracted (tick labels are absolute)
\begin{axis}[%
width=\figWidth,
height=\figHeight,
at={(0\figWidth,0\figHeight)},
scale only axis,
xmin=-0.00075,
xmax=0.02575,
xlabel style={font=\color{white!15!black}},
xlabel={$1 \big/ D$},
ylabel={$\Re\left(\lcda\right)$},
ymin=-0.000365219172991,
ymax=0.000362702221191,
ytick={-0.0002, 0, 0.0002},
yticklabels={0.5506, 0.5508, 0.5510},
ylabel style={font=\color{white!15!black}},
ylabel style={{rotate=-90}}, scaled ticks=false, tick label style={/pgf/number format/fixed},
axis background/.style={fill=white},
xmajorgrids,
ymajorgrids,
legend style={at={(0,-0.2)}, anchor=north west, legend cell align=left, cells={align=left}, align=left, draw=white!15!black},
xticklabel style={yshift=-.5ex, /pgf/number format/fixed, /pgf/number format/precision=4},
yticklabel style={xshift=-.5ex, /pgf/number format/fixed, /pgf/number format/precision=6}
]
\addplot [myline2, only marks, forget plot]
  table[row sep=crcr]{%
0.025	0.000329614885091867\\
};
\addplot [myline3, only marks, forget plot]
  table[row sep=crcr]{%
0.0166666666666667	-6.77823955508305e-06\\
};
\addplot [myline1, only marks, forget plot, error bar legend]
 plot [error bars/.cd, y dir=both, y explicit, error bar style={line width=\tikzlinewidth}, error mark options={line width=\tikzlinewidth, mark size=\tikzmarksize, rotate=90, mycolor1}]
 table[row sep=crcr, y error plus index=2, y error minus index=3]{%
0.0125	-1.25847589993455e-06	0.000330873360991801	0.000330873360991801\\
};
\addplot [myline4, only marks, forget plot]
  table[row sep=crcr]{%
0.00833333333333333	-1.00468698899236e-05\\
};
\addplot [myline8, only marks, forget plot]
  table[row sep=crcr]{%
0	-0.000247177383999042\\
};
\end{axis}
\end{tikzpicture}%
	}
	\hfill
	\subfloat[Imaginary part of LCDA for $\uscale = 0.3$\label{sfig:LCDA_imag_D_u3}]{%
		%!TEX root =  ../Schwinger_PDF_LCDA.tex
% Generated by writeTikzPlot.m -- do not edit data by hand.
\begin{tikzpicture}
% y values written with offset y0 = -0.13235 subtracted (tick labels are absolute)
\begin{axis}[%
width=\figWidth,
height=\figHeight,
at={(0\figWidth,0\figHeight)},
scale only axis,
xmin=-0.00075,
xmax=0.02575,
xlabel style={font=\color{white!15!black}},
xlabel={$1 \big/ D$},
ylabel={$\Im\left(\lcda\right)$},
ymin=-0.00013133328366,
ymax=0.000128427782545,
ytick={-0.0001, -5e-05, 0, 5e-05, 0.0001},
yticklabels={-0.13245, -0.13240, -0.13235, -0.13230, -0.13225},
ylabel style={font=\color{white!15!black}},
ylabel style={{rotate=-90}}, scaled ticks=false, tick label style={/pgf/number format/fixed},
axis background/.style={fill=white},
xmajorgrids,
ymajorgrids,
legend style={at={(0,-0.2)}, anchor=north west, legend cell align=left, cells={align=left}, align=left, draw=white!15!black},
xticklabel style={yshift=-.5ex, /pgf/number format/fixed, /pgf/number format/precision=4},
yticklabel style={xshift=-.5ex, /pgf/number format/fixed, /pgf/number format/precision=6}
]
\addplot [myline2, only marks, forget plot]
  table[row sep=crcr]{%
0.025	-0.000115814840972395\\
};
\addplot [myline3, only marks, forget plot]
  table[row sep=crcr]{%
0.0166666666666667	-4.56838636481161e-06\\
};
\addplot [myline1, only marks, forget plot, error bar legend]
 plot [error bars/.cd, y dir=both, y explicit, error bar style={line width=\tikzlinewidth}, error mark options={line width=\tikzlinewidth, mark size=\tikzmarksize, rotate=90, mycolor1}]
 table[row sep=crcr, y error plus index=2, y error minus index=3]{%
0.0125	-1.45275055757343e-06	0.000118073211911096	0.000118073211911096\\
};
\addplot [myline4, only marks, forget plot]
  table[row sep=crcr]{%
0.00833333333333333	3.53666151264587e-05\\
};
\addplot [myline8, only marks, forget plot]
  table[row sep=crcr]{%
0	0.000116620461353523\\
};
\end{axis}
\end{tikzpicture}%
	}
	\caption{Dependence of matrix elements and LCDA on $D$ at $\mass=10$ and with base parameters as shown in \cref{tab:sim_params}. \protect\subref{sfig:LCDA_M_real_D_d14} and \protect\subref{sfig:LCDA_M_imag_D_d14}: Examples of matrix elements at a fixed light-front distance $\zlight$ and extrapolations to $1/D \rightarrow 0$. \protect\subref{sfig:LCDA_real_D_u5} -- \protect\subref{sfig:LCDA_imag_D_u3} LCDA at selected values of \uscale with symmetric error bars at the base point $D^b=80$ that include all data points and the extrapolation.}
	\label{fig:D_dependence}
\end{figure*}

The same bond dimension $D$ is used in our calculations for both the initial state and the time-evolved states. We vary $D$ and ensure that the results are not significantly affected by the MPS approximation. The real and imaginary parts of the matrix elements are extrapolated independently to $D \rightarrow \infty$ by linear fits in $\frac{1}{D}$. The PDFs and LCDAs are computed from the matrix elements for each value of the bond dimension and for the extrapolated results. The uncertainty due to the finite bond dimension $D^b$ is then approximated by the maximal spread between PDFs or LCDAs at the base parameter $D^b$ [see \cref{tab:sim_params}] and all other results, including the extrapolated values. Symmetric error bars are assigned. Examples of matrix elements and LCDAs at different $D$, extrapolations and assigned error bars are shown in \cref{fig:D_dependence}.

\subsection{Truncation of electric flux}

\begin{figure*}[htbp]
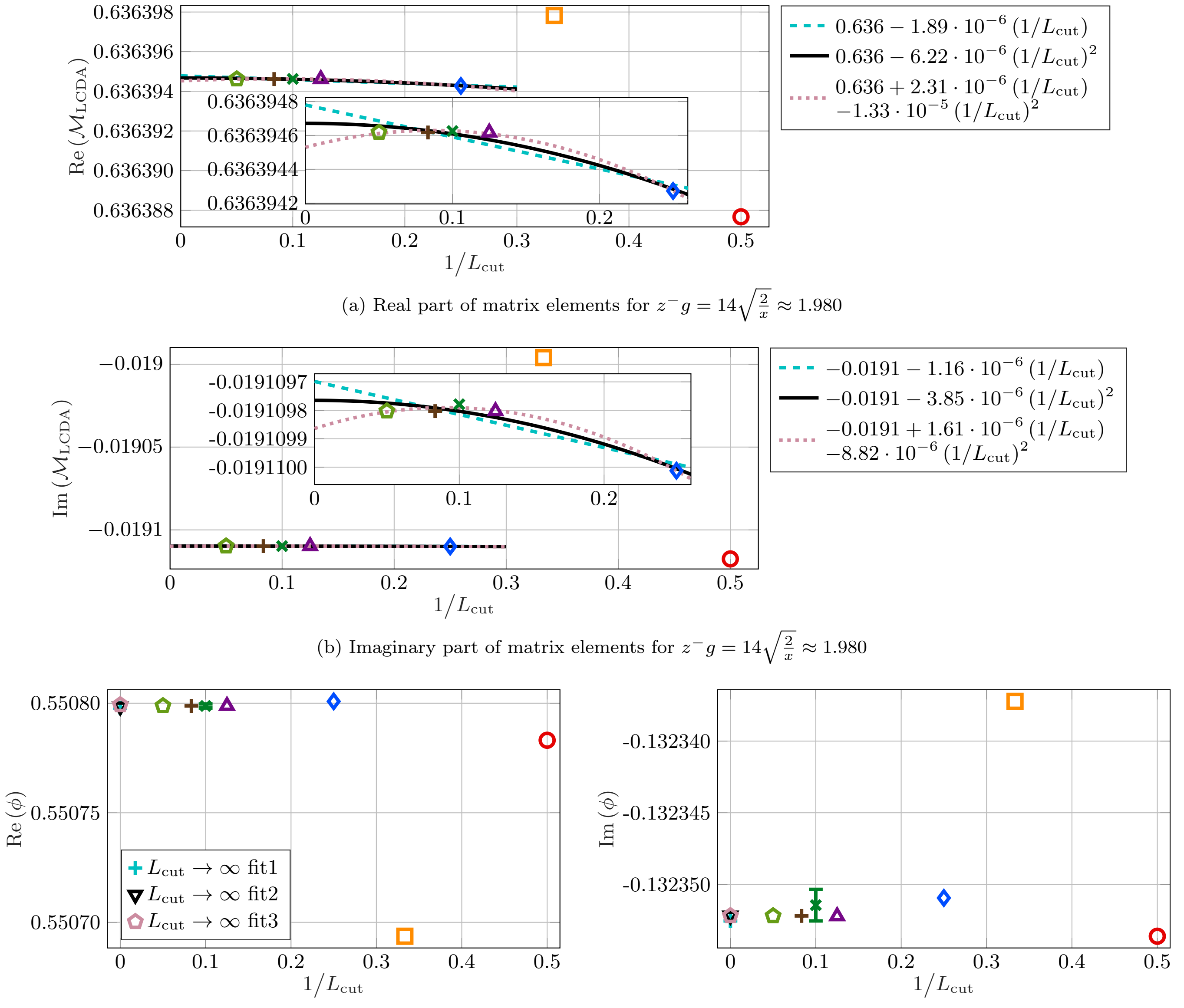

	\centering
	\def\figWidth{0.5*\textwidth}
	\def\figHeight{\textheight/7}
	\subfloat[Real part of matrix elements for $\zlight g = 14 \sqrt{\frac{2}{x}} \approx 1.980$\label{sfig:M_real_Lcut_d14}]{%
		\input{tikz/LCDA_M_real_Lcut_d14_m10}
	}
	\\
	\subfloat[Imaginary part of matrix elements for $\zlight g = 14 \sqrt{\frac{2}{x}} \approx 1.980$\label{sfig:M_imag_Lcut_d14}]{%
		\input{tikz/LCDA_M_imag_Lcut_d14_m10}
	}
	\\
	\def\figWidth{\textwidth/2.6}
	\def\figHeight{\textheight/6}
	\subfloat[LCDA for $\uscale = 0.3$\label{sfig:LCDA_real_Lcut_u3}]{%
		%!TEX root =  ../Schwinger_PDF_LCDA.tex
% Generated by writeTikzPlot.m -- do not edit data by hand.
\begin{tikzpicture}
% y values written with offset y0 = 0.55075 subtracted (tick labels are absolute)
\begin{axis}[%
width=\figWidth,
height=\figHeight,
at={(0\figWidth,0\figHeight)},
scale only axis,
xmin=-0.015,
xmax=0.515,
xlabel style={font=\color{white!15!black}},
xlabel={$1 \big/ \Lcut$},
ylabel={$\Re\left(\lcda\right)$},
ymin=-6.16988402724e-05,
ymax=5.61908140659e-05,
ytick={-5e-05, 0, 5e-05},
yticklabels={0.55070, 0.55075, 0.55080},
ylabel style={font=\color{white!15!black}},
ylabel style={{rotate=-90}}, scaled ticks=false, tick label style={/pgf/number format/fixed},
axis background/.style={fill=white},
xmajorgrids,
ymajorgrids,
legend pos=south west,
legend style={legend cell align=left, cells={align=left}, align=left, draw=white!15!black},
xticklabel style={yshift=-.5ex, /pgf/number format/fixed, /pgf/number format/precision=3},
yticklabel style={xshift=-.5ex, /pgf/number format/fixed, /pgf/number format/precision=6}
]
\addplot [myline2, only marks, forget plot]
  table[row sep=crcr]{%
0.5	3.30951996130313e-05\\
};
\addplot [myline3, only marks, forget plot]
  table[row sep=crcr]{%
0.333333333333333	-5.63402196206431e-05\\
};
\addplot [myline4, only marks, forget plot]
  table[row sep=crcr]{%
0.25	5.08321934141742e-05\\
};
\addplot [myline5, only marks, forget plot]
  table[row sep=crcr]{%
0.125	4.88063501333791e-05\\
};
\addplot [myline1, only marks, forget plot, error bar legend]
 plot [error bars/.cd, y dir=both, y explicit, error bar style={line width=\tikzlinewidth}, error mark options={line width=\tikzlinewidth, mark size=\tikzmarksize, rotate=90, mycolor1}]
 table[row sep=crcr, y error plus index=2, y error minus index=3]{%
0.1	4.87415241000599e-05	9.27865746458423e-07	9.27865746458423e-07\\
};
\addplot [myline6, only marks, forget plot]
  table[row sep=crcr]{%
0.0833333333333333	4.87704153309121e-05\\
};
\addplot [myline7, only marks, forget plot]
  table[row sep=crcr]{%
0.05	4.87703942693152e-05\\
};
\addplot [myline8, only marks]
  table[row sep=crcr]{%
0	4.78136583536015e-05\\
};
\addlegendentry{$\Lcut \rightarrow \infty$ fit1}
\addplot [myline9, only marks]
  table[row sep=crcr]{%
0	4.84627324838227e-05\\
};
\addlegendentry{$\Lcut \rightarrow \infty$ fit2}
\addplot [myline10, only marks]
  table[row sep=crcr]{%
0	4.92431725191977e-05\\
};
\addlegendentry{$\Lcut \rightarrow \infty$ fit3}
\end{axis}
\end{tikzpicture}%
	}
	\hfill
	\subfloat[Imaginary part of LCDA for $\uscale = 0.3$\label{sfig:LCDA_imag_Lcut_u3}]{%
		%!TEX root =  ../Schwinger_PDF_LCDA.tex
% Generated by writeTikzPlot.m -- do not edit data by hand.
\begin{tikzpicture}
% y values written with offset y0 = -0.132345 subtracted (tick labels are absolute)
\begin{axis}[%
width=\figWidth,
height=\figHeight,
at={(0\figWidth,0\figHeight)},
scale only axis,
xmin=-0.015,
xmax=0.515,
xlabel style={font=\color{white!15!black}},
xlabel={$1 \big/ \Lcut$},
ylabel={$\Im\left(\lcda\right)$},
ymin=-9.44291509422e-06,
ymax=8.59446622473e-06,
ytick={-5e-06, 0, 5e-06},
yticklabels={-0.132350, -0.132345, -0.132340},
ylabel style={font=\color{white!15!black}},
ylabel style={{rotate=-90}}, scaled ticks=false, tick label style={/pgf/number format/fixed},
axis background/.style={fill=white},
xmajorgrids,
ymajorgrids,
legend style={at={(0,-0.2)}, anchor=north west, legend cell align=left, cells={align=left}, align=left, draw=white!15!black},
xticklabel style={yshift=-.5ex, /pgf/number format/fixed, /pgf/number format/precision=3},
yticklabel style={xshift=-.5ex, /pgf/number format/fixed, /pgf/number format/precision=7}
]
\addplot [myline2, only marks, forget plot]
  table[row sep=crcr]{%
0.5	-8.62303412518806e-06\\
};
\addplot [myline3, only marks, forget plot]
  table[row sep=crcr]{%
0.333333333333333	7.77458525569319e-06\\
};
\addplot [myline4, only marks, forget plot]
  table[row sep=crcr]{%
0.25	-5.95134897379834e-06\\
};
\addplot [myline5, only marks, forget plot]
  table[row sep=crcr]{%
0.125	-7.22168944897184e-06\\
};
\addplot [myline1, only marks, forget plot, error bar legend]
 plot [error bars/.cd, y dir=both, y explicit, error bar style={line width=\tikzlinewidth}, error mark options={line width=\tikzlinewidth, mark size=\tikzmarksize, rotate=90, mycolor1}]
 table[row sep=crcr, y error plus index=2, y error minus index=3]{%
0.1	-6.45275055757843e-06	1.10139930395436e-06	1.10139930395436e-06\\
};
\addplot [myline6, only marks, forget plot]
  table[row sep=crcr]{%
0.0833333333333333	-7.20446573121425e-06\\
};
\addplot [myline7, only marks, forget plot]
  table[row sep=crcr]{%
0.05	-7.2062088592173e-06\\
};
\addplot [myline8, only marks, forget plot]
  table[row sep=crcr]{%
0	-7.5541498615328e-06\\
};
\addplot [myline9, only marks, forget plot]
  table[row sep=crcr]{%
0	-7.18776348573869e-06\\
};
\addplot [myline10, only marks, forget plot]
  table[row sep=crcr]{%
0	-7.19714107427771e-06\\
};
\end{axis}
\end{tikzpicture}%
	}
	\caption{Dependence of matrix elements and LCDA on the cutoff in the electric field $\Lcut$ at $\mass=10$ and with base parameters as shown in \cref{tab:sim_params}. Three different fit models are applied to extrapolate to $\Lcut \rightarrow \infty$; fit1: linear; fit2: quadratic; fit3: linear plus quadratic. Data points with $\Lcut=2$ and $\Lcut=3$ are shown but do not enter the data analysis or fits. Symmetric error bars are assigned to the LCDA at the base point $\Lcut^b=10$ that include all data points with $\Lcut \ge \Lcut^b$, including the extrapolations.}
	\label{fig:Lcut_dependence}
\end{figure*}

The possible values of the electric flux are truncated during the time evolution to the range from $-\Lcut$ to $\Lcut$~\cite{Schwinger_thermalEvolution}. As shown in \cref{fig:Lcut_dependence}, we vary $\Lcut$ and extrapolate to the limit $\Lcut \rightarrow \infty$ (linear, quadratic, and linear plus quadratic terms in $\frac{1}{\Lcut}$). We only include results with $\Lcut \ge 4$ in the extrapolations. The uncertainty due to the truncation is estimated by the largest deviation between the base value $\Lcut^b$ [see \cref{tab:sim_params}] and results with larger values of $\Lcut$, including the extrapolations. The effect of the gauge-field truncation is not significant for $\Lcut \ge 2$ (notice the narrow y-axis range in \cref{fig:Lcut_dependence}) with the fermion masses considered in this work.

\subsection{Trotterization}

\begin{figure*}[htbp]
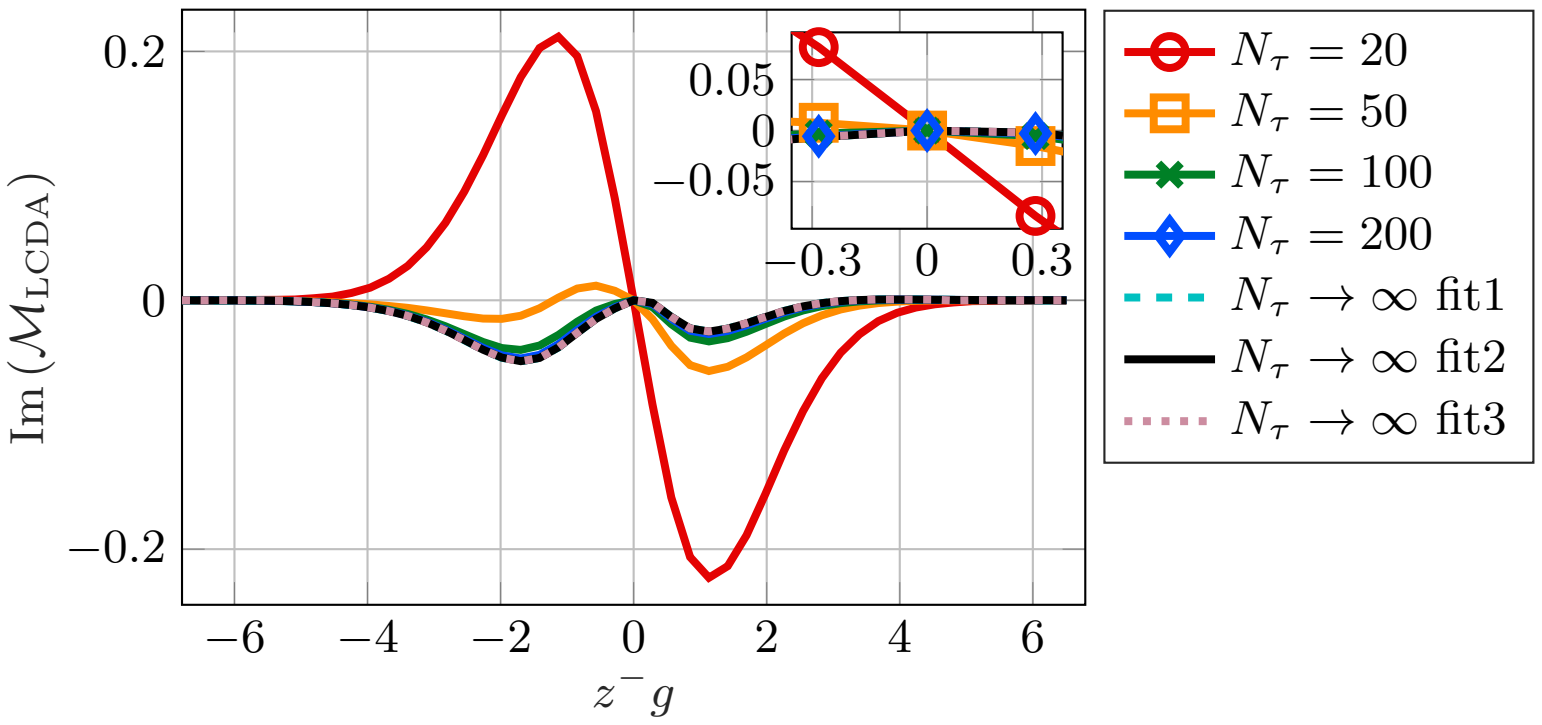
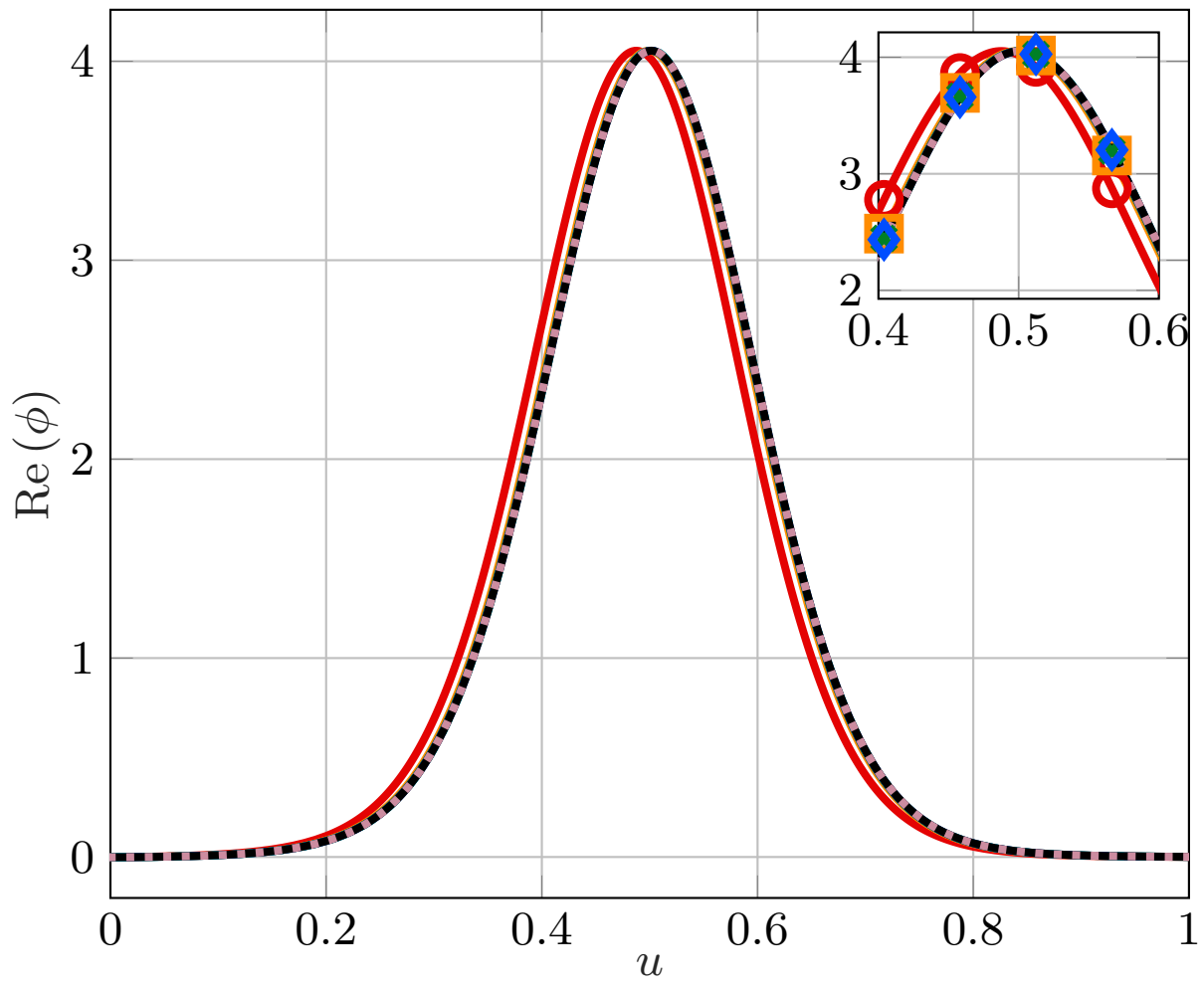
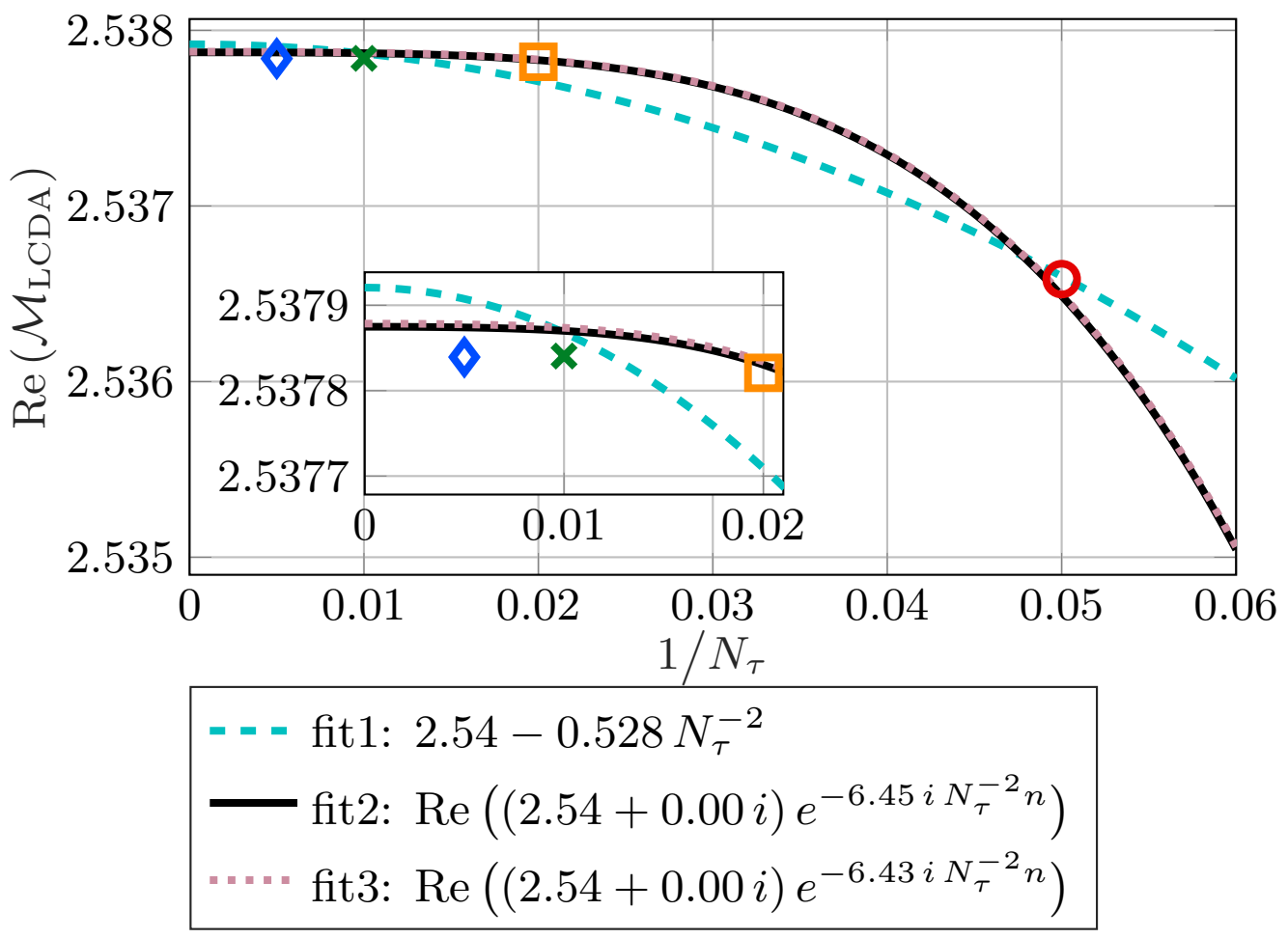
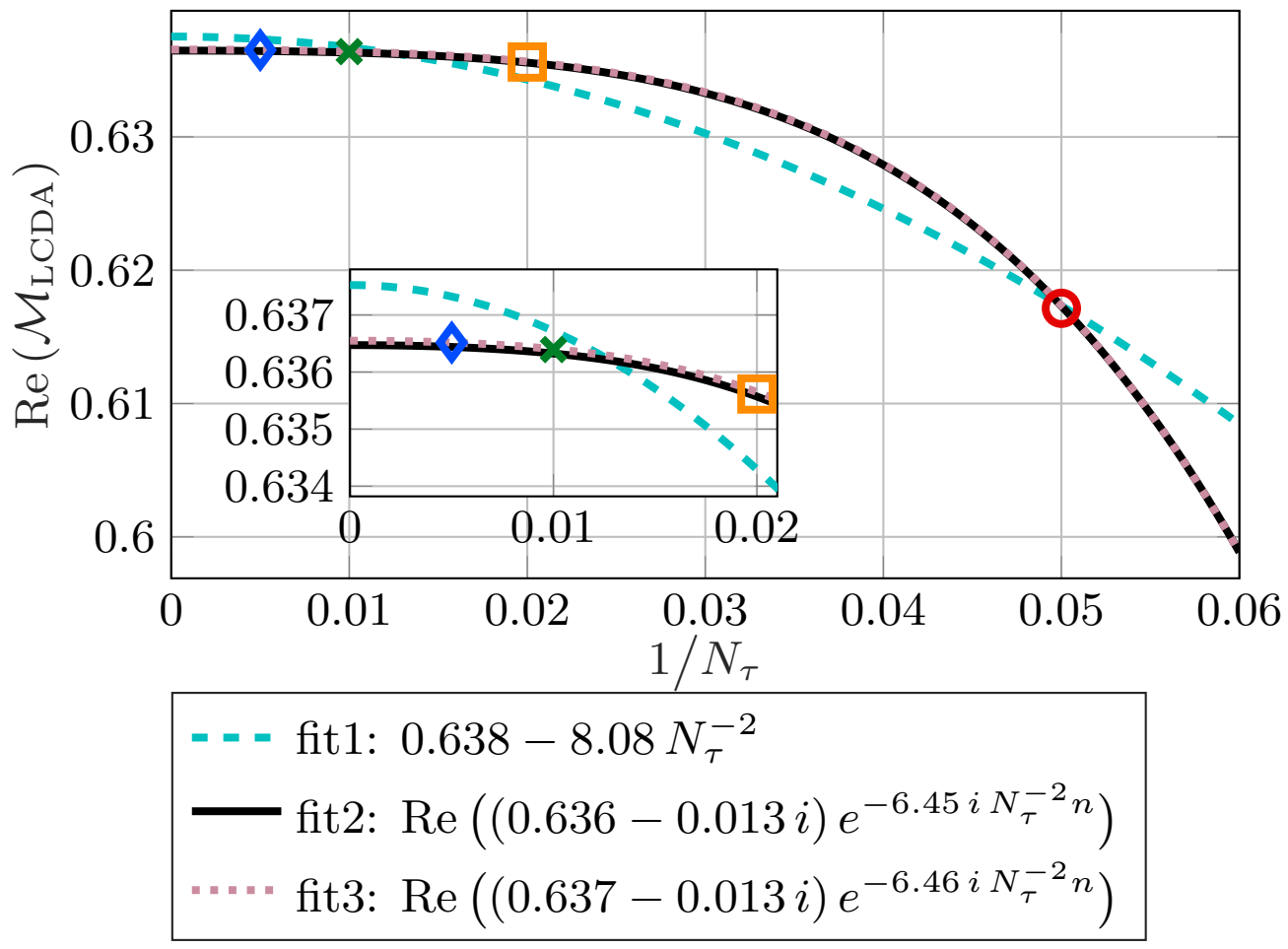
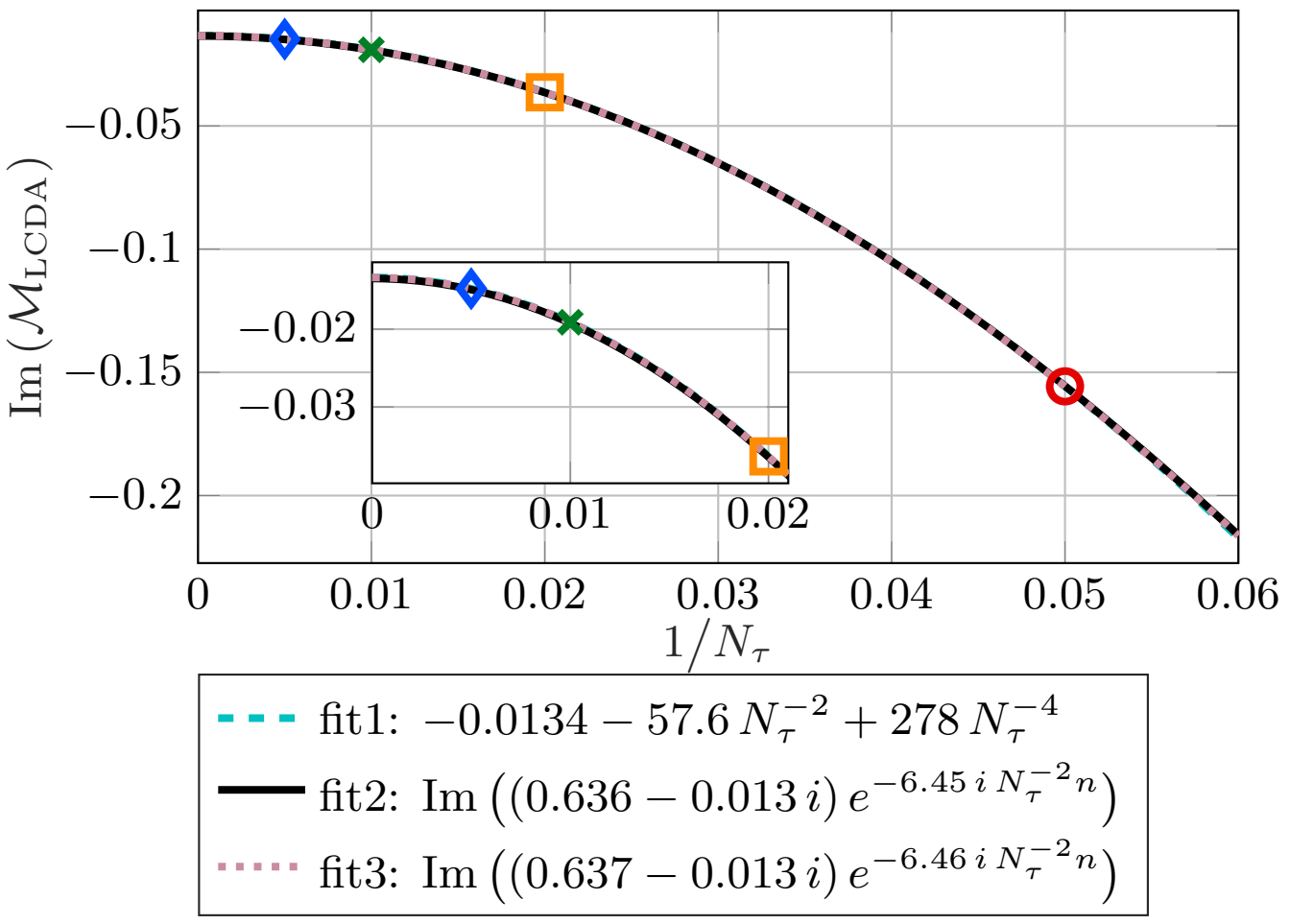

	\centering
	\def\figWidth{\textwidth/3}
	\def\figHeight{\textheight/6}
	\subfloat[Real part of matrix elements\label{sfig:M_real_tau}]{%
		\input{tikz/LCDA_M_real_tau_m10}
	}
	\hfill
	\subfloat[Imaginary part of matrix elements\label{sfig:M_imag_tau}]{%
		\input{tikz/LCDA_M_imag_tau_m10}
	}
	\\
	\def\figWidth{\textwidth/2.5}
	\def\figHeight{\textheight/4}
	\subfloat[LCDA\label{sfig:LCDA_real_tau}]{%
		\input{tikz/LCDA_real_tau_m10}
	}
	\hfill
	\def\figHeight{\textheight/6.2}
	\subfloat[Real part of matrix elements for $\zlight g = 2 \sqrt{\frac{2}{x}} \approx 0.2828$\label{sfig:M_real_tau_d2}]{%
		\input{tikz/LCDA_M_real_tau_d2_m10}
	}
	\\
	\subfloat[Real part of matrix elements for $\zlight g = 14 \sqrt{\frac{2}{x}} \approx 1.980$\label{sfig:M_real_tau_d14}]{%
		\input{tikz/LCDA_M_real_tau_d14_m10}
	}
	\hfill
	\subfloat[Imaginary part of matrix elements for $\zlight g = 14 \sqrt{\frac{2}{x}} \approx 1.980$\label{sfig:M_imag_tau_d14}]{%
		\input{tikz/LCDA_M_imag_tau_d14_m10}
	}
	\caption{Dependence of matrix elements and LCDA on the number of Trotter steps \Nt at $\mass=10$ and with base parameters as shown in \cref{tab:sim_params}. Three different fit models are applied to extrapolate to $\Nt \rightarrow \infty$; fit1 applies a polynomial fit for the real and imaginary parts of the matrix elements independently for each distance $\z$ (spatial distance in lattice units); fit2 assumes a complex, distance-dependent prefactor and a global phase factor, where the factor in the phase exponent is fitted globally from all distances $\z$ together; fit3 makes use of the same functional form, but the phase factor is assumed to be \z-dependent and fitted for each $\z$ independently. The continuous lines in \protect\subref{sfig:M_real_tau} and \protect\subref{sfig:M_imag_tau} are shown for clarity; all data points are discrete. Inset of \protect\subref{sfig:LCDA_real_tau}: data points from a discrete Fourier transform are shown on a curve obtained by a continuous Fourier transform.}
	\label{fig:tau_dependence}
\end{figure*}

The time evolution in each zigzag step of the stepwise Wilson line [see \cref{fig:zigzag}] is subdivided into $\Nt$ second-order Trotter steps. The continuum in the time-direction is recovered for $\Nt \rightarrow \infty$. We vary $\Nt$ and fit several functional forms for extrapolating the matrix elements to the $\Nt \rightarrow \infty$ limit, as shown in \cref{fig:tau_dependence}.

We observe that a fine Trotterization with $\Nt \ge 20$ is required for accurate results. The real part of matrix elements is fitted for each spatial distance $\z$ (in lattice units) with a polynomial of the form $a_r(\z) + b_r(\z) \frac{1}{\Nt^2}$, and the imaginary part is independently fitted with $a_i(\z) + b_i(\z) \frac{1}{\Nt^2} + c_i(\z) \frac{1}{\Nt^4}$ (fit1). Furthermore, we observe that the Trotter error is dominated by a phase error of the matrix elements. We therefore fit a complex prefactor $c_0(\z)$ that does not depend on $\Nt$ and a phase factor $c_p$, leading to the form $c_0 e^{i c_p \Nt^{-2} \z}$ for the second-order Trotterization used in this work (fit2 and fit3). In fit2, the value $c_p$ is obtained from a global fit for all $\z$ simultaneously, while fit3 extracts individual values $c_p(\z)$ for each $\z$. The $c_0$ are $\z$-dependent in both cases. The results of the exponential fits agree well with each other. They describe the data significantly better than the polynomial fit, which is an indication of a phase shift as the dominant error source. We take all three fits into account for a conservative error estimation.

In the limit $\Nt \rightarrow \infty$, the imaginary part of the LCDA matrix elements is reduced to the level of remaining lattice artifacts, which vanish in the continuum limit [see \cref{sfig:M_imag_tau}]. The LCDA becomes symmetric around $\uscale = 0.5$ when $\Nt \rightarrow \infty$ [see \cref{sfig:LCDA_real_tau}], as expected and derived in \cref{app:symmetries}.

The obtained factors $c_p$ in fit2 and fit3 are similar, indicating that the same phase error appears in each zigzag step. The effect of this error is a shift of the PDF and LCDA to smaller $\bjorx$ and $\uscale$, respectively, as shown in \cref{sfig:LCDA_real_tau} for the LCDA. This systematic behavior of the error indicates that a coarser Trotterization could be used if the phase factor is mitigated. In particular, the requirements of symmetry around $\uscale = 0.5$ and vanishing imaginary part of the matrix elements could be leveraged to determine the parameter $c_p$ in the exponent and to mitigate the error. This can be particularly helpful for quantum simulations, where circuit depth is a crucial limitation, and a coarser Trotterization with mitigated phase errors could reduce circuit depths.

\subsection{Lattice spacing}

The effect of a finite lattice spacing is examined by varying $x$, as shown in \cref{fig:x_dependence}. The continuum limit corresponds to $x \rightarrow \infty$. The stepwise structure of the Wilson line [see \cref{fig:zigzag}] introduces $\order{a} \equiv \order{\frac{1}{\sqrt{x}}}$ lattice effects. We therefore extrapolate PDFs and LCDAs at each value of \bjorx and \uscale, respectively, to the continuum limit with polynomial fits in $\frac{1}{\sqrt{x}}$ (linear and up to quadratic). The PDF uncertainty is estimated by the maximum spread between the base value $x^b$ [see \cref{tab:sim_params}] and results with larger $x$ (including extrapolated results).

For the LCDA, we choose a conservative extrapolation to the continuum limit. For each value of $\uscale$, the midpoint between the smallest and largest LCDA with $x \ge x^b$ (including extrapolations) is taken. The uncertainty related to $x$ is assigned to cover the spread between these points. The midpoint is reported as the LCDA in \cref{fig:LCDA_fermionmass}. It is further shown in \cref{fig:x_dependence} as olive pentagons/curves, with error bars/bands obtained by this midpoint procedure (instead of the uncertainty estimation at the base point for the other parameters). This way, the obtained LCDAs are compatible within uncertainties with the value at $x^b$ and with all extrapolations.

The imaginary part of the PDFs and LCDAs vanishes as expected in the continuum limit within uncertainties, with small deviations for the LCDA close to the boundaries of the physical $\uscale$ range.

\begin{figure*}[htbp]
	\centering
	\def\figWidth{\textwidth/3}
	\def\figHeight{\textheight/6}
	\subfloat[LCDA\label{sfig:LCDA_real_x}]{%
		\input{tikz/LCDA_real_x_m10}
	}\hfill
	\subfloat[Imaginary part of LCDA\label{sfig:LCDA_imag_x}]{%
		\input{tikz/LCDA_imag_x_m10}
	}\\
	\def\figWidth{\textwidth/2.5}
	\subfloat[LCDA for $\uscale = 0.5$\label{sfig:LCDA_real_x_u5}]{%
		%!TEX root =  ../Schwinger_PDF_LCDA.tex
% Generated by writeTikzPlot.m -- do not edit data by hand.
\begin{tikzpicture}
\begin{axis}[%
width=\figWidth,
height=\figHeight,
at={(0\figWidth,0\figHeight)},
scale only axis,
xmin=-0.003,
xmax=0.13416407865,
xlabel style={font=\color{white!15!black}},
xlabel={$1 \big/ \sqrt{x}$},
ylabel={$\Re\left(\lcda\right)$},
ymin=4.00226733977,
ymax=4.07445509561,
ylabel style={font=\color{white!15!black}},
ylabel style={{rotate=-90}}, scaled ticks=false, tick label style={/pgf/number format/fixed},
axis background/.style={fill=white},
xmajorgrids,
ymajorgrids,
legend style={at={(0,-0.2)}, anchor=north west, legend cell align=left, cells={align=left}, align=left, draw=white!15!black},
xticklabel style={yshift=-.5ex, /pgf/number format/fixed, /pgf/number format/precision=3},
yticklabel style={xshift=-.5ex, /pgf/number format/fixed, /pgf/number format/precision=4}
]
\addplot [myline2, only marks, forget plot]
  table[row sep=crcr]{%
0.111803398874989	4.0590130450909\\
};
\addplot [myline1, only marks, forget plot]
  table[row sep=crcr]{%
0.1	4.05320501259695\\
};
\addplot [myline3, only marks, forget plot]
  table[row sep=crcr]{%
0.0707106781186548	4.03786885558341\\
};
\addplot [myline4, only marks, forget plot]
  table[row sep=crcr]{%
0.0577350269189626	4.0318523965971\\
};
\addplot [myline5, only marks, forget plot]
  table[row sep=crcr]{%
0.0447213595499958	4.02803038371427\\
};
\addplot [myline6, only marks, forget plot]
  table[row sep=crcr]{%
0.0408248290463863	4.02587284865788\\
};
\addplot [myline11, only marks, forget plot]
  table[row sep=crcr]{%
0.0316227766016838	4.01995389565116\\
};
\addplot [myline8, mark=none]
  table[row sep=crcr]{%
0	4.00554860139661\\
0.00135519271363624	4.00619137770864\\
0.00271038542727247	4.00683415402068\\
0.00406557814090871	4.00747693033271\\
0.00542077085454494	4.00811970664474\\
0.00677596356818118	4.00876248295677\\
0.00813115628181742	4.00940525926881\\
0.00948634899545365	4.01004803558084\\
0.0108415417090899	4.01069081189287\\
0.0121967344227261	4.0113335882049\\
0.0135519271363624	4.01197636451694\\
0.0149071198499986	4.01261914082897\\
0.0162623125636348	4.013261917141\\
0.0176175052772711	4.01390469345303\\
0.0189726979909073	4.01454746976507\\
0.0203278907045435	4.0151902460771\\
0.0216830834181798	4.01583302238913\\
0.023038276131816	4.01647579870116\\
0.0243934688454522	4.0171185750132\\
0.0257486615590885	4.01776135132523\\
0.0271038542727247	4.01840412763726\\
0.028459046986361	4.01904690394929\\
0.0298142396999972	4.01968968026133\\
0.0311694324136334	4.02033245657336\\
0.0325246251272697	4.02097523288539\\
0.0338798178409059	4.02161800919742\\
0.0352350105545421	4.02226078550946\\
0.0365902032681784	4.02290356182149\\
0.0379453959818146	4.02354633813352\\
0.0393005886954508	4.02418911444555\\
0.0406557814090871	4.02483189075759\\
0.0420109741227233	4.02547466706962\\
0.0433661668363595	4.02611744338165\\
0.0447213595499958	4.02676021969368\\
0.046076552263632	4.02740299600572\\
0.0474317449772683	4.02804577231775\\
0.0487869376909045	4.02868854862978\\
0.0501421304045407	4.02933132494181\\
0.051497323118177	4.02997410125385\\
0.0528525158318132	4.03061687756588\\
0.0542077085454494	4.03125965387791\\
0.0555629012590857	4.03190243018994\\
0.0569180939727219	4.03254520650198\\
0.0582732866863582	4.03318798281401\\
0.0596284793999944	4.03383075912604\\
0.0609836721136306	4.03447353543807\\
0.0623388648272669	4.03511631175011\\
0.0636940575409031	4.03575908806214\\
0.0650492502545393	4.03640186437417\\
0.0664044429681756	4.0370446406862\\
0.0677596356818118	4.03768741699824\\
0.069114828395448	4.03833019331027\\
0.0704700211090843	4.0389729696223\\
0.0718252138227205	4.03961574593433\\
0.0731804065363568	4.04025852224637\\
0.074535599249993	4.0409012985584\\
0.0758907919636292	4.04154407487043\\
0.0772459846772654	4.04218685118246\\
0.0786011773909017	4.0428296274945\\
0.0799563701045379	4.04347240380653\\
0.0813115628181742	4.04411518011856\\
0.0826667555318104	4.04475795643059\\
0.0840219482454466	4.04540073274263\\
0.0853771409590829	4.04604350905466\\
0.0867323336727191	4.04668628536669\\
0.0880875263863553	4.04732906167872\\
0.0894427190999916	4.04797183799076\\
0.0907979118136278	4.04861461430279\\
0.092153104527264	4.04925739061482\\
0.0935082972409003	4.04990016692685\\
0.0948634899545365	4.05054294323889\\
0.0962186826681728	4.05118571955092\\
0.097573875381809	4.05182849586295\\
0.0989290680954452	4.05247127217498\\
0.100284260809081	4.05311404848702\\
0.101639453522718	4.05375682479905\\
0.102994646236354	4.05439960111108\\
0.10434983894999	4.05504237742311\\
0.105705031663626	4.05568515373515\\
0.107060224377263	4.05632793004718\\
0.108415417090899	4.05697070635921\\
0.109770609804535	4.05761348267124\\
0.111125802518171	4.05825625898328\\
0.112480995231808	4.05889903529531\\
0.113836187945444	4.05954181160734\\
0.11519138065908	4.06018458791937\\
0.116546573372716	4.06082736423141\\
0.117901766086353	4.06147014054344\\
0.119256958799989	4.06211291685547\\
0.120612151513625	4.0627556931675\\
0.121967344227261	4.06339846947954\\
0.123322536940897	4.06404124579157\\
0.124677729654534	4.0646840221036\\
0.12603292236817	4.06532679841563\\
0.127388115081806	4.06596957472767\\
0.128743307795442	4.0666123510397\\
0.130098500509079	4.06725512735173\\
0.131453693222715	4.06789790366376\\
0.132808885936351	4.0685406799758\\
0.134164078649987	4.06918345628783\\
};
\addlegendentry{fit1: $4.01+0.474 / \sqrt{x}$}
\addplot [myline9, mark=none]
  table[row sep=crcr]{%
0	4.00850914058383\\
0.00135519271363624	4.00902094632909\\
0.00271038542727247	4.009535224951\\
0.00406557814090871	4.01005197644955\\
0.00542077085454494	4.01057120082474\\
0.00677596356818118	4.01109289807659\\
0.00813115628181742	4.01161706820508\\
0.00948634899545365	4.01214371121021\\
0.0108415417090899	4.012672827092\\
0.0121967344227261	4.01320441585043\\
0.0135519271363624	4.0137384774855\\
0.0149071198499986	4.01427501199723\\
0.0162623125636348	4.0148140193856\\
0.0176175052772711	4.01535549965061\\
0.0189726979909073	4.01589945279228\\
0.0203278907045435	4.01644587881059\\
0.0216830834181798	4.01699477770554\\
0.023038276131816	4.01754614947715\\
0.0243934688454522	4.01809999412539\\
0.0257486615590885	4.01865631165029\\
0.0271038542727247	4.01921510205183\\
0.028459046986361	4.01977636533002\\
0.0298142396999972	4.02034010148486\\
0.0311694324136334	4.02090631051634\\
0.0325246251272697	4.02147499242447\\
0.0338798178409059	4.02204614720925\\
0.0352350105545421	4.02261977487067\\
0.0365902032681784	4.02319587540874\\
0.0379453959818146	4.02377444882345\\
0.0393005886954508	4.02435549511481\\
0.0406557814090871	4.02493901428282\\
0.0420109741227233	4.02552500632748\\
0.0433661668363595	4.02611347124878\\
0.0447213595499958	4.02670440904673\\
0.046076552263632	4.02729781972132\\
0.0474317449772683	4.02789370327256\\
0.0487869376909045	4.02849205970045\\
0.0501421304045407	4.02909288900499\\
0.051497323118177	4.02969619118617\\
0.0528525158318132	4.03030196624399\\
0.0542077085454494	4.03091021417847\\
0.0555629012590857	4.03152093498959\\
0.0569180939727219	4.03213412867736\\
0.0582732866863582	4.03274979524177\\
0.0596284793999944	4.03336793468283\\
0.0609836721136306	4.03398854700054\\
0.0623388648272669	4.03461163219489\\
0.0636940575409031	4.03523719026589\\
0.0650492502545393	4.03586522121354\\
0.0664044429681756	4.03649572503783\\
0.0677596356818118	4.03712870173877\\
0.069114828395448	4.03776415131636\\
0.0704700211090843	4.03840207377059\\
0.0718252138227205	4.03904246910148\\
0.0731804065363568	4.039685337309\\
0.074535599249993	4.04033067839317\\
0.0758907919636292	4.04097849235399\\
0.0772459846772654	4.04162877919146\\
0.0786011773909017	4.04228153890557\\
0.0799563701045379	4.04293677149633\\
0.0813115628181742	4.04359447696374\\
0.0826667555318104	4.04425465530779\\
0.0840219482454466	4.04491730652849\\
0.0853771409590829	4.04558243062584\\
0.0867323336727191	4.04625002759983\\
0.0880875263863553	4.04692009745047\\
0.0894427190999916	4.04759264017775\\
0.0907979118136278	4.04826765578169\\
0.092153104527264	4.04894514426227\\
0.0935082972409003	4.04962510561949\\
0.0948634899545365	4.05030753985336\\
0.0962186826681728	4.05099244696388\\
0.097573875381809	4.05167982695105\\
0.0989290680954452	4.05236967981486\\
0.100284260809081	4.05306200555532\\
0.101639453522718	4.05375680417242\\
0.102994646236354	4.05445407566617\\
0.10434983894999	4.05515382003657\\
0.105705031663626	4.05585603728362\\
0.107060224377263	4.05656072740731\\
0.108415417090899	4.05726789040765\\
0.109770609804535	4.05797752628463\\
0.111125802518171	4.05868963503826\\
0.112480995231808	4.05940421666854\\
0.113836187945444	4.06012127117546\\
0.11519138065908	4.06084079855903\\
0.116546573372716	4.06156279881925\\
0.117901766086353	4.06228727195612\\
0.119256958799989	4.06301421796963\\
0.120612151513625	4.06374363685978\\
0.121967344227261	4.06447552862659\\
0.123322536940897	4.06520989327004\\
0.124677729654534	4.06594673079013\\
0.12603292236817	4.06668604118688\\
0.127388115081806	4.06742782446027\\
0.128743307795442	4.0681720806103\\
0.130098500509079	4.06891880963699\\
0.131453693222715	4.06966801154032\\
0.132808885936351	4.07041968632029\\
0.134164078649987	4.07117383397692\\
};
\addlegendentry{fit2: $4.01+0.377 / \sqrt{x}+0.673 / x$}
\addplot [myline7, solid, only marks, error bar legend]
 plot [error bars/.cd, y dir=both, y explicit, error bar style={line width=\tikzlinewidth}, error mark options={line width=\tikzlinewidth, mark size=\tikzmarksize, rotate=90, mycolor7}]
 table[row sep=crcr, y error plus index=2, y error minus index=3]{%
0	4.02937680699678	0.0238282056001689	0.0238282056001697\\
};
\addlegendentry{extrapolation $x \rightarrow \infty$}
\end{axis}
\end{tikzpicture}%
	}
	\hfill
	\subfloat[Imaginary part of LCDA for $\uscale = 0.5$\label{sfig:LCDA_imag_x_u5}]{%
		\input{tikz/LCDA_imag_x_u5_m10}
	}
	\\
	\subfloat[LCDA for $\uscale = 0.3$\label{sfig:LCDA_real_x_u3}]{%
		\input{tikz/LCDA_real_x_u3_m10}
	}
	\hfill
	\subfloat[Imaginary part of LCDA for $\uscale = 0.3$\label{sfig:LCDA_imag_x_u3}]{%
		\input{tikz/LCDA_imag_x_u3_m10}
	}
	\caption{Dependence of LCDA on $x$ at $\mass=10$ and with base parameters as shown in \cref{tab:sim_params}. Linear and quadratic fits in $\frac{1}{\sqrt{x}}$ (proportional to lattice spacing $a$) extrapolate to $x \rightarrow \infty$. The midpoint between the smallest and largest values with $x \ge x^b$ (including extrapolations) is shown as olive dotted curves in \protect\subref{sfig:LCDA_real_x} and \protect\subref{sfig:LCDA_imag_x}, and as pentagons at $\frac{1}{\sqrt{x}}=0$ in \protect\subref{sfig:LCDA_real_x_u5} -- \protect\subref{sfig:LCDA_imag_x_u3}. Error bars/bands are assigned such that they are compatible with all values $x \ge x^b$. Insets of \protect\subref{sfig:LCDA_real_x} and \protect\subref{sfig:LCDA_imag_x}: data points from a discrete Fourier transform are shown on a curve obtained by a continuous Fourier transform.}
	\label{fig:x_dependence}
\end{figure*}

\subsection{Volume dependence}

\begin{figure*}[htbp]
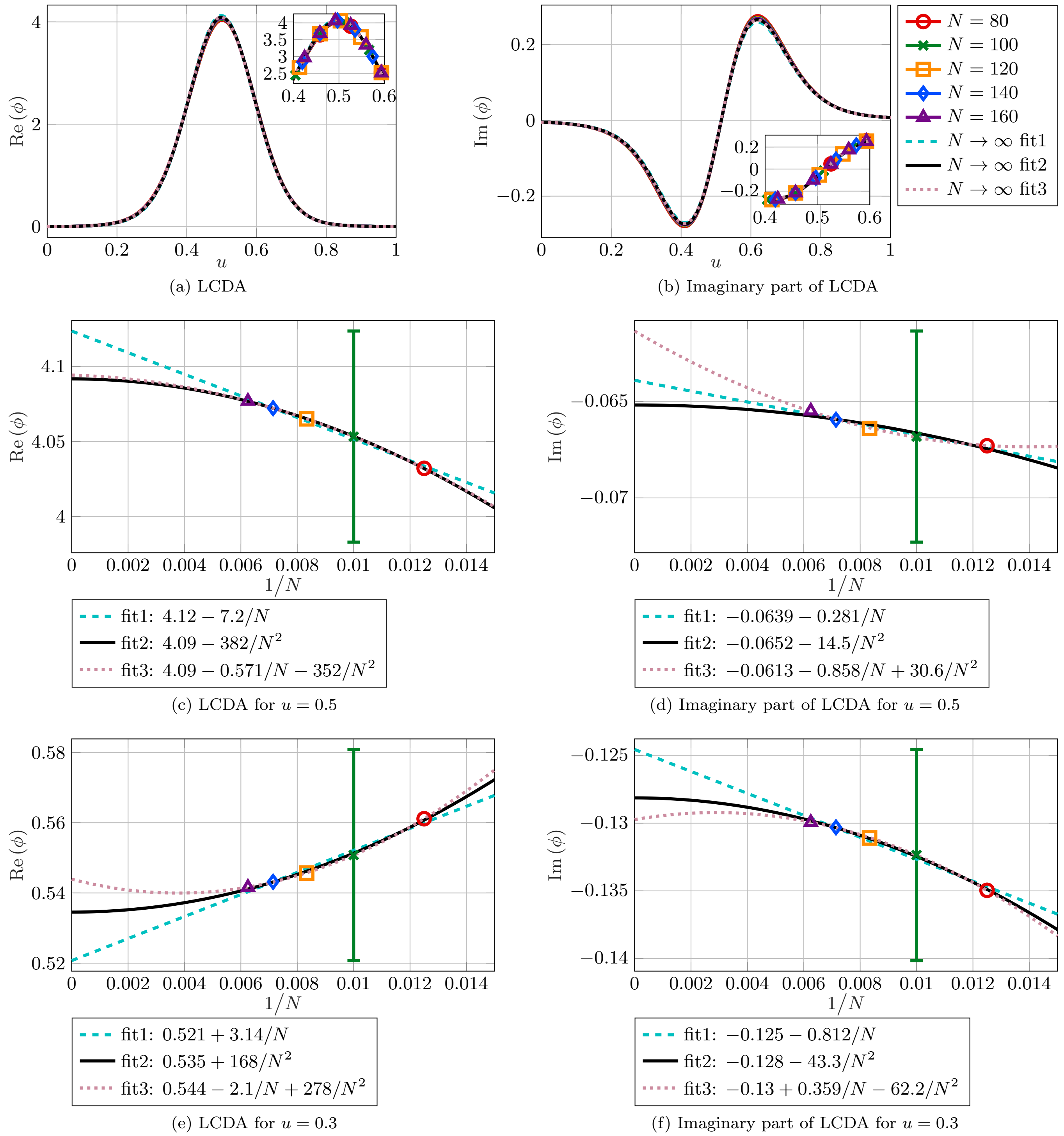

	\centering
	\def\figWidth{.33*\textwidth}
	\def\figHeight{\textheight/6}
	\subfloat[LCDA\label{sfig:LCDA_real_N}]{%
		\input{tikz/LCDA_real_N_m10}
	}
	\hfill
	\subfloat[Imaginary part of LCDA\label{sfig:LCDA_imag_N}]{%
		\input{tikz/LCDA_imag_N_m10}
	}
	\\
	\def\figWidth{\textwidth/2.5}
	\subfloat[LCDA for $\uscale = 0.5$\label{sfig:LCDA_real_N_u5}]{%
		\input{tikz/LCDA_real_N_u5_m10}
	}
	\hfill
	\subfloat[Imaginary part of LCDA for $\uscale = 0.5$\label{sfig:LCDA_imag_N_u5}]{%
		\input{tikz/LCDA_imag_N_u5_m10}
	}
	\\
	\subfloat[LCDA for $\uscale = 0.3$\label{sfig:LCDA_real_N_u3}]{%
		\input{tikz/LCDA_real_N_u3_m10}
	}
	\hfill
	\subfloat[Imaginary part of LCDA for $\uscale = 0.3$\label{sfig:LCDA_imag_N_u3}]{%
		\input{tikz/LCDA_imag_N_u3_m10}
	}
	\caption{Dependence of LCDA on the number of lattice sites $N$ at fixed lattice spacing for $\mass=10$ and with base parameters as shown in \cref{tab:sim_params}. Fits in $\frac{1}{N}$ extrapolate to $N \rightarrow \infty$ (fit1: linear; fit2: quadratic; fit3: linear and quadratic terms). Symmetric error bars at the base point $N^b=100$ are assigned that include all data points with $N \ge N^b$ and the extrapolations. Insets of \protect\subref{sfig:LCDA_real_N} and \protect\subref{sfig:LCDA_imag_N}: data points from a discrete Fourier transform are shown on a curve obtained by a continuous Fourier transform.}
	\label{fig:N_dependence}
\end{figure*}

We vary the number of lattice sites at fixed lattice spacing as shown in \cref{fig:N_dependence} and extrapolate PDFs and LCDAs to the infinite volume limit (fits in $\frac{1}{N}$; fit1: linear; fit2: quadratic; fit3: linear plus quadratic). The LCDA uncertainty at the base value of \vol [see \cref{tab:sim_params}] is estimated from the maximal spread between the base value and values at larger volumes, including the extrapolated points. Since the volume effects are more significant for PDFs compared to LCDAs, we report PDF results in \cref{fig:PDF_fermionmass}, which follow the same midpoint procedure explained for the $x$-dependence of LCDAs above.

\subsection{Total error}

All uncertainties are combined in quadrature to obtain the error bands in \cref{fig:PDF_fermionmass,fig:LCDA_fermionmass}. Both real and imaginary parts of the uncertainties are included, such that deviations from a real PDF or LCDA are included in the uncertainties. The individual contributions to the uncertainties are shown in \cref{fig:PDF_errors} for the PDFs and in \cref{fig:LCDA_errors} for the LCDAs. Finite volume effects and lattice artifacts dominate the errors for our choice of parameters. The Trotterization has a sub-leading effect, while the gauge truncation is negligible. The finite bond dimension is only relevant for the antifermion PDF with $\mass = 2.5$. The LCDAs are more sensitive to the lattice spacing compared to the PDFs, which depend more strongly on the volume (notice that the uncertainties related to \vol and $x$ of the PDFs and LCDAs, respectively, are reduced by the midpoint procedure in \cref{fig:PDF_errors,fig:LCDA_errors}; they would be up to twice as large at the base point).

\begin{figure*}[htbp]
	\centering
	\def\figWidth{0.9*\textwidth}
	\def\figHeight{\textheight/12.2}
	\subfloat[$\mass = 1$\label{sfig:PDF_errors_m1}]{%
		\input{tikz/PDF_errors_m1}
	}\\
	\subfloat[$\mass = 2.5$\label{sfig:PDF_errors_m2_5}]{%
		\input{tikz/PDF_errors_m2_5}
	}\\
	\subfloat[$\mass = 5$\label{sfig:PDF_errors_m5}]{%
		\input{tikz/PDF_errors_m5}
	}\\
	\subfloat[$\mass = 10$\label{sfig:PDF_errors_m10}]{%
		\input{tikz/PDF_errors_m10}
	}\\
	\subfloat[$\mass = 20$\label{sfig:PDF_errors_m20}]{%
		\input{tikz/PDF_errors_m20}
	}\\
	\subfloat[$\mass = 40$\label{sfig:PDF_errors_m40}]{%
		\input{tikz/PDF_errors_m40}
	}
	\caption{Contributions to the uncertainties of PDFs due to finite bond dimension $D$, gauge-field truncation $\Lcut$, Trotterization $\Nt$, lattice spacing ($\propto \frac{1}{\sqrt{x}}$), and volume $\vol$ for different fermion masses $\mass$. The total uncertainties are obtained as the square root of the sum of squares. The $\vol$ contribution is obtained from the midpoint procedure, while the other uncertainties are estimated at the base point.}
	\label{fig:PDF_errors}
\end{figure*}

\begin{figure*}[htbp]
	\centering
	\def\figWidth{0.9*\textwidth}
	\def\figHeight{\textheight/11.1}
	\subfloat[$\mass = 1$\label{sfig:LCDA_errors_m1}]{%
		\input{tikz/LCDA_errors_m1}
	}\\
	\subfloat[$\mass = 2.5$\label{sfig:LCDA_errors_m2_5}]{%
		\input{tikz/LCDA_errors_m2_5}
	}\\
	\subfloat[$\mass = 5$\label{sfig:LCDA_errors_m5}]{%
		\input{tikz/LCDA_errors_m5}
	}\\
	\subfloat[$\mass = 10$\label{sfig:LCDA_errors_m10}]{%
		\input{tikz/LCDA_errors_m10}
	}\\
	\subfloat[$\mass = 20$\label{sfig:LCDA_errors_m20}]{%
		\input{tikz/LCDA_errors_m20}
	}\\
	\subfloat[$\mass = 40$\label{sfig:LCDA_errors_m40}]{%
		\input{tikz/LCDA_errors_m40}
	}
	\caption{Contributions to the uncertainties of LCDAs due to finite bond dimension $D$, gauge-field truncation $\Lcut$, Trotterization $\Nt$, lattice spacing ($\propto \frac{1}{\sqrt{x}}$), and volume $\vol$ for different fermion masses $\mass$. The total uncertainties are obtained as the square root of the sum of squares. The $x$ contribution is obtained from the midpoint procedure, while the other uncertainties are estimated at the base point.}
	\label{fig:LCDA_errors}
\end{figure*}

\subsection{Decay constant, norms and average momenta}

The scalar observables reported in \cref{tab:scalars} are obtained from the same data as the PDFs and LCDAs and similar extrapolations. The decay constant is proportional to $\mel_{\text{LCDA}}(0)$, and the uncertainty estimates of matrix elements for the $D$-, $\Lcut$-, and $\Nt$-dependence described above are applied. The $x$- and $\vol$-dependencies of the decay constant are treated the same way as those of the LCDAs: fits extrapolate in $x$ (linear, quadratic in the lattice spacing). The midpoint procedure is applied to obtain the reported value and the uncertainty of this continuum limit. Furthermore, fits in $\frac{1}{\vol}$ are performed (linear, quadratic, both linear and quadratic). The uncertainty due to finite volume is assigned by symmetric error bars that are compatible with all $\vol \ge \vol^b$, including extrapolations. All uncertainties are combined in quadrature.

The PDF norms $\normpsipsibar$ and average parton momenta $P_{\psi / \overline{\psi}}$ are calculated directly from the matrix elements by using the analytic forms of the integrals $\int_0^1 d\bjorx e^{\kappa \siteeven \bjorx}$ and $\int_0^1 d\bjorx e^{\kappa \siteeven \bjorx} \bjorx$ with $\kappa = -i \frac{\energy}{2x}$:
\begin{align}
	\normpsipsibar &= \int_0^1 d\bjorx \pdfantipdf , \nonumber\\
	\normpsi &= \frac{1}{\pi \sqrt{2x}} \sum_{\siteeven=\text{even}} \mel_{\text{PDF}}(\siteeven) \int_0^1 d\bjorx e^{\kappa \siteeven \bjorx} \nonumber\\
	&= \frac{1}{\pi \sqrt{2x}} \sum_{\siteeven=\text{even}, \siteeven \ne 0} \mel_{\text{PDF}}(\siteeven) \frac{e^{\kappa \siteeven} - 1}{\kappa \siteeven} , \\
	\normpsibar &= \frac{1}{\pi \sqrt{2x}} \sum_{\siteeven=\text{even}} \mel_{\text{PDF}}(\siteeven) \left( -\int_0^1 d\bjorx e^{-\kappa \siteeven \bjorx} \right) \nonumber\\
	&= \frac{1}{\pi \sqrt{2x}} \sum_{\siteeven=\text{even}, \siteeven \ne 0} \mel_{\text{PDF}}(\siteeven) \frac{e^{-\kappa \siteeven} - 1}{\kappa \siteeven} ,
\end{align}
\begin{align}
	P_{\psi / \overline{\psi}} &= \int_0^1 d\bjorx \pdfantipdf \bjorx , \nonumber\\
	P_{\psi} &= \frac{1}{\pi \sqrt{2x}} \sum_{\siteeven=\text{even}, \siteeven \ne 0} \mel_{\text{PDF}}(\siteeven) \frac{e^{\kappa \siteeven} \left( 1-\frac{1}{\kappa \siteeven} \right)+\frac{1}{\kappa \siteeven}}{\kappa \siteeven} , \\
	P_{\overline{\psi}} &= \frac{1}{\pi \sqrt{2x}} \sum_{\siteeven=\text{even}, \siteeven \ne 0} \mel_{\text{PDF}}(\siteeven) \frac{e^{-\kappa \siteeven} \left( 1+\frac{1}{\kappa \siteeven} \right)-\frac{1}{\kappa \siteeven}}{\kappa \siteeven} .
\end{align}

The matrix elements $\mel_{\text{PDF}}(0)$ vanish in the exact theory because of symmetry, and to a good approximation numerically, and are therefore excluded in the sums.

Each of these scalar observables is calculated at fixed parameters and from the extrapolated matrix elements to obtain the uncertainties in $D$, \Lcut and \Nt. The uncertainties due to the lattice spacing and volume effects are obtained similarly to those for the decay constant. However, the PDF-related scalar observables $\normpsipsibar$ and $P_{\psi / \overline{\psi}}$ are obtained by the midpoint extrapolation in the volume, while this procedure is applied in $x$ for the LCDA-related decay constant. The reported final values are these midpoints with the uncertainties due to all simulation parameters combined in quadrature.

\end{document}